\documentclass[conference]{llncs}

\usepackage{bsymb,b2latex}

\usepackage[top=2.5cm, bottom=2.0cm, left=2.5cm, right=2.5cm]{geometry}

\usepackage{cite}
\usepackage{amsmath,amssymb,amsfonts}
\usepackage{algorithmic}
\usepackage{graphicx}
\usepackage{oz}
\usepackage{textcomp}
\def\BibTeX{{\rm B\kern-.05em{\sc i\kern-.025em b}\kern-.08em
    T\kern-.1667em\lower.7ex\hbox{E}\kern-.125emX}}

\usepackage{mathtools}

\usepackage{listings}

\usepackage[utf8]{inputenc}
\usepackage[T1]{fontenc}
\usepackage{microtype}

\usepackage{color}
\definecolor{gray}{rgb}{0.4,0.4,0.4}
\definecolor{darkblue}{rgb}{0.0,0.0,0.6}
\definecolor{cyan}{rgb}{0.0,0.6,0.6}
\definecolor{keycolor}{rgb}{0,0,0.8}     
\definecolor{labelcolor}{rgb}{0,0.4,0.8} 
\definecolor{codecolor}{rgb}{0,0,0}      
\definecolor{inhcolor}{rgb}{0.6,0.2,0}   
\definecolor{cmtcolor}{rgb}{0,0.4,0}     

\usepackage{listings}

\usepackage{color}
\definecolor{gray}{rgb}{0.4,0.4,0.4}
\definecolor{darkblue}{rgb}{0.0,0.0,1.0}
\definecolor{cyan}{rgb}{0.0,0.6,0.6}

\lstdefinelanguage{MPS}
{
  morestring=[b]",
  morestring=[s]
  ,
  morecomment=[s]{/*}{*/},
  stringstyle=\color{black},
  identifierstyle=\color{black},
  keywordstyle=\color{darkblue},
  morekeywords={domain,model,parent,concepts,concept,is,variable,
  individuals,deduced,attribute,domain,dom,relations,relation,attributes,
  enumerated,elements,
  range,functional,total,maplets,custom,data,sets,set,values,value,type,lexical,form,
  predicates,p1,p2,not}
  ,
  otherkeywords = {=,&,(,),\{,\},>,<,",:,?},
}

\makeatletter
\newcommand\footnoteref[1]{\protected@xdef\@thefnmark{\ref{#1}}\@footnotemark}
\makeatother

\usepackage[bookmarks=false,hidelinks]{hyperref}

\usepackage{breqn}
\usepackage{longtable}

\ExplSyntaxOn
\cs_set_eq:NN \int_eval:w \__int_eval:w
\cs_set_eq:NN \int_eval_end: \__int_eval_end:
\ExplSyntaxOff

\usepackage[linewidth=1pt]{mdframed}
\usepackage{multicol}

\usepackage{flushend}

\begin{document}
%

\pagestyle{headings}

\mainmatter

\title{SysML/KAOS Domain Models and B System Specifications}
\titlerunning{SysML/KAOS Domain Models and B System Specifications}  
%
\author{Steve Jeffrey Tueno Fotso\inst{1, 3} \and Marc Frappier\inst{3}  \and Amel Mammar\inst{2} \and Régine Laleau\inst{1} }
\authorrunning{Steve Tueno et al.} 
%
\tocauthor{Steve Tueno,  Marc Frappier, Amel Mammar, Régine Laleau}
\institute{Université Paris-Est Créteil, LACL, Créteil, France,\\
\email{steve.tuenofotso@univ-paris-est.fr},
\email{laleau@u-pec.fr}
\and
Télécom SudParis, SAMOVAR-CNRS, Evry, France, \\
\email{amel.mammar@telecom-sudparis.eu}
\and
Université de Sherbrooke, GRIL, Québec, Canada, \\
\email{Marc.Frappier@usherbrooke.ca}
\email{Steve.Jeffrey.Tueno.Fotso@USherbrooke.ca}}

\maketitle              

\begin{abstract}
In this paper, we use a combination of the \textit{SysML/KAOS} requirements engineering method, an extension of \textit{SysML}, with concepts of the \textit{KAOS} goal model, and of the \textit{B System} formal method. Translation rules from a SysML/KAOS goal model to a B System specification have  been defined. They allow to obtain a skeleton of the B System specification. To complete it, we have defined a language  to express the domain model associated to the goal model. 
The translation of this domain model gives the structural part of the B System specification. The contribution of this paper is the description of translation rules from SysML/KAOS domain models to B System specifications.  We also present the formal verification of these rules and we describe an open source tool that implements the languages and the rules. Finally, we provide a  review of the application of the SysML/KAOS method on case studies such as for the formal specification of the \textit{hybrid ERTMS/ETCS level 3} standard.

\end{abstract}



%

\section{Context}

\subsection{SysML/KAOS}
Requirements engineering focuses on elicitation, analysis, verification and validation of requirements. 
The \textit{KAOS} method \cite{DBLP:books/daglib/0025377}
  proposes to represent the requirements in the form of goals described through five sub-models of which the two main ones are:
  the \textbf{goal model} for the representation of  requirements to be satisfied by the system and of expectations with regard to the environment through a hierarchy of goals  and   
 the \textbf{object model} which uses the \textit{UML} class diagram for the representation of the domain  vocabulary.
The hierarchy is built through a succession of refinements using two main operators: \textbf{\textit{AND}} and  \textbf{\textit{OR}}.
An \textit{\textbf{AND refinement}} decomposes a goal into subgoals, and all of them must be achieved to realise the parent goal.  Dually, an \textit{\textbf{OR refinement}} decomposes a goal into subgoals
such that the achievement of only one of them is sufficient for the accomplishment of the parent goal.
Requirements and expectations correspond to the lowest level goals of the  model.


KAOS  offers no mechanism to maintain a strong traceability between    requirements and  design deliverables, making it difficult to validate them against the needs formulated.
 \textit{SysML/KAOS} \cite{DBLP:conf/inforsid/GnahoS10,DBLP:conf/isola/MammarL16} addresses this issue by adding to KAOS the \textit{SysML} UML profile specially designed by the Object Management Group (OMG)  for the analysis and specification of complex systems. 
\textit{SysML}  allows for the capturing of requirements and the maintaining of   traceability links between those requirements and  design deliverables, but it does  not define  a precise syntax for requirements specification.  SysML/KAOS therefore proposes to extend the SysML metamodel in order  to allow the  representation of   
  requirements  using the KAOS  expressivity.



\begin{figure}[!h]
\begin{center}
\includegraphics[width=0.7\textwidth]{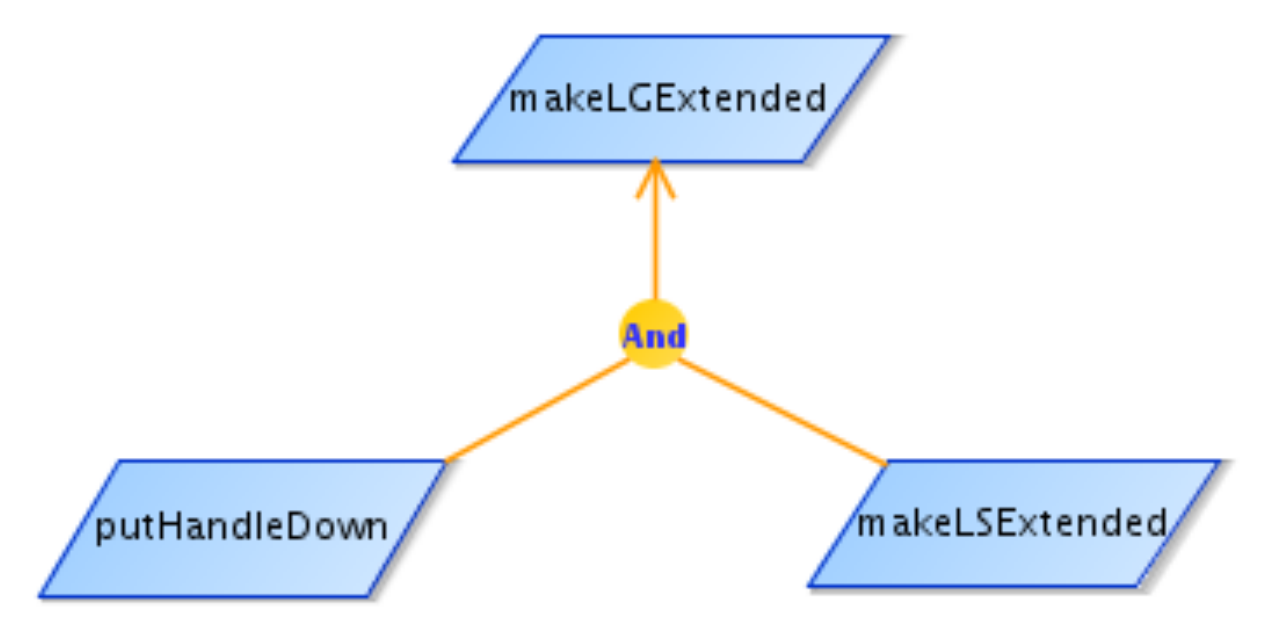}
\end{center}
\caption{\label{lgsystem_goal_model_makelgextended} Excerpt from the landing gear system goal diagram}
\end{figure}

Our case study deals with the landing gear system  of an aircraft which can be retracted (respectively extended) through a handle \cite{Boniol2014}.
Figure \ref{lgsystem_goal_model_makelgextended} is an excerpt from its goal diagram focused on the purpose of  landing gear expansion (\textbf{makeLGExtended}). To achieve it, the handle must be put down  (\textbf{putHandleDown}) and  landing gear sets must be extended  (\textbf{makeLSExtended}).  We assume that each aircraft has one landing gear system.

\subsection{Event-B and B System} \label{event_b_description_section}
\textit{Event-B} is an industrial-strength formal method 
for \emph{system modeling} \cite{DBLP:books/daglib/0024570}. 
It is used to
incrementally construct a system specification, using refinement, and to 
 prove   properties. 
%
%
An \textit{Event-B} model includes a static part called \textsf{context} and a dynamic part called \textsf{machine}.
The \textsf{context} contains
 the declarations of abstract and enumerated sets, 
  constants and 
   axioms. 
 The   \textsf{machine} contains variables, invariants and events. 
A machine can refine another machine,  a context can extend others contexts and a machine can see contexts. 
Gluing invariants are invariants that capture
links between variables  defined within a machine and those appearing in more abstract ones.
\textit{B System} is  an  \textit{Event-B} syntactic variant   proposed by \textit{ClearSy}, an industrial partner in the \textit{FORMOSE} project \cite{anr_formose_reference_link}, and supported by   \textit{Atelier B} \cite{clearsy_b_system_link}.

\begin{figure*}[!h]
\begin{center}
\includegraphics[width=1.\textwidth]{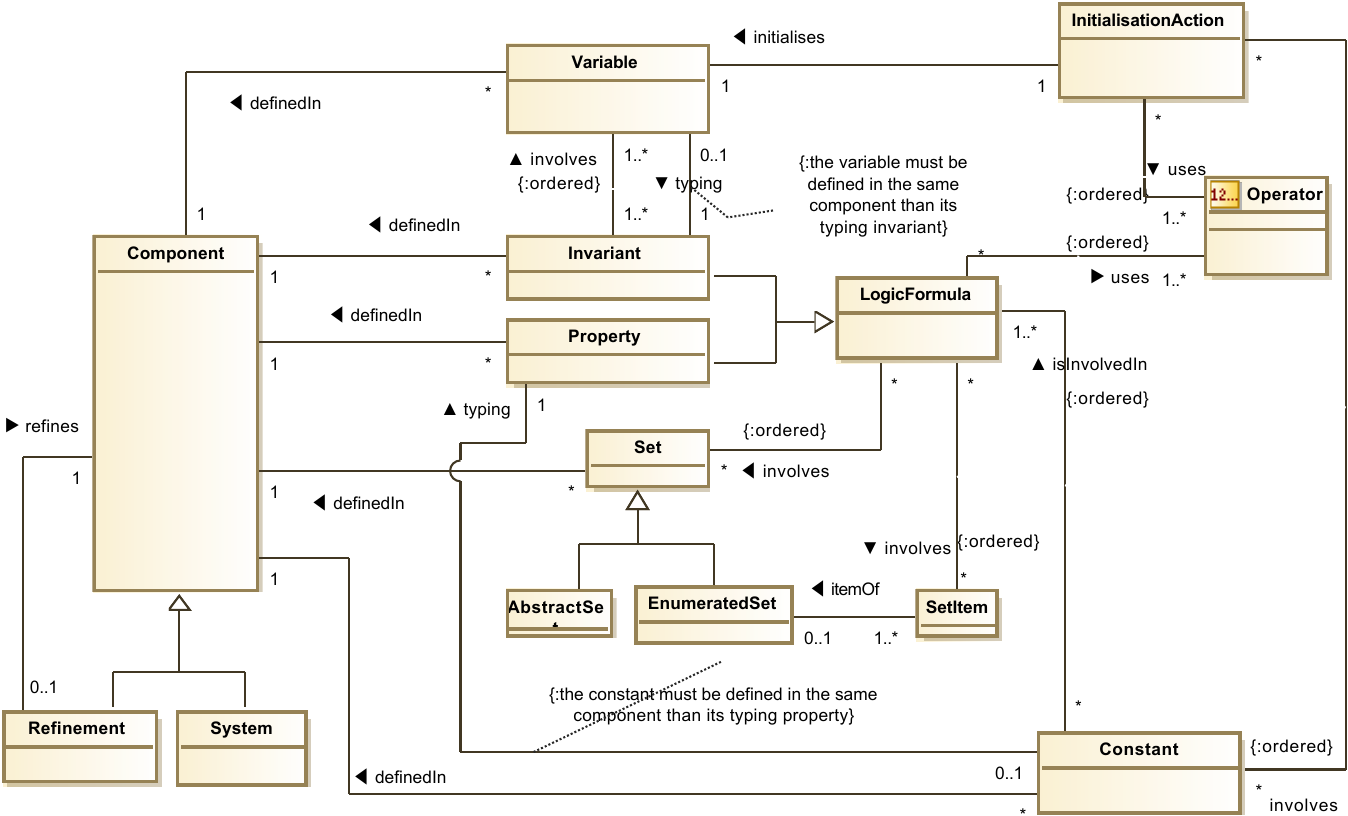}
\end{center}
\caption{\label{eventb_metamodel} Excerpt from a metamodel of the \textit{B System} specification language}
\end{figure*}

 Figure \ref{eventb_metamodel} is 
a metamodel of the \textit{B System} language restricted to  concepts that are relevant to us.
A \textit{B System} specification
consists of components (instances of \textsf{Component}). Each component  can be either a system or a refinement and it   may define static or dynamic elements. A refinement is a  component which refines another one in order  to access the elements defined in it and to reuse them for new constructions. 
Constants, abstract and enumerated sets, and their properties, constitute the static part.
 The dynamic part includes the representation of the  system state   using variables constrained through invariants and initialised through initialisation actions.  Properties and invariants can be categorised as instances of \textsf{LogicFormula}. 
 In our case, it is sufficient to consider  that logic formulas are    successions of operands in relation through operators. Thus,  
  an instance of  \textsf{LogicFormula} references  
its operators (instances of \textsf{Operator}) and its operands that may be instances of \textsf{Variable}, \textsf{Constant},  \textsf{Set} or \textsf{SetItem}. 
Operators   include, but not limited to \footnote{The full list can be found in \cite{1712.07406}},  \textbf{\textit{Inclusion_OP}}  which is used to assert that the first operand is a subset of the second operand ($(Inclusion\_OP, [op_1, op_2]) \Leftrightarrow op_1 \subseteq op_2$) and  \textbf{\textit{Belonging_OP}} which  is used to assert that the first operand is an element of the second operand ($(Belonging\_OP, [op_1, op_2]) \Leftrightarrow op_1 \in op_2$)  and \textbf{\textit{BecomeEqual2SetOf_OP}} which is used to initialize a variable as a set of elements ($(BecEq2Set\_OP,  [va, op_2, ...,  op_n]) \Leftrightarrow va :=  \{op_2, ..., op_n\}$).
\\
In the rest of this paper, \textit{target} is used in place of  \textit{B System}. 

\subsection{Formalisation of SysML/KAOS Goal Models} \label{goal_model_formalisation}
The formalisation of SysML/KAOS goal models   is the focus of the work done by \cite{DBLP:conf/iceccs/MatoussiGL11}. The proposed rules allow the generation of a formal model whose structure reflects the hierarchy of the SysML/KAOS goal diagram :  one  component is associated with each hierarchy level; this component defines one event for each goal. 
The semantics of refinement links between goals is expressed in the formal specification with a  set of  proof obligations which complement the standard  proof obligations for \textit{invariant preservation} and for \textit{event actions feasibility} \cite{DBLP:books/daglib/0024570}.
Regarding the new proof obligations, they depend on the goal refinement  operator  used.  
For an abstract goal $G$ and two concrete goals $G_1$ and $G_2$ : \footnote{For an event \texttt{G}, \texttt{G\_Guard} represents the guards of G and \texttt{G\_Post} represents the post condition of its actions.}
\begin{itemize}
\item For the \textit{AND} operator, the proof obligations are
\begin{itemize}
\item[•] $G_1\_Guard \Rightarrow G\_Guard$  
\item[•] $G_2\_Guard \Rightarrow G\_Guard$
\item[•] $(G_1\_Post \wedge G_2\_Post) \Rightarrow G\_Post$
\end{itemize}
\item For the \textit{OR} operator, they are
\begin{itemize}
\item[•] $G_1\_Guard \Rightarrow G\_Guard$
\item[•] $G_2\_Guard \Rightarrow G\_Guard$
\item[•] $G_1\_Post \Rightarrow G\_Post$
\item[•] $G_2\_Post \Rightarrow G\_Post$
\item[•] $G_1\_Post \Rightarrow \neg G_2\_Guard$
\item[•] $G_2\_Post \Rightarrow \neg G_1\_Guard$
\end{itemize}
\item For the \textit{MILESTONE} operator, they are
\begin{itemize}
\item[•] $G_1\_Guard \Rightarrow G\_Guard$
\item[•]  $G_2\_Post \Rightarrow G\_Post$
\item[•] $\square(G1\_Post \Rightarrow \lozenge G2\_Guard)$ (each system state, corresponding to the post condition of $G\_1$, must be followed, at least once in the future, by a system state enabling $G\_2$)
\end{itemize}
\end{itemize}

Nevertheless, the generated  specification does not contain the system structure, composed of variables, constrained by an invariant, and constants, constrained by properties.

\subsection{The SysML/KAOS Domain Modeling Language}
Domain models in SysML/KAOS are represented using ontologies.  These ontologies are expressed using
the SysML/KAOS domain modeling language \cite{1710.00903,sysml_kaos_domain_modeling},   built  
 based on \textit{OWL} \cite{DBLP:reference/snam/SenguptaH14} and \textit{PLIB} \cite{DBLP:conf/ifip/Pierra04}, two well-known and complementary ontology modeling formalisms.

\begin{figure}[!h]
\begin{center}
\includegraphics[width=0.5\textwidth]{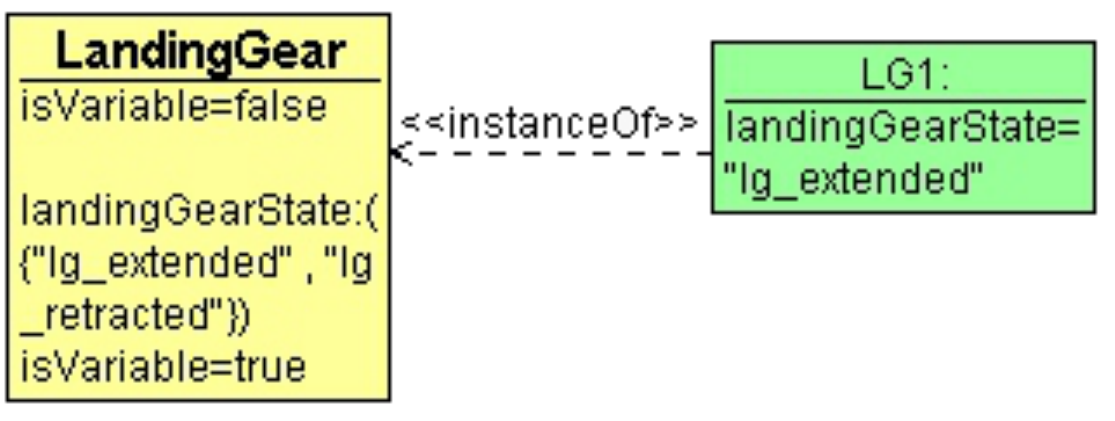}
\caption{\label{lgsystem_refinment_0_ontology} \textit{\textbf{lg\_system\_ref\_0}}: ontology associated with the root level of the landing gear goal model}
\end{center}
\end{figure}


\begin{figure}[!h]
\begin{center}
\includegraphics[width=1.\textwidth]{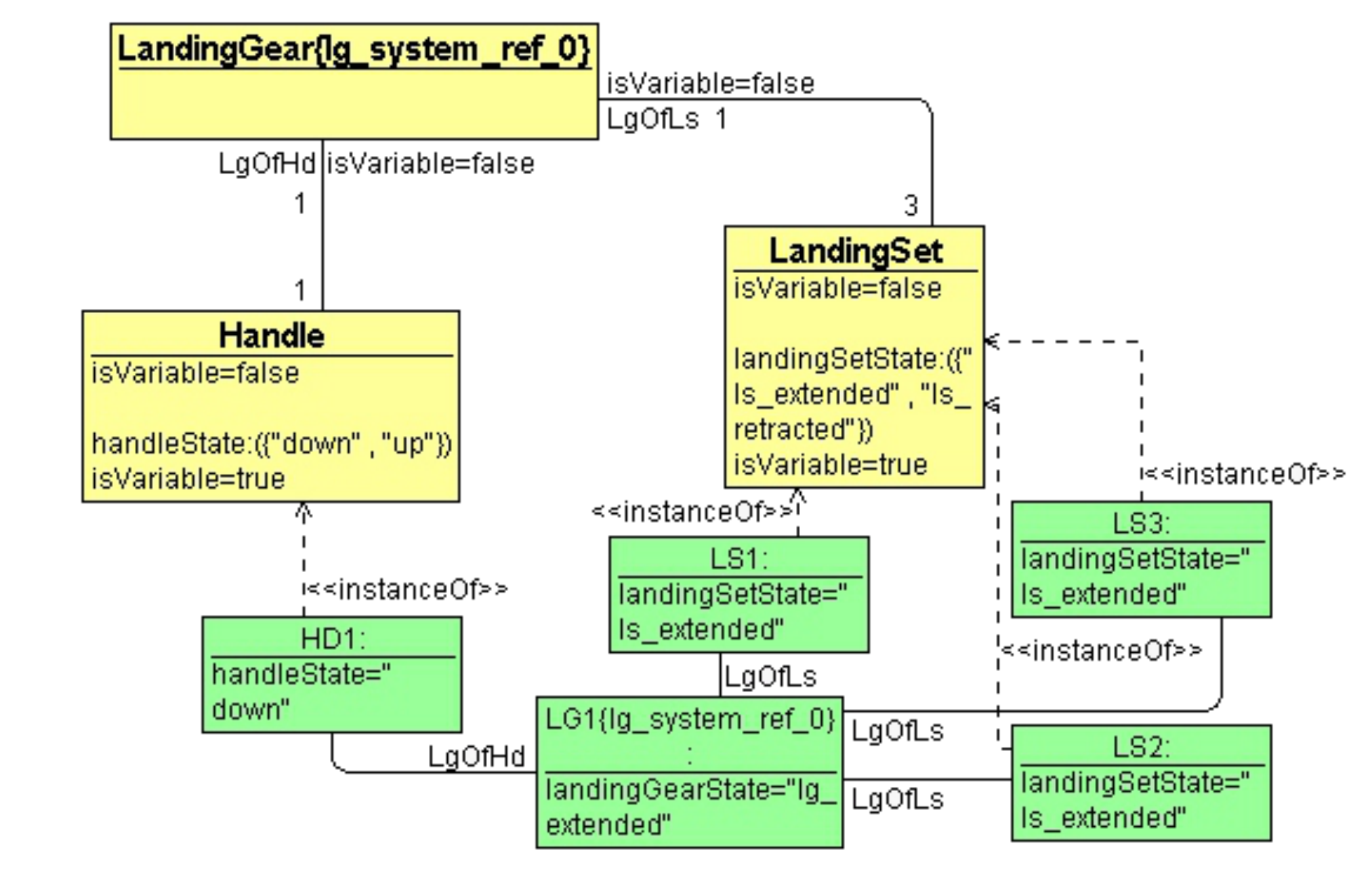}
\caption{\label{lgsystem_refinment_1_ontology}
\textit{\textbf{lg\_system\_ref\_1}}: ontology associated with the first   refinement level of the landing gear goal model}
\end{center}

\end{figure}

\begin{figure*}[!h]
\begin{center}
\includegraphics[width=1.05\textwidth]{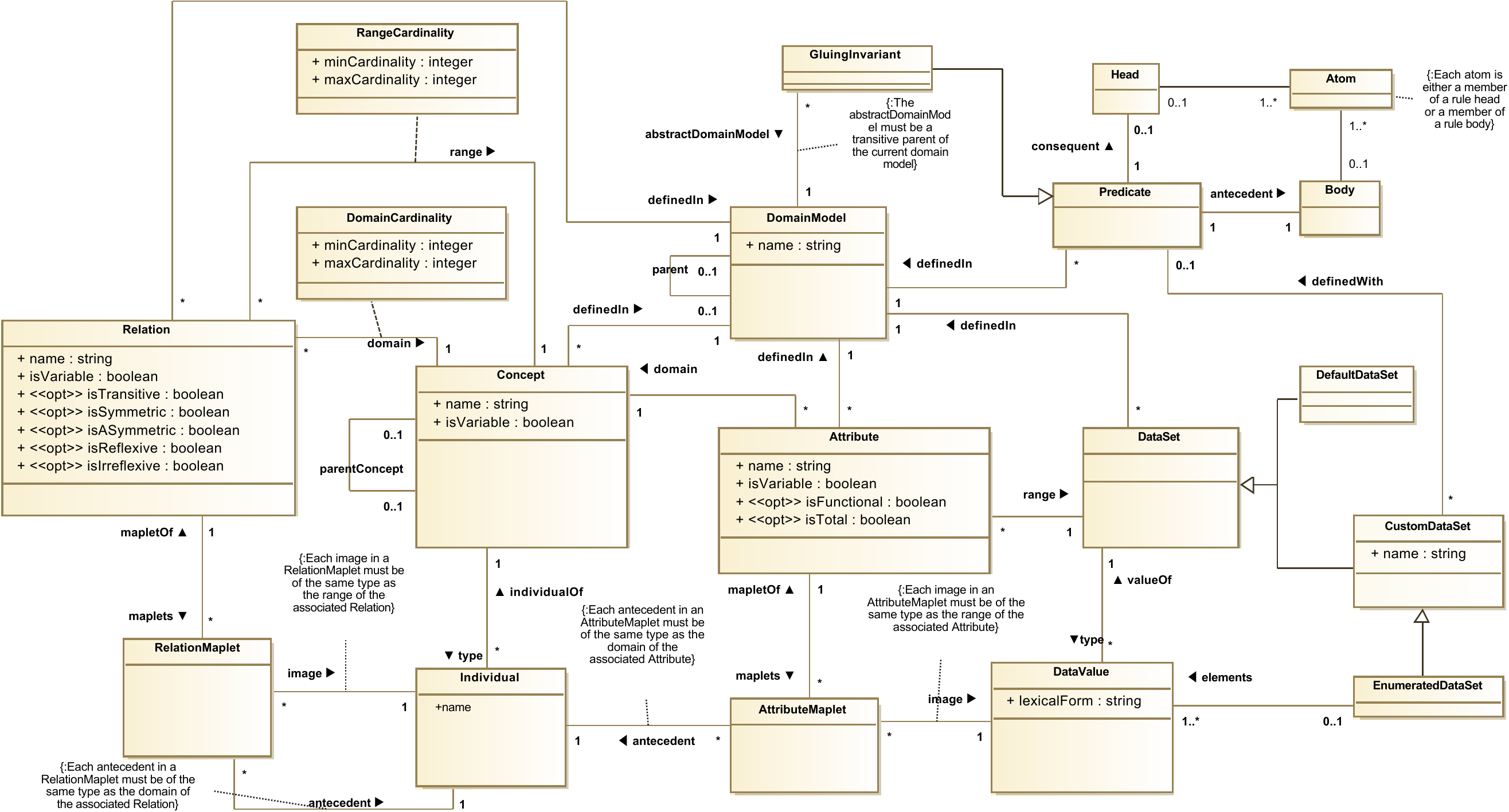}
\end{center}
\caption{\label{our_businessdomain_metamodel} Metamodel associated with the SysML/KAOS domain modeling language}
\end{figure*}

 
 Figure \ref{lgsystem_refinment_0_ontology}    represents  the SysML/KAOS domain model associated with the root level of the landing gear system goal model of Fig. \ref{lgsystem_goal_model_makelgextended},  and Fig. \ref{lgsystem_refinment_1_ontology} represents the one associated with  the first refinement level. 
They are illustrated using the syntax proposed by   \textit{OWLGred} \cite{owlgred_reference_link}
 and,  for readability purposes, we have decided to hide optional characteristics representation. 
It should be noted that the \textit{individualOf} association is illustrated, through  \textit{OWLGred}, as a stereotyped link with the tag \textit{<<instanceOf>>}. 
 
 Figure \ref{our_businessdomain_metamodel} is  an excerpt from the metamodel associated with the SysML/KAOS  domain modeling language. 
 Each domain model is associated with a level of refinement of the SysML/KAOS goal diagram and is likely to have as its parent, through the \textit{parent} association, another domain model. 
 For example,  the domain model  \texttt{lg\_system\_ref\_1} (Fig. \ref{lgsystem_refinment_1_ontology}) refines  \texttt{lg\_system\_ref\_0} (Fig. \ref{lgsystem_refinment_0_ontology}). 
 We use the notion of \textit{concept} (instance of  \textsf{Concept}) to designate an instantiable universal or a collection of individuals with common properties.
A \textit{concept} can be declared \textit{variable} (\textit{isVariable=TRUE}) when the set of its individuals can be updated by adding or deleting individuals.. Otherwise, it is considered to be \textit{constant} (\textit{isVariable=FALSE}).
For example, in \texttt{lg\_system\_ref\_0},  the landing gear entity is modeled as an instance of \textsf{Concept} named \texttt{LandingGear}. As in the case study adding or deleting a landing gear is not considered,  the property \textsf{isVariable} of  \texttt{LandingGear} is set to \texttt{false}.
Instances of \textsf{Relation} are used to capture  links between concepts, and instances of \textsf{Attribute} capture  links between concepts and data sets.
The most basic way to build an instance of \textsf{DataSet} is by listing its elements. This can be done through the \textsf{DataSet} specialization   called \textsf{EnumeratedDataSet}. 
A relation   or an attribute can be declared \textit{variable} if the list of maplets related to it  is likely to change over time. Otherwise, it is considered to be \textit{constant}. 
For example, 
the possible states of a landing gear are modeled by an instance of \textsf{Attribute} named \texttt{landingGearState}, having \texttt{LandingGear} as domain and as range an instance of \textsf{EnumeratedDataSet} containing two instances of  \textsf{DataValue} of type \textsf{STRING}: \texttt{lg\_extended} for the extended state and \texttt{lg\_retracted} for the retracted state. Its \texttt{isVariable} property is set to \textit{true}, since it is possible to dynamically change the state of a landing gear.
Furthermore, 
the association between  landing sets and landing gears, in \texttt{lg\_system\_ref\_1},  is modeled as an instance of \textsf{Relation} named \texttt{LgOfLs}. Since the association of a landing set to a landing gear cannot be changed dynamically, the property \textsf{isVariable} of \texttt{LgOfLs} is set to \texttt{false}.

 Each instance of \textsf{DomainCardinality} (respectively \textsf{RangeCardinality}) makes it possible to define, for an instance of \textsf{Relation} \texttt{re}, the minimum and maximum limits of the number of instances of \textsf{Individual}, having the domain (respectively range) of \texttt{re} as \textsf{type}, that can be put in relation with one instance of \textsf{Individual}, having the range (respectively domain) of \texttt{re} as \textsf{type}.
 For example, in \texttt{lg\_system\_ref\_1}, the instance of \textsf{DomainCardinality} associated with \texttt{LgOfLs} has its \textsf{minCardinality}  and  \textsf{maxCardinality} properties set to \textit{1}. 
Instances of \textsf{RelationMaplet} are used to define associations between instances of \textsf{Individual} through  instances of \textsf{Relation}. Instances of \textsf{AttributeMaplet} play the same role for attributes.
For example, in \texttt{lg\_system\_ref\_1}, 
there are three instances of \textsf{RelationMaplet} to model the association of the landing gear  \texttt{LG1} to the landing sets  \texttt{LS1}, \texttt{LS2} and \texttt{LS3},  each having as \textsf{image} \texttt{LG1} and as \textsf{antecedent} the corresponding \texttt{LandingSet} individual.

The notion of \textsf{Predicate}   is used to represent constraints between different elements of the domain model in the form of \textit{Horn clauses}: each predicate has a body which represents its \textit{antecedent} and a head which represents its \textit{consequent}, body and head designating conjunctions of atoms.
A data set can be declared abstractly, as an instance of \textsf{CustomDataSet}, and defined with a predicate.
\textsf{GluingInvariant}, specialization of \textsf{Predicate}, is used to represent links between variables and constants defined within a domain model and those appearing in more abstract domain models,  transitively linked to it through the \textit{parent} association. 
 Gluing invariants are  extremely important because they capture relationships between abstract and concrete data during refinement and are used to discharge  proof obligations. 
The following gluing invariant is associated with our case study: if there is at least one landing set having the retracted state, then the state of \texttt{LG1} is retracted
\begin{dmath}[number={inv1}]
landingGearState(LG1, "lg\_retracted") 
 \leftarrow  LandingSet(?ls) \wedge landingSetState(?ls, "ls\_retracted") 
 \label{inv1}
\end{dmath}

\section{Existing Approaches for the Formalization  of Domain Models }
In \cite{DBLP:conf/icfem/WangDS10}, an approach is proposed for the automatic extraction of  domain knowledge, as \textit{OWL} ontologies, from \textit{Z/Object-Z (OZ)} models \cite{DBLP:journals/stt/Doberkat01} :  OZ types and classes are transformed into OWL classes. Relations and functions are transformed into OWL properties, with the \textit{cardinality} restricted to \textit{1} for total functions and the \textit{maxCardinality} restricted to \textit{1} for partial functions. OZ constants are translated into OWL individuals. Rules are also proposed for subsets and  state schemas.  A similar approach is proposed in \cite{DBLP:conf/icfem/DongSW02}, for the extraction of \textit{DAML}  ontologies \cite{van2001reference} from \textit{Z} models.
These approaches are interested in  correspondence links between formal methods and ontologies, but their rules are restricted to the extraction of domain model elements from formal specifications.
Furthermore, all elements extracted from a formal model are defined within a single ontology component, while in our approach, we work on the opposite direction: each ontology refinement level is used to generate a formal model component and links between domain models give links between formal components.

 In \cite{DBLP:conf/ifip2/PoernomoU09}, domain is modeled by defining agents, business entities and relations between them. The paper proposes rules to translate domain models so designed into \textit{Event-B} specifications: agents are transformed into machines, business entities are transformed into sets, and relations are transformed into \textit{Event-B} variable relations. These rules are certainly sufficient for domain models of  interest for \cite{DBLP:conf/ifip2/PoernomoU09}, but they are very far from covering the extent of the SysML/KAOS domain modeling language.
 

 Some rules for passing from an \textit{OWL} ontology  representing a domain model to \textit{Event-B} specifications are proposed in \cite{DBLP:journals/scp/AlkhammashBFC15}, in \cite{h.Alkhammash} and   through a case study in \cite{DBLP:conf/isola/MammarL16}.
 In \cite{h.Alkhammash}, the proposed rules requires the generation of an \textit{ACE (Attempto Controlled English)} version of the OWL ontology which serves as the basis for the development of the \textit{Event-B} specification. This is done through a step called OWL verbalization. 
The verbalization method transforms  OWL individuals into capitalized proper names, classes into common names, and properties into active and passive verbs. Once the verbalization process has been completed, \cite{h.Alkhammash} proposes a set of rules for obtaining the \textit{Event-B} specification: classes are translated to \textit{Event-B} sets, properties are translated to relations, etc. 
In \cite{DBLP:journals/scp/AlkhammashBFC15}, domain properties are described through  data-oriented requirements for concepts, attributes and associations and through  constraint-oriented requirements for axioms. Possible states of a variable element are represented using UML state machines.
Concepts, attributes and associations arising from  data-oriented requirements are modeled as UML class diagrams and translated to \textit{Event-B} using  \textit{UML-B} \cite{Snook:2006:UFM:1125808.1125811}: 
nouns and attributes are represented as UML classes and relationships between nouns are represented as UML associations. \textit{UML-B} is also used for the translation of state machines to \textit{Event-B} variables, invariants and events.
The approaches  in \cite{DBLP:journals/scp/AlkhammashBFC15} and 
\cite{h.Alkhammash}  require a manual transformation of the ontology before the possible application of   translation rules to obtain the formal specifications:  In \cite{DBLP:journals/scp/AlkhammashBFC15}, it is necessary to convert OWL ontologies into UML diagrams. In \cite{h.Alkhammash}, the proposal requires the generation of a controlled English 
version of the \textit{OWL} ontology.
Furthermore, since the OWL formalism supports weak typing and multiple inheritance, the approaches define a unique \textit{Event-B} abstract set named \textit{Thing}. Thus, all sets, corresponding to OWL classes, are defined as subsets of \textit{Thing}.
Our formalism, on the other hand, imposes strong typing and simple inheritance; which makes it possible to translate some concepts into \textit{Event-B} abstract sets.
In \cite{DBLP:conf/isola/MammarL16}, the case study reveals
three translation rules: 
  each ontology class, having no individual, is modeled as an \textit{Event-B} abstract set. If the class has individuals, then it is modeled as an enumerated set. Finally, each object property between two classes is modeled as a constant defined  as a relation.
  Several shortcomings are common to these approaches: 
the provided rules  do not take into account the refinement links between model parts. Furthermore, they have not been implemented or formally verified  and they
are provided in an informal way that does not allow the assesment of their quality and consistency. 
Finally, the approaches are  far from covering the extent of  the SysML/KAOS domain modeling language and they are 
  only interested  in static domain knowledge (they do not distinguish what gives rise to formal constants or variables).

Several works have been done on the translation of UML diagrams into B specifications such as \cite{DBLP:conf/kbse/LaleauM00,Snook:2006:UFM:1125808.1125811}. They have obviously  inspired many of our rules, like those dealing with the translation of classes  (concepts) and of associations (attributes and relations). But,  our work differs from them because of the distinctions between ontologies and UML diagrams: 
 within an ontology,  concepts or classes and their instances are represented within the same model as well as the predicates defining domain constraints. Moreover, these studies are most often interested in the translation of model elements and not really in handling links between models.
Since our domain models are associated with SysML/KAOS goal model  refinement levels, the  hierarchy between domain models is converted into  refinement links between formal components. Moreover, the predicates linking the elements of concrete models to those of  abstract models give   gluing invariants.
Taking into account links between models guarantees a better scalability,  readability and reusability of rules and models.  
 Finally,  in the case of the SysML/KAOS domain modeling language, the changeability properties (properties characterising  the belonging of an element to the static or dynamic knowledge, materialised with the \textsf{isVariable} property in classes \textsf{Concept}, \textsf{Relation} and \textsf{Attribute})  are first-class citizens, as well as association characteristics (such as \textsf{isTransitive} of the class \textsf{Relation} and  \textsf{isFunctional} or the class \textsf{Attribute}), in order to produce a strongly expressive formal specification. As a result, they are explicitly represented.


\section{Translation Rules from Domain Models to B System Specifications}
In the following, we describe a set of rules that allow to obtain a formal specification from domain models associated with refinement levels of a SysML/KAOS goal model. 
The rules are fully described in \cite{translation_rules_event_b_specification_full_code_link,1712.07406}.

Table \ref{tableau_recapitulatif_correspondances} summarises the  translation rules, from domain models with or without  parents to concepts with or without parents, including relations, individuals or attributes. 
It should be noted that \textit{o\_x} designates the result of the translation of \textit{x} and that the \textit{abstract} qualifier is used for  "without parent".

\begin{scriptsize}
\begin{center}
\begin{longtable}{|p{2.5cm}|p{1.1cm}|p{6.0cm}|p{1.1cm}|p{6.0cm}|}
\caption{\label{tableau_recapitulatif_correspondances} Summary  of the translation rules}\\
\hline
&\multicolumn{2}{|c|}{\textbf{Domain Model}} & \multicolumn{2}{|c|}{\textbf{B System}} \\
\hline
\textbf{Translation Of} &\textbf{Element} & \textbf{Constraint} & \textbf{Element} & \textbf{Constraint} \\
\hline
\textbf{Abstract domain model}  & DM & $DM \in \textsf{DomainModel}$ \newline \textit{DM} is not associated with a parent domain model & o\_DM & $o\_DM \in \textsf{System}$ \\
\hline
\textbf{Domain model with parent} & DM PDM & $\{DM,PDM\} \subseteq \textsf{DomainModel} $ \newline \textit{DM} is associated with \textit{PDM} through the \textit{parent} association and \textit{PDM} has already been translated & o\_DM & $o\_DM \in \textsf{Refinement} $ \newline
\textit{o\_DM} refines \textit{o\_PDM} \\
\hline
\textbf{Abstract concept} & CO & $CO \in \textsf{Concept}$ \newline \textit{CO} is not associated with a parent concept & o\_CO & $o\_CO \in \textsf{AbstractSet}$ \\
\hline
\textbf{Concept with parent} & CO PCO & $\{CO,PCO\} \subseteq \textsf{Concept} $ \newline \textit{CO} is associated with \textit{PCO} through the \textit{parentConcept} association and \textit{PCO} has already been translated & o\_CO & $o\_CO \in \textsf{Constant}$  \newline $o\_CO \subseteq o\_PCO$\\
\hline
\textbf{Relation} &
RE 
CO1 
CO2 
& $\{CO1,CO2\} \subseteq \textsf{Concept} $ 
 \newline $RE \in \textsf{Relation}$
  \newline $CO1$ is the \textit{domain} of \textit{RE} 
  \newline $CO2$ is the \textit{range} of \textit{RE} 
  \newline $Relation\_DomainCardinality\_maxCardinality(RE)=da$
    \newline $Relation\_DomainCardinality\_minCardinality(RE)=di$
        \newline $Relation\_RangeCardinality\_maxCardinality(RE)=ra$
            \newline $Relation\_RangeCardinality\_minCardinality(RE)=ri$
  \newline \textit{CO1} and \textit{CO2} have already been translated 
 & 
 T\_RE
o\_RE 
 &  
 $T\_RE \in  \textsf{Constant}$  
  \textbf{IF} the \textit{isVariable} property of \textit{RE} is set to  $FALSE$
 \newline \textbf{THEN} $o\_RE \in  \textsf{Constant}$
  \newline \textbf{ELSE} $o\_RE \in  \textsf{Variable}$
  \newline \textbf{END}
 \newline $o\_RE \in T\_RE$
  \newline \textbf{IF}  $\{ra,ri,da,di\} = \{1\}$ \newline \textbf{THEN} $T\_RE = o\_CO1 \bij o\_CO2$
    \newline \textbf{ELSE IF}  $\{ra,ri,da\} = \{1\}$ \newline  \textbf{THEN} $T\_RE = o\_CO1 \tinj o\_CO2$
      \newline \textbf{ELSE IF}  $\{ra,ri,di\} = \{1\}$ \newline \textbf{THEN} $T\_RE = o\_CO1 \tsur o\_CO2$
     \begin{scriptsize}
      \newline \textbf{ELSE IF}  $\{ra,di\} = \{1\}$ \newline \textbf{THEN} $T\_RE = o\_CO1 \psur o\_CO2$
       \newline \textbf{ELSE IF}  $\{ra,da\} = \{1\}$ \newline \textbf{THEN} $T\_RE = o\_CO1 \pinj o\_CO2$
   \newline \textbf{ELSE IF}  $\{ra,ri\} = \{1\}$ \newline \textbf{THEN} $T\_RE = o\_CO1 \longrightarrow o\_CO2$
  \newline \textbf{ELSE IF}  $ra=1$  \textbf{THEN} $T\_RE = o\_CO1 \pfun o\_CO2$
     \end{scriptsize}
    \newline \textbf{ELSE} 
    	\newline \hspace*{0.20in} 	$T\_RE = o\_CO1 \leftrightarrow o\_CO2$ 
    	\newline \hspace*{0.20in}	$\wedge \forall x. (x \in CO2 \Rightarrow card(o\_RE^{-1}[\{x\}])  \in di..da)$
    	\newline \hspace*{0.20in} $\wedge \forall x. (x \in CO1 \Rightarrow card(o\_RE[\{x\}])  \in ri..ra)$
 \\
 \hline
 \textbf{Attribute} &
AT 
CO 
DS 
& $CO \in \textsf{Concept} $ 
 \newline $DS \in \textsf{DataSet}$
 \newline $AT \in \textsf{Attribute}$
  \newline $CO$ is the \textit{domain} of \textit{AT} 
  \newline $DS$ is the \textit{range} of \textit{AT} 
  \newline \textit{CO} and \textit{DS} have already been translated 
 & 
o\_AT
 &  
  \textbf{IF}  the \textit{isVariable} property of \textit{AT} is set to  $FALSE$
 \newline \textbf{THEN} $o\_AT \in  \textsf{Constant}$
  \newline \textbf{ELSE} $o\_AT \in  \textsf{Variable}$
  \newline \textbf{END}
 \newline \textbf{IF}   \textit{isFunctional} and \textit{isTotal}  are set to  $TRUE$ 
  \newline \textbf{THEN} $o\_AT \in o\_CO \tfun o\_DS$ 
  \newline \textbf{ELSE}  \textbf{IF}   \textit{isFunctional}  is set to  $TRUE$
   \newline \textbf{THEN} $o\_AT \in o\_CO \pfun o\_DS$ 
     \newline \textbf{ELSE} 
       $o\_AT \in o\_CO \leftrightarrow o\_DS$ 
  \newline \textbf{END}
 \\
\hline
\textbf{Concept changeability} &
CO & $CO \in \textsf{Concept}$ 
\newline  the \textit{isVariable} property of \textit{CO} is set to  $TRUE$
\newline \textit{CO} has already been translated
& X\_CO &$X\_CO \in  \textsf{Variable}$
  \newline $X\_CO \subseteq o\_CO$ \\
\hline
\textbf{Individual} &
Ind CO & $Ind \in \textsf{Individual}$ ~~~
 $CO \in \textsf{Concept}$ 
\newline $Ind$ is an individual of \textit{CO} 
\newline \textit{CO} has already been translated
& o\_Ind & $o\_Ind \in \textsf{Constant}$  
\newline $o\_Ind \in o\_CO$\\
\hline
\textbf{Data value} &
Dva DS & $Dva \in \textsf{DataValue}$ ~~~
 $DS \in \textsf{DataSet}$ 
\newline $Dva$ is a value of \textit{DS}
\newline \textit{DS} has already been translated
& o\_Dva & $o\_Dva \in \textsf{Constant}$  
\newline $o\_Dva \in o\_DS$\\
\hline
\textbf{Relation transitivity} &
RE & $RE \in \textsf{Relation}$ 
 \newline the \textit{isTransitive} property of \textit{RE} is set to  $TRUE$
\newline \textit{RE} has already been translated
&  & $(o\_RE ~ ; ~ o\_RE) \subseteq o\_RE$ \\
\hline
\textbf{Relation symmetry} &
RE & $RE \in \textsf{Relation}$ 
\newline $Relation\_isSymmetric(RE)=TRUE$
\newline \textit{RE} has already been translated
&  &  $o\_RE^{-1}  = o\_RE$ \\
\hline
\textbf{Relation asymmetry} &
RE CO & $RE \in \textsf{Relation}$ 
\newline $Relation\_isASymmetric(RE)=TRUE$
  \newline $Relation\_domain\_Concept(RE)=CO$ 
\newline \textit{RE} and \textit{CO} have already been translated
&  &  $(o\_RE^{-1}  \cap o\_RE) \subseteq id(o\_CO)$ \\
\hline
\textbf{Relation reflexivity }  &
RE CO & $RE \in \textsf{Relation}$ 
\newline $Relation\_isReflexive(RE)=TRUE$
  \newline $Relation\_domain\_Concept(RE)=CO$ 
\newline \textit{RE} and \textit{CO} have already been translated
&  &  $id(o\_CO) \subseteq o\_RE$  \\
\hline
\textbf{Relation irreflexivity} &
RE CO & $RE \in \textsf{Relation}$ 
\newline $Relation\_isIrreflexive(RE)=TRUE$
  \newline $Relation\_domain\_Concept(RE)=CO$ 
\newline \textit{RE} and \textit{CO} have already been translated
&  &  $id(o\_CO) \cap o\_RE = \emptyset$ \\
\hline
\textbf{Relation maplets} &
RE  {\scriptsize $(M_j)_{j=1..n}$} $(a_j, \newline i_j)_{j=1..n}$ & 
$RE \in \textsf{Relation}$
\begin{scriptsize}
\newline $(M_j)_{j=1..n}$ are \textit{maplets} of \textit{RE} 
  \newline $\forall j \in 1..n, a_j$ is the antecedent of $M_j$
  \newline $\forall j \in 1..n, i_j$ is the image of $M_j$
\end{scriptsize}
\newline \textit{RE} and $(a_j, i_j)_{j=1..n}$ have already been translated
&  &  
  \textbf{IF} the \textit{isVariable} property of \textit{RE} is set to  $FALSE$
 \newline \textbf{THEN} $o\_RE  =  \{(o\_a_j, o\_i_j)_{j=1..n}\}$ (Property)
  \newline \textbf{ELSE} $o\_RE :=  \{(o\_a_j, o\_i_j)_{j=1..n}\}$ {\footnotesize (Initialisation)}
    \newline \textbf{END} 
 \\
\hline
\textbf{Attribute maplets} &
AT  {\scriptsize $(M_j)_{j=1..n}$} $(a_j, \newline i_j)_{j=1..n}$ & $AT \in \textsf{Attribute}$ 
\begin{scriptsize}
\newline $(M_j)_{j=1..n}$ are \textit{maplets} of \textit{AT} 
  \newline $\forall j \in 1..n, a_j$ is the antecedent of $M_j$
  \newline $\forall j \in 1..n, i_j$ is the image of $M_j$
\end{scriptsize}
\newline \textit{AT} and $(a_j, i_j)_{j=1..n}$ have already been translated
&  &  
  \textbf{IF}  the \textit{isVariable} property of \textit{AT} is set to  $FALSE$
 \newline \textbf{THEN} $o\_AT  =  \{(o\_a_j, o\_i_j)_{j=1..n}\}$
  \newline \textbf{ELSE} $o\_AT :=  \{(o\_a_j, o\_i_j)_{j=1..n}\}$
    \newline \textbf{END}
 \\
\hline
\end{longtable}
\end{center}

\end{scriptsize}

\begin{figure}[h]
\begin{footnotesize}
\begin{mdframed}
   \it SYSTEM
\hspace*{0.20in}\it lg\_system\_ref\_0

\textbf{SETS}
\hspace*{0.20in} LandingGear; DataSet\_1= \{lg\_extended, lg\_retracted\}

\textbf{CONSTANTS}
\hspace*{0.20in} T\_landingGearState, LG1

 \textbf{PROPERTIES}

\textsf{(0.1)} \hspace*{0.1in} LG1 $\in$ LandingGear 

\textsf{(0.2)} \hspace*{0.10in} $\wedge$  LandingGear=\{LG1\} 

\textsf{(0.3)} \hspace*{0.10in} $\wedge$  T\_landingGearState = LandingGear $\longrightarrow$ DataSet\_1

 \textbf{VARIABLES}
\hspace*{0.20in} landingGearState


 \textbf{INVARIANT}

\textsf{(0.4)} \hspace*{0.10in} \hspace*{0.20in}  landingGearState $\in$ T\_landingGearState


 \textbf{INITIALISATION}

\textsf{(0.5)} \hspace*{0.10in} \hspace*{0.20in} landingGearState := \{LG1 $\mapsto$ lg\_extended \}


END

\end{mdframed}
\caption{\label{lgsystem_event_b_model_refinment_0} Formalization of the Root Level of the Landing Gear System Domain Model}
\end{footnotesize}
\end{figure}

Figures \ref{lgsystem_event_b_model_refinment_0} and \ref{lgsystem_event_b_model_refinment_1}  represent  respectively
  the \textit{B System} specifications associated with the 
  root level of the landing gear system domain model illustrated  in Fig. \ref{lgsystem_refinment_0_ontology} and that associated with the first refinement level domain model illustrated in Fig. \ref{lgsystem_refinment_1_ontology}.

\begin{figure}[h]

\begin{footnotesize}
\begin{mdframed}

   \it REFINEMENT
\hspace*{0.20in} lg\_system\_ref\_1

 \textbf{REFINES}
\hspace*{0.20in} lg\_system\_ref\_0

\textbf{SETS}
\hspace*{0.20in} Handle; LandingSet; DataSet\_2=\{ls\_extended, ls\_retracted\}; DataSet\_3=\{down, up\}

\textbf{CONSTANTS}
\hspace*{0.20in} T\_LgOfHd,  LgOfHd,  T\_LgOfLs, LgOfLs,  T\_landingSetState, T\_handleState, HD1, LS1, LS2, LS3

 \textbf{PROPERTIES}

\textsf{(1.1)} \hspace*{0.10in} HD1 $\in$ Handle 

\textsf{(1.2)} \hspace*{0.10in} $\wedge$  Handle=\{HD1\} 

\textsf{(1.3)} \hspace*{0.10in} $\wedge$  LS1 $\in$ LandingSet 

\textsf{(1.4)} \hspace*{0.10in} $\wedge$  LS2 $\in$ LandingSet 

\textsf{(1.5)} \hspace*{0.10in} $\wedge$  LS3 $\in$ LandingSet 

\textsf{(1.6)} \hspace*{0.10in} $\wedge$  LandingSet=\{LS1, LS2, LS3\} 

\textsf{(1.7)} \hspace*{0.10in} $\wedge$  T\_LgOfHd = Handle $\leftrightarrow$ LandingGear 

\textsf{(1.8)} \hspace*{0.10in} $\wedge$  LgOfHd $\in$ T\_LgOfHd 

\textsf{(1.9)} \hspace*{0.10in} $\wedge$  $\forall$ xx.(xx $\in$ Handle $\Rightarrow$ card(LgOfHd[\{xx\}])=1) 

\textsf{(1.10)} \hspace*{0.10in} $\wedge$  $\forall$xx.(xx $\in$ LandingGear $\Rightarrow$ card(LgOfHd$^{-1}$[\{xx\}])=1) 

\textsf{(1.11)} \hspace*{0.10in} $\wedge$  LgOfHd = \{HD1 $\mapsto$ LG1 \} 

\textsf{(1.12)} \hspace*{0.10in} $\wedge$  T\_LgOfLs = LandingSet $\leftrightarrow$ LandingGear 

\textsf{(1.13)} \hspace*{0.10in} $\wedge$  LgOfLs $\in$ T\_LgOfLs 

\textsf{(1.14)} \hspace*{0.10in} $\wedge$  $\forall$xx.(xx $\in$ LandingSet $\Rightarrow$ card(LgOfLs[\{xx\}])=1) 

\textsf{(1.15)} \hspace*{0.10in} $\wedge$  $\forall$xx.(xx $\in$ LandingGear $\Rightarrow$ card(LgOfLs$^{-1}$[\{xx\}])=3) 

\textsf{(1.16)} \hspace*{0.10in} $\wedge$  LgOfLs = \{LS1 $\mapsto$ LG1, LS2 $\mapsto$ LG1, LS3 $\mapsto$ LG1 \} 

\textsf{(1.17)} \hspace*{0.10in} $\wedge$ T\_landingSetState = LandingSet $\longrightarrow$ DataSet\_2 

\textsf{(1.18)} \hspace*{0.10in} $\wedge$  T\_handleState = Handle $\longrightarrow$ DataSet\_3 

 \textbf{VARIABLES}
\hspace*{0.20in} landingSetState, handleState


 \textbf{INVARIANT}

\textsf{(1.19)} \hspace*{0.10in}  landingSetState $\in$ T\_landingSetState 

\textsf{(1.20)} \hspace*{0.10in} $\wedge$   handleState $\in$ T\_handleState

\textsf{(1.21)} \hspace*{0.10in} $\wedge$ $ \forall ls. (ls \in LandingSet \wedge landingSetState(ls, ls\_extended)\Rightarrow \\ landingGearState(LG1, lg\_extended))$


 \textbf{INITIALISATION}

\textsf{(1.22)} \hspace*{0.10in} landingSetState := \{LS1 $\mapsto$ ls\_extended, LS2 $\mapsto$ ls\_extended, LS3 $\mapsto$ ls\_extended \} 

\textsf{(1.23)} \hspace*{0.10in} || \hspace*{0.20in} handleState := \{HD1 $\mapsto$ down \}


END

\end{mdframed}
\caption{\label{lgsystem_event_b_model_refinment_1} Formalization of the First Refinement Level of the Landing Gear System Domain Model}
\end{footnotesize}
\end{figure}


%
%
%
\subsection{Generation of B System Components}
%
%

Any domain model that is not associated with another domain model 
 through the \textit{parent} association, gives  a \textsf{System} component (line 1 of  Table \ref{tableau_recapitulatif_correspondances}).
This is illustrated in Fig. \ref{lgsystem_event_b_model_refinment_0} where the root level domain model is translated into a  system named   \textbf{\texttt{lg\_system\_ref\_0}}. 

A domain model associated with another one representing its parent 
  gives  a \textsf{Refinement} component (line 2 of  Table \ref{tableau_recapitulatif_correspondances}).
    This component   refines the one corresponding  to the parent domain model. 
     This is illustrated in Fig. \ref{lgsystem_event_b_model_refinment_1} where the first refinement level domain model is translated into a  refinement  named   \textbf{\texttt{lg\_system\_ref\_1}} refining \textbf{\texttt{lg\_system\_ref\_0}}.

\subsection{Generation of B System Sets}
Any  concept  that is not associated with another one
through the \textsf{parentConcept} association, gives  an abstract set (line 3 of  Table \ref{tableau_recapitulatif_correspondances}).
  For example, in Fig. \ref{lgsystem_event_b_model_refinment_0}, abstract set named  \textbf{\texttt{LandingGear}} appears because of \textsf{Concept} instance \texttt{LandingGear}.

Any instance of \textsf{CustomDataSet}, defined through an enumeration (instance of \textsf{EnumeratedDataSet}), gives  a \textit{B System} enumerated set. 
Otherwise, if it is defined with an instance of \textsf{Predicate} \texttt{P}, then it gives  a constant for which the typing axiom is the result of the translation of \texttt{P}. Finally, it gives  an abstract set if no typing predicate is provided.
  For example, in Fig. \ref{lgsystem_event_b_model_refinment_0},
   the data set \texttt{\{lg\_extended, lg\_retracted\}}, defined in  Fig.   \ref{lgsystem_refinment_0_ontology},  gives  the enumerated set \textbf{\texttt{DataSet\_1=\{lg\_extended, lg\_retracted\}}}.

 Any instance of \textsf{DefaultDataSet} is mapped directly to a \textit{B System} default  set: \textsf{NATURAL}, \textsf{INTEGER}, \textsf{FLOAT}, \textsf{STRING} or \textsf{BOOL}.
 
\subsection{Generation of B System Constants}
Any  concept  associated with another one 
 through the \textsf{parentConcept} association, gives  a constant
   typed as a subset of the \textit{B System} element corresponding to the parent concept (line 4 of  Table \ref{tableau_recapitulatif_correspondances}).

Each relation   gives  a \textit{B System} constant representing the type of its corresponding   element and defined as the set of relations between the \textit{B System} element corresponding to the relation domain and the one corresponding to the relation range.  Moreover, if the relation has its \textsf{isVariable} property  set to \textit{FALSE},  a second constant is added (line 5 of  Table \ref{tableau_recapitulatif_correspondances}).
This is illustrated in Fig. \ref{lgsystem_event_b_model_refinment_1} where \texttt{LgOfHd}, for which \textsf{isVariable} is set to \textit{FALSE}, is translated into  a  constant named   \textbf{\texttt{LgOfHd}} and having as type \textbf{\texttt{T\_LgOfHd}} defined as the set of relations between \textbf{\texttt{Handle}} and \textbf{\texttt{LandingGear}} (assertions \texttt{'1.7)} and \texttt{(1.8)}).

Similarly to relations, each attribute gives  a constant representing the type of its corresponding element and, in the case where \textsf{isVariable} is set to \textit{FALSE},  to another constant (line 6 of  Table \ref{tableau_recapitulatif_correspondances}).
 However, when the \textsf{isFunctional} property is set to \textit{TRUE}, the constant representing the type is defined as the set of functions between the \textit{B System} element corresponding to the attribute domain and the one corresponding to the attribute range. The  element corresponding to the attribute is then typed as a function.
Furthermore, when \textsf{isFunctional} is set to \textit{TRUE}, the  \textsf{isTotal} property is used to assert if the function is total (\textsf{isTotal}=TRUE) or partial (\textsf{isTotal}=FALSE).
  For example, in Fig. \ref{lgsystem_event_b_model_refinment_0}, \textbf{\texttt{landingGearState}} is typed as a function (assertions \texttt{(0.3)} and \texttt{(0.4)}), since its type is the set of functions between \textbf{\texttt{LandingGear}} and \textbf{\texttt{DataSet\_1}} (\textbf{\texttt{DataSet\_1}}=\{lg\_extended, lg\_retracted\}).

 Finally, each individual (or data value) gives  a constant (lines 8 and 9 of  Table \ref{tableau_recapitulatif_correspondances}).
    For example, in Fig. \ref{lgsystem_event_b_model_refinment_1}, the constant named \textbf{\texttt{HD1}} is the correspondent of the individual \textit{HD1}.

\subsection{Generation of B System Variables}

An instance of \textsf{Relation}, of \textsf{Concept} or of \textsf{Attribute}, having its \textsf{isVariable} property set to \textit{TRUE} gives  a variable.
 For a concept, the variable represents the set of \textit{B System}  elements having this concept as type (line 7 of  Table \ref{tableau_recapitulatif_correspondances}).  For a relation or an attribute, it represents the set of pairs  between individuals (in case of relation) or between individuals and data values (in case of attribute) defined through it (lines 5 and 6 of  Table \ref{tableau_recapitulatif_correspondances}). For example, in Fig. \ref{lgsystem_event_b_model_refinment_1}, the variables named \textbf{\texttt{landingSetState}} and \textbf{\texttt{handleState}} appear because of the  \textsf{Attribute} instances \textit{landingSetState} and \textit{handleState} for which the \textsf{isVariable} property is set to \textit{TRUE} (Fig. \ref{lgsystem_refinment_1_ontology}).

\subsection{Generation of B System Invariants and Properties}
In this section, we are interested in translation rules between  domain models and  \textit{B System} specifications that  give   \textit{invariants} (instances of the \textsf{Invariant} class) or \textit{properties} (instances of the \textsf{Property} class).
Throughout this section, we will denote by \textit{logic formula} (instance of the \textsf{LogicFormula} class) any invariant or property, knowing that a logic formula is a property when it involves only constant  elements.
  Any other logic formula is an invariant. 
  It should be noted that when the logic formula relates variables defined within the model and those defined within more abstract models, it is a \textit{gluing invariant}.
  
  When the \textsf{isTransitive} property of an instance of \textsf{Relation} \textit{re} is set to \textit{TRUE}, the logic formula  
$(re ~ ; ~ re) \subseteq re$  must appear in the \textit{B System} component corresponding to the domain model, knowing that \textit{";"} is the  composition operator for relations (line 10 of  Table \ref{tableau_recapitulatif_correspondances}).
 For the \textsf{isSymmetric} property, the logic formula is  $re^{-1}  = re$. For the \textsf{isASymmetric} property, the logic formula is  $(re^{-1}  \cap re) \subseteq id(dom(re))$. For the \textsf{isReflexive} property, the logic formula is  $id(dom(re)) \subseteq re$  and for the \textsf{isIrreflexive} property, the logic formula is  $id(dom(re)) \cap re = \emptyset$, knowing that \textit{"id"} is the \textit{identity} function and \textit{"dom"} is  an operator that gives the \textit{domain} of a relation (\textit{"ran"} is the operator that gives the \textit{range}).

 An instance of \textsf{DomainCardinality} (respectively \textsf{RangeCardinality}) associated with an instance of \textsf{Relation} \textit{re}, with bounds \textsf{minCardinality} and \textsf{maxCardinality} ($maxCardinality \geq 0$), gives  the logic formula $ \forall x. (x \in ran(re) \Rightarrow card(re^{-1}[\{x\}])  \in minCardinality..maxCardinality)   $ (respectively $ \forall x. (x \in dom(re) \Rightarrow card(re[\{x\}])  \in minCardinality..maxCardinality) $). \\
 
 When $minCardinality = maxCardinality$, then the logic formula is $\forall x. (x \in ran(re) \Rightarrow card(re^{-1}[\{x\}]) = minCardinality)$ (respectively $\forall x. (x \in dom(re) \Rightarrow card(re[\{x\}])  = minCardinality) $). \\
 
  Finally, when $maxCardinality = \infty$, then the logic formula is $\forall x. (x \in ran(re) \Rightarrow card(re^{-1}[\{x\}]) \geq minCardinality)$ (respectively $\forall x. (x \in dom(re) \Rightarrow card(re[\{x\}])  \geq minCardinality) $). \\
  
    For example, in Fig. \ref{lgsystem_event_b_model_refinment_1},  logic formula \texttt{(1.9)} and \texttt{(1.10)}   appear because of instances of \textsf{RangeCardinality} and \textsf{DomainCardinality} associated with the instance of \textsf{Relation} \texttt{LgOfHd}  (Fig. \ref{lgsystem_refinment_0_ontology}). 
    
    The dual version of  the previous rule allows the processing of instances of \textsf{RangeCardinality}.
 
 Instances of \textsf{RelationMaplet} (respectively \textsf{AttributeMaplet}) associated with an instance of \textsf{Relation} (respectively \textsf{Attribute}) \textit{RE} give rise, in the case where the \textsf{isVariable} property of \textit{RE} is set to \textit{FALSE}, to the property $RE = \{a_1  \mapsto i_1, a_2  \mapsto i_2, ..., a_j  \mapsto i_j, ...,  a_n  \mapsto i_n\}$, where $a_j$ designates the instance of \textsf{Individual} linked to the j-th instance of \textsf{RelationMaplet} (respectively \textsf{AttributeMaplet}), through the \textsf{antecedent} association, and $i_j$ designates the instance of \textsf{Individual} (respectively \textsf{DataValue}) linked  through the \textsf{image} association (line 11 of  Table \ref{tableau_recapitulatif_correspondances}).
  When the \textsf{isVariable} property of \textit{RE} is set to \textit{TRUE}, it is the substitution  $RE := \{a_1  \mapsto i_1, a_2  \mapsto i_2, ..., a_j  \mapsto i_j, ...,  a_n  \mapsto i_n\}$ which is rather defined in the \textit{INITIALISATION} clause of the \textit{B System} component (lines 12 and 13 of  Table \ref{tableau_recapitulatif_correspondances}). 
For example, in Fig. \ref{lgsystem_event_b_model_refinment_1},  
the property \texttt{(1.11)}   appears because of the association between \textbf{\texttt{LG1}} and \textbf{\texttt{HD1}} through \textbf{\texttt{LgOfHd}}  (Fig. \ref{lgsystem_refinment_1_ontology}). 
Furthermore, the substitution \texttt{(1.23)} appears in the \textit{INITIALISATION} clause because the \texttt{handleState} attribute, for which  \textsf{isVariable}  is  \textit{TRUE}, is set to \textit{down}, for the individual \texttt{HD1} (through an instance of \textsf{AttributeMaplet}).
 
Finally, any instance of \textsf{Predicate} gives  a \textit{B System} logic formula.
When the predicate is an instance of \textsf{GluingInvariant}, the logic formula  is a \textit{B System} gluing invariant.
For example, in Fig. \ref{lgsystem_event_b_model_refinment_1},   assertion \texttt{(1.21)}   appears because of the gluing invariant (\ref{inv1}).


\subsection{The SysML/KAOS Domain Modeling  Tool}\label{the_tool_description_section}
\begin{figure*}[!h]
\begin{center}
\includegraphics[width=1\textwidth]{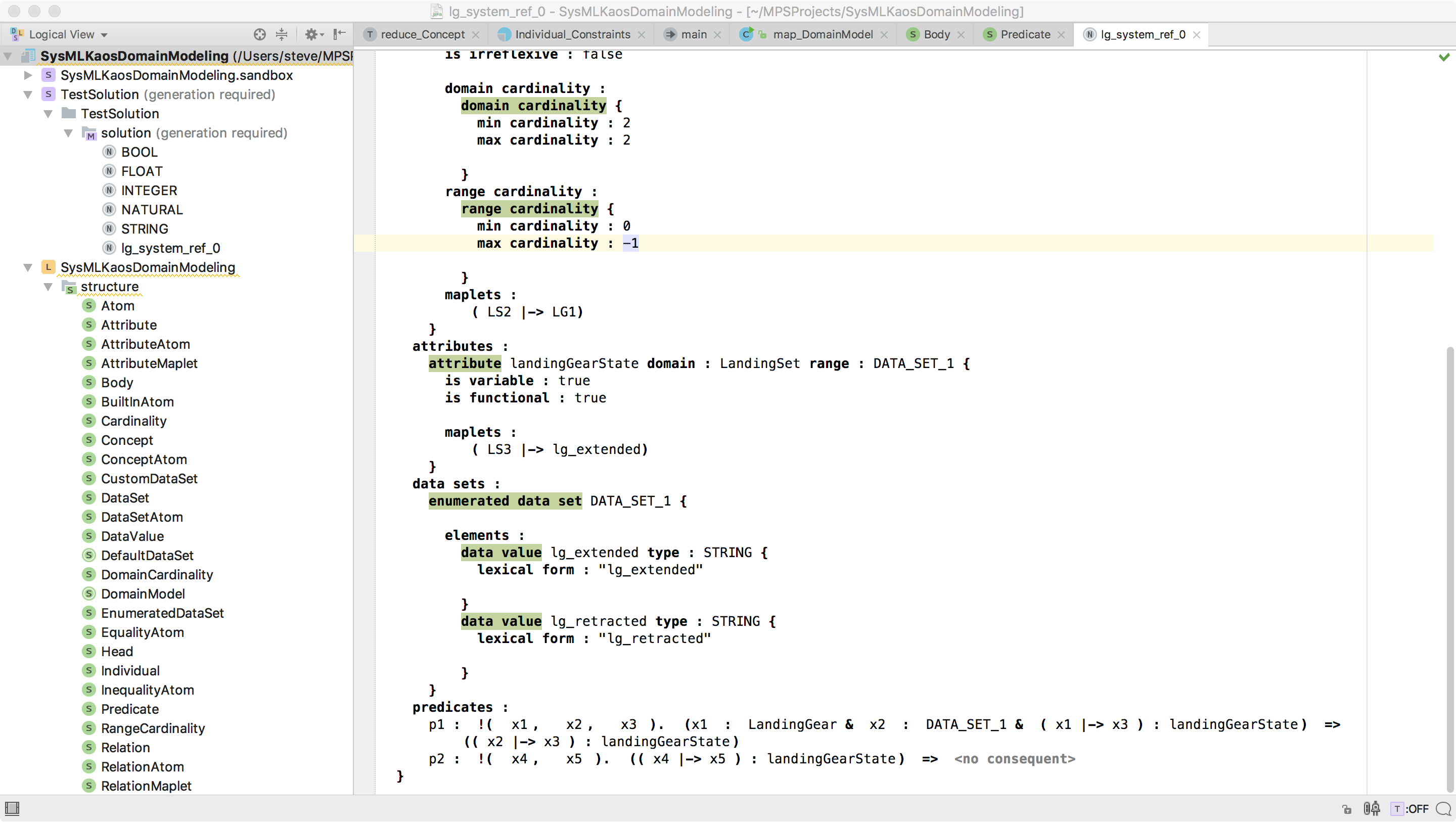}
\end{center}
\caption{\label{symlkaos_domain_modeling_tool_capture1} Main screen of the   SysML/KAOS domain modeling  tool}
\end{figure*}

The translation rules outlined here have been implemented within an open source tool  \cite{SysML_KAOS_Domain_Model_Parser_link}. It allows the construction of domain  ontologies \textit{(Fig. \ref{symlkaos_domain_modeling_tool_capture1})}  and generates  the corresponding \textit{B System} specifications (Fig. \ref{lgsystem_event_b_model_refinment_0} and \ref{lgsystem_event_b_model_refinment_1}). It is build through \textit{Jetbrains Meta Programming System} (MPS) \cite{jetbrains_mps}, a tool to design domain specific languages using language-oriented programming. The SysML/KAOS domain modeling language is build using 28 MPS concepts by defining for each the properties, childrens, references, constraints, behaviours and  custom editors. Each MPS concept represents a class of the  SysML/KAOS domain metamodel. For each  concept, the \textit{properties} clause is used to define the attributes. The \textit{Childrens} clause is used to define, for a concept \texttt{C},   the  concepts that are parts of \texttt{C}. Finally, the  \textit{references} clause is used to define the linked concepts. For example, the MPS concept representing the class \textsf{DomainModel}  defines all concepts representing the domain model elements as childrens and has a reference named \textit{parentDomainModel} linking it to its parent domain model. Each new domain model is defined in a MPS solution  using the SysML/KAOS domain modeling language.
A MPS solution is an instantiation of a MPS language.
 We have also defined a language for the \textit{B System} method. Thus, SysML/KAOS domain model solutions give   \textit{B System} solutions and  traceability links that can be used to propagate updates performed on a solution into the paired solution. 
However, the update propagation feature is not currently supported by the tool and is a next step in our work.

\section{Back Propagation Rules from B System Specifications to Domain Models}
The work done on  case studies  \cite{abz18_formose_on_ertms_case_study,sysml_kaos_domain_models_case_studies_link}
 reveals that, very often,  new elements need to be added to the structural part of the formal specification. These additions may be required during the specification of the body of events  or during the verification   and validation  of the formal model (e.g. to define an invariant or a theorem required to discharge a proof obligation).
 These lead us to the definition of a set of rules allowing the back propagation, within the domain model, of the new elements introduced in the structural part of the \textit{B System} specification. 
They prevent these additions from introducing inconsistencies between a domain model and its \textit{B System} specification.

We choose to support only the most repetitive additions that can be performed within the formal specification,   the domain model remaining the one to be updated in case of any major changes such as the addition or the deletion of a refinement level. 
 Table \ref{tableau_recapitulatif_correspondances} summarises the  most relevant back propagation rules.  
 Each rule defines its inputs (elements added to the \textit{B System} specification) and  constraints that each input must fulfill.
 It also defines its outputs (elements introduced within  domain models as a result of the application of the rule) and  their respective constraints.
It should be noted that for an element \textit{b\_x} of the \textit{B System} specification, \textit{x} designates the domain model element corresponding to \textit{b\_x}. In addition, when used, qualifier \textit{abstract}  denotes "without parent".

\begin{scriptsize}
\begin{center}
\begin{longtable}{|p{0.1cm}|p{3cm}|p{0.8cm}|p{5.7cm}|p{0.8cm}|p{7.5cm}|}
\caption{\label{tableau_recapitulatif_correspondances} back propagation rules in case of addition of an element in the \textit{B System} specification}\\
\hline
& &\multicolumn{2}{|c|}{\textbf{B System}} & \multicolumn{2}{|c|}{\textbf{Domain Model}} \\
\hline
& \textbf{Addition Of} &\textbf{Input} & \textbf{Constraint} & \textbf{Output} & \textbf{Constraint} \\
\hline
1 & \textbf{Abstract set}  & b\_CO & $b\_CO \in{} \textsf{AbstractSet} $ & CO & $CO \in \textsf{Concept}$\newline $Concept\_isVariable(CO) =  FALSE$
\newline { Knowing that an abstract set introduced can correspond to a concept or to a custom data set, to avoid non-determinism, we choose to define \textit{CO} as an instance of \textsf{Concept}. The user may subsequently change his type.} 
\\
\hline
2 & \textbf{Variable typed as subset of the correspondent of a concept} & x\_CO b\_CO &$b\_CO \in \textsf{Variable} $\newline $b\_CO \in ran(Concept\_corresp\_AbstractSet) \vee  b\_CO \in ran(Concept\_corresp\_Constant)$ \newline$x\_CO \subseteq b\_CO$  &  & 
$Concept\_isVariable(CO) =  TRUE$
 \\ 
 \hline
  3 &
 \textbf{Constant (resp. Variable) typed as a relation with the range corresponding to a data set} & b\_AT b\_CO b\_DS &$b\_AT \in \textsf{Constant} ~ (resp. ~ \textsf{Variable}) $\newline $b\_CO \in ran(Concept\_corresp\_AbstractSet) \cup ran(Concept\_corresp\_Constant)$\newline
 $b\_DS \in ran(DataSet\_corresp\_Set)$
  \newline$b\_AT \in  b\_CO \leftrightarrow b\_DS$  & AT & 
$AT \in \textsf{Attribute} $ \newline
$Attribute\_domain\_Concept(AT) =  CO$
\newline
$Attribute\_range\_DataSet(AT) =  DS$
\newline
$Attribute\_isVariable(AT) =  FALSE$ \newline
 (The \textit{isVariable} property is set to \textit{TRUE} if $b\_AT \in \textsf{Variable}$)
\newline
{ The properties of \textit{AT} such as \textit{isFunctional} are set according to the type of \textit{b\_AT} (partial/total function, ...).}
 \\ 
 \hline
4 &  \textbf{Constant (resp. Variable) typed as a relation with the range corresponding to a concept} & b\_RE b\_CO1 b\_CO2 &$b\_RE \in \textsf{Constant}~ (resp.~ \textsf{Variable})$ \newline $\{b\_CO1, b\_CO2\} \subset ran(Concept\_corresp\_AbstractSet) \cup ran(Concept\_corresp\_Constant)$ \newline$b\_RE \in  b\_CO1 \leftrightarrow b\_CO2$  & RE & 
$RE \in \textsf{Relation} $ \newline
$Relation\_domain\_Concept(RE) =  CO1$
\newline
$Relation\_range\_Concept(RE) =  CO2$
\newline
$Relation\_isVariable(RE) =  FALSE$\newline
 (The \textit{isVariable} property is set to \textit{TRUE} if $b\_RE \in \textsf{Variable}$)
\newline
{ As usual, the cardinalities of \textit{RE} are set according to the type of \textit{b\_RE} (\textit{function},  \textit{injection}, ...).}
 \\ 
  \hline
5 & \textbf{Constant typed as subset of the correspondent of a concept} & b\_CO b\_PCO &$b\_CO \in \textsf{Constant} $\newline $b\_PCO \in ran(Concept\_corresp\_AbstractSet) \vee  b\_PCO \in ran(Concept\_corresp\_Constant)$ \newline$b\_CO \subseteq b\_PCO$  & CO & 
$CO \in \textsf{Concept} $ \newline
$Concept\_parentConcept\_\-Concept(CO) = PCO$ \newline
$Concept\_isVariable(CO) =  FALSE$
 \\ 
 \hline
 6 & \textbf{Set item} & b\_elt b\_ES & $b\_elt \in{} SetItem$
\newline
$b\_ES = SetItem\_itemOf\_EnumeratedSet(b\_elt)$
\newline
\textit{b\_ES} has a domain model correspondent  & elt & 
$elt \in \textsf{DataValue} $ \newline
$DataValue\_elements\_\-EnumeratedDataSet(elt) =  ES$
 \\
\hline
 7 & \textbf{Constant typed as element of the correspondent of a concept} & b\_ind b\_CO &$b\_ind \in \textsf{Constant} $\newline $b\_CO \in ran(Concept\_corresp\_AbstractSet) \vee  b\_CO \in ran(Concept\_corresp\_Constant)$ \newline$b\_ind \in  b\_CO$  & ind & 
$ind \in \textsf{Individual} $ \newline
$Individual\_individualOf\_\-Concept(ind) =  CO$
 \\ 
 \hline
 8 &  \textbf{Constant typed as element of the correspondent of a data set} & b\_dva b\_DS &$b\_dva \in \textsf{Constant} $\newline $b\_DS \in ran(DataSet\_corresp\_Set)$ \newline$b\_dva \in  b\_DS$  & dva & 
$dva \in \textsf{DataValue} $ \newline
$DataValue\_valueOf\_DataSet(dva) =  DS$
 \\ 
\hline

\end{longtable}
\end{center}
\end{scriptsize}

The addition of a non typing logic formula (logic formula that does not contribute to the definition of the type of a formal element) in the \textit{B System} specification is propagated through the definition of the same formula  in the corresponding domain model, since both languages use first-order logic notations. This back propagation is limited  to a syntactic translation.

In what follows, we provide a description of some relevant rules.
These rules have been chosen to make explicit the formalism used in  Table \ref{tableau_recapitulatif_correspondances}.

\subsection{Addition of Abstract Sets} \label{addition_abstract_set}
An abstract set \textit{b\_CO} (instance of class \textsf{AbstractSet} of the metamodel of  Fig. \ref{eventb_metamodel}) introduced in the \textit{B System} specification gives a concept \textit{CO} (instance of class \textsf{Concept} of the metamodel of  Fig. \ref{our_businessdomain_metamodel}) having its property \textit{isVariable} set to \textit{FALSE} (line 1  of Table \ref{tableau_recapitulatif_correspondances}). 
If \textit{b\_CO} is set as  the superset of a variable \textit{x\_CO}, then it is possible to dynamically add/remove individuals from concept \textit{CO}: thus, property \textit{isVariable} of \textit{CO} must be set to \textit{TRUE} (line 2  of Table \ref{tableau_recapitulatif_correspondances}).

\subsection{Addition of Constants or Variables typed as relations}
The introduction in the \textit{B System} specification  of a constant typed as a relation can be back propagated, within the domain model,  with the definition of a constant attribute (instance of class \textsf{Attribute}) or  relation (instance of class \textsf{Relation}): (1) if the range of the constant is the correspondence of a data set (instance of class \textsf{DataSet}), then the element added within the domain model must be an attribute (line  3  of Table \ref{tableau_recapitulatif_correspondances});  (2) however, if the range is the correspondence of a concept (instance of class \textsf{Concept}), then the element added within the domain model must be a relation (line  4  of Table \ref{tableau_recapitulatif_correspondances}).
When the \textit{B System} relation is a variable, then property \textit{isVariable} of the relation or attribute introduced in the domain model is set to \textit{true}.

\subsection{Addition of   Subsets  of  Correspondences of  concepts}
A constant \textit{b\_CO} introduced in the \textit{B System} specification and defined as a subset of \textit{b\_PCO}, the correspondent of a concept \textit{PCO}, gives a concept \textit{CO} having \textit{PCO} as its parent concept (association \textsf{parentConcept} of the metamodel of  Fig. \ref{our_businessdomain_metamodel}) (line 5   of Table \ref{tableau_recapitulatif_correspondances}).
If \textit{b\_CO} is set as  the superset of a variable \textit{x\_CO}, then it is possible to dynamically add/remove individuals from concept \textit{CO}: thus, property \textit{isVariable} of \textit{CO} must be set to \textit{TRUE} (line 2  of Table \ref{tableau_recapitulatif_correspondances}).

\subsection{Addition of Set Items}


An item \textit{b\_elt} (instance of class \textsf{SetItem} of the metamodel of  Fig. \ref{eventb_metamodel}) added to a set \textit{b\_ES} gives a data value \textit{elt} (instance of class \textsf{DataValue} of the metamodel of  Fig. \ref{our_businessdomain_metamodel}) linked to the enumerated dataset corresponding to \textit{b\_ES} with the  association \textsf{element}  (line 6  of Table \ref{tableau_recapitulatif_correspondances}). 


\section{Specification of  Rules in Event-B}

\subsection{Specification of Source and Target Metamodels}
\begin{figure*}[!h]
\begin{center}
\includegraphics[width=1.0\textwidth]{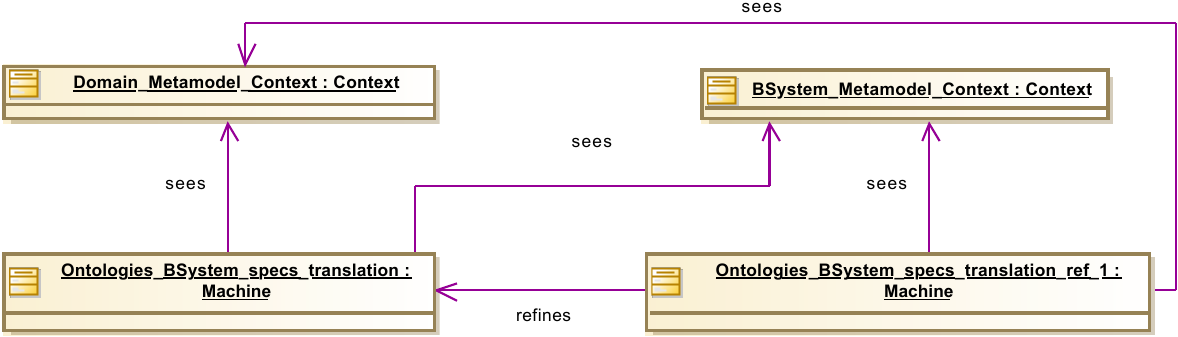}
\end{center}
\caption{\label{Rodin_SysMLKAOSDomainModelRules_structure} Structure of the \textit{Event-B} specification}
\paragraph{}
\end{figure*}

As we have chosen \textit{Event-B} to express and verify the translation rules between the source and target metamodels, the first step is to specified them in \textit{Event-B}. 
This  also allows us to formally define the semantics of  SysML/KAOS domain models.
Figure \ref{Rodin_SysMLKAOSDomainModelRules_structure} represents the structure of the whole \textit{Event-B} specification.  
 This specification can only be splitten into two abstraction levels because all the translation rules use the class \textsf{LogicFormula}, except those related to the class \textsf{DomainModel}. 
The first machine, \texttt{Ontologies\_BSystem\_specs\_translation}, 
contains the rules for the translation of instances of \textsf{DomainModel} into instances of \textsf{Component}.  
The other rules are defined in the machine \texttt{Ontologies\_BSystem\_specs\_\-translation\_ref\_1}.
We have defined static elements of the  target metamodel in a context named \texttt{BSystem\_Meta\-model\_Context} and static elements 
 of the source metamodel in the one named  \texttt{Domain\_Metamodel\_Context}. 
The two machines have access to the definitions of  the contexts.
  For the sake of concision, we provide only an illustrative excerpt of these \textit{Event-B} specifications.  For instance, the model
\texttt{Ontologies\_\-BSystem\_specs\_translation\_ref\_1}
   contains more than a hundred variables, a hundred invariants and fifty events and 
   it gives rise to a
   thousand proof obligations. 
   The full version can be found in \cite{translation_rules_event_b_specification_full_code_link,1712.07406}.
  
  For the translation of some metamodel elements, we have followed the rules  proposed in   \cite{DBLP:conf/kbse/LaleauM00,Snook:2006:UFM:1125808.1125811}, such as :
 classes which are not subclasses give rise to abstract sets, each class gives rise to a variable typed as a subset and containing its instances and each association or property gives rise to a variable typed as a relation.
 For example, in the following specification,   class \textsf{DomainModel}   of the source metamodel and class \textsf{Component} of the target metamodel give rise to abstract sets representing all their possible instances. 
Variables are introduced and typed (\texttt{inv0\_1, inv0\_2} and \texttt{inv0\_3}) to represent sets of defined instances.

\begin{scriptsize}

\textcolor{keycolor}{\textbf{CONTEXT}}
Domain\_Metamodel\_Context

\textcolor{keycolor}{\textbf{SETS}}  DomainModel\_Set

\textcolor{keycolor}{\textbf{END}}

\noindent \hspace*{0.1in} \textcolor{keycolor}{\textbf{CONTEXT}}
BSystem\_Metamodel\_Context

\textcolor{keycolor}{\textbf{SETS}}  Component\_Set

\textcolor{keycolor}{\textbf{END}}

\noindent \hspace*{-0.08in} \textcolor{keycolor}{\textbf{MACHINE}}
Ontologies\_BSystem\_specs\_translation

\noindent \textcolor{keycolor}{\textbf{VARIABLES}}  Component 
	System 
	Refinement 
	
\noindent 	\hspace*{0.7in} DomainModel
	
\noindent 	\textcolor{keycolor}{\textbf{INVARIANT}}

 \noindent \hspace*{0.1in} \textcolor{labelcolor}{\small{\texttt{inv0\_1}}}: $Component\subseteq{}Component\_Set$
 
\noindent  \hspace*{0.1in} \textcolor{labelcolor}{\small{\texttt{inv0\_2}}}: {\tiny $partition(Component, System, Refinement)$}
 
 \noindent \hspace*{0.1in} \textcolor{labelcolor}{\small{\texttt{inv0\_3}}}: $DomainModel \subseteq{} DomainModel\_Set$
 
 \noindent \textcolor{keycolor}{\textbf{END}}
\end{scriptsize}
 
 UML enumerations are represented as \textit{Event-B} enumerated sets. For example, in the following specification, defined in \texttt{BSystem\_Metamodel\_Context},  class \textsf{Operator} of the target metamodel is represented as an enumerated set containing the constants \textit{Inclusion\_OP}, 	\textit{Belonging\_OP} and 	\textit{BecEq2Set\_OP}.
 
\begin{scriptsize}

\noindent\textcolor{keycolor}{\textbf{SETS}} Operator\\
\noindent \textcolor{keycolor}{\textbf{CONSTANTS}} Inclusion\_OP
	Belonging\_OP
	BecEq2Set\_OP
	\\	
\noindent	\textcolor{keycolor}{\textbf{AXIOMS}}
\hspace*{0.1in} \textcolor{labelcolor}{\small{\texttt{axiom1}}}: $partition(Operator, \{Inclusion\_OP\}, \{Belonging\_OP\}, \{BecEq2Set\_OP\})$
\end{scriptsize}

 Variables are also used to represent attributes and associations \cite{DBLP:conf/kbse/LaleauM00,Snook:2006:UFM:1125808.1125811} such as    
the attribute \textsf{isVariable} of the class \textsf{Concept} in the
source metamodel (\texttt{inv1\_5}) and  the association \textsf{definedIn} between the classes \textsf{Constant} and \textsf{Component} in the target metamodel (\texttt{inv1\_7}).  
To avoid ambiguity, 
we have prefixed and suffixed  each  element name with that of the class to which it is attached  (\textbf{e.g.} \texttt{Concept_isVariable} or \texttt{Constant_definedIn_Component}).
Furthermore, for a better readability of the specification, we have chosen to add \textit{"s"} to the name of all \textit{Event-B} relations for which an image is a set  (\textbf{e.g.} \texttt{Constant\_isInvolvedIn\_\-LogicFormulas} or \texttt{Invariant\_involves\_Variables}).

\begin{scriptsize}

\noindent \hspace*{-0.08in} \textcolor{keycolor}{\textbf{MACHINE}}
Ontologies\_BSystem\_specs\_translation\_ref\_1

\noindent \textcolor{keycolor}{\textbf{VARIABLES}}  Concept\_isVariable
	Constant\_definedIn\_Component
	Invariant\_involves\_Variables
	
	\noindent 	\hspace*{0.7in}
	Constant\_isInvolvedIn\_LogicFormulas
		
\noindent 	\textcolor{keycolor}{\textbf{INVARIANT}}

 \noindent \hspace*{0.1in} \textcolor{labelcolor}{\small{\texttt{inv1\_5}}}: $Concept\_isVariable  \in{} Concept \tfun{} BOOL$
 
\noindent  \hspace*{0.1in} \textcolor{labelcolor}{\small{\texttt{inv1\_7}}}: $Constant\_definedIn\_Component \in{}  Constant \tfun{} Component$
 
 \noindent \hspace*{0.1in} \textcolor{labelcolor}{\small{\texttt{inv1\_11}}}: $Invariant\_involves\_Variables \in{}  Invariant \tfun{} (\natn{}\pfun{}Variable)$
 
  \noindent \hspace*{0.1in} \textcolor{labelcolor}{\small{\texttt{inv1\_12}}}: $ran(union(ran(Invariant\_involves\_Variables)))=Variable$
  
   \noindent \hspace*{0.1in} \textcolor{labelcolor}{\small{\texttt{inv1\_13}}}: $Constant\_isInvolvedIn\_LogicFormulas \in{}  Constant \tfun{} \pown{}(\natn{}\cprod{}LogicFormula)$
   
    \noindent \hspace*{0.1in} \textcolor{labelcolor}{\small{\texttt{inv1\_14}}}:  $\forall{}co\qdot{}(co\in{}Constant\limp{}ran(Constant\_isInvolvedIn\_LogicFormulas(co))\binter{}\\
\noindent 	\hspace*{0.7in}    
    Property\neq{}\emptyset{})$
 
 \noindent \textcolor{keycolor}{\textbf{END}}

\end{scriptsize}

\noindent An  association \textit{r} from a class \textit{A} to a class \textit{B} to which the \textit{ordered} constraint is attached is represented as a variable \textit{r} typed through the invariant  $r \in (A \tfun{} (\natn{}\pfun{} B))$. This is for example the case of the association \textit{Invariant\_involves\_Variables} of the target metamodel (\texttt{inv1\_11}). 
If  instances of \textit{B}  have the same sequence number, then the invariant becomes $r \in (A \tfun{} \pown{}(\natn{}\cprod{}B))$. This is for example the case of the association \textit{Constant\_isInvolvedIn\_LogicFormulas} of the target metamodel (\texttt{inv1\_13}).  Invariant \texttt{inv1\_12} ensures that each variable is involved in at least one invariant and \texttt{inv1\_14} ensures the same constraint for constants.

\subsection{Event-B Specification of Translation Rules}
The correspondence links between instances of a class \textsf{A} of the source metamodel and instances of a class \textsf{B} of the target metamodel are captured in a variable named \textit{A\_corresp\_B} typed by the invariant $A\_corresp\_B \in A \pinj B$. It is  an  injection because  each  instance, on both sides,  must have at most one correspondence. The injection is partial because all the  elements are not translated at the same time.
Thus, it is possible that at an intermediate state of the system, there are elements not yet translated.
 For example, correspondence links between instances of \textsf{Concept} and instances of \textsf{AbstractSet}
are captured as follows

\begin{footnotesize}
\noindent \textcolor{keycolor}{\textbf{INVARIANTS}}
\hspace*{0.1in} \textcolor{labelcolor}{\small{\texttt{inv1\_8}}}: $Concept\_corresp\_AbstractSet \in{} Concept \pinj{} AbstractSet$

\end{footnotesize}

\noindent Translation rules have been modeled as  \textit{convergent} events. 
Each event execution translates an element of the source into the target.
 Variants and event guards and type  have been defined such that when the system  reaches a state where no transition is possible (deadlock state), all translations are done.
Up to fifty   events have been specified.
 The rest of this section  provides an overview of the specification of some of these events in order to illustrate the formalisation process and some of its benefits and difficulties. The full specification can be found in \cite{translation_rules_event_b_specification_full_code_link,1712.07406}.

\subsubsection{Translating a Domain Model with Parent (line 2 of  table \ref{tableau_recapitulatif_correspondances})} \label{sect_domain_model_with_parent}

The corresponding event  is called \textit{domain\_model\_with\_parent\_to\_component}.
It states that a domain model, associated with another one representing its parent,  gives rise to a refinement component.

\begin{scriptsize}

\noindent \hspace*{-0.08in} \textcolor{keycolor}{\textbf{MACHINE}} Ontologies\_BSystem\_specs\_translation

\noindent 	\textcolor{keycolor}{\textbf{INVARIANT}}

 \noindent \hspace*{0.1in} \textcolor{labelcolor}{\small{\texttt{inv0\_6}}}: $Refinement\_refines\_Component  \in{} Refinement \tinj{} Component$
 
\noindent  \hspace*{0.1in} \textcolor{labelcolor}{\small{\texttt{inv0\_7}}}: $\forall{}xx,px\qdot{}(~(~  	xx\in{}dom(DomainModel\_parent\_DomainModel)~  						\land{} ~ px=DomainModel\_\-parent\_DomainModel(xx) ~   						\land{}  px \in{} dom(DomainModel\_corresp\_Component)~  						\land{} xx \notin{} dom(Domain\-Model\_corresp\_Component) ~   					)   					\limp{}DomainModel\_corresp\_Component(px) \notin{} ran(Refinement\_\-refines\_Component )~ )$
 
\hspace*{-0.10in}{\footnotesize \textcolor{keycolor}{\textbf{Event}}} domain\_model\_with\_parent\_to\_component  \textcolor{keycolor}{\small{$\langle$convergent$\rangle$}} $\defi$

 \hspace*{-0.05in} \textcolor{keycolor}{\textbf{any}} DM PDM o\_DM

 \hspace*{-0.05in}\textcolor{keycolor}{\textbf{where}}

 \hspace*{0.in} \textcolor{labelcolor}{\small{\texttt{grd0}}}: {\tiny $dom(DomainModel\_parent\_DomainModel) \setminus{}  dom(DomainModel\_corresp\_Component) \neq{} \emptyset{}$}

 \hspace*{0.in} \textcolor{labelcolor}{\small{\texttt{grd1}}}: {\tiny $DM \in{} dom(DomainModel\_parent\_DomainModel) \setminus{}  dom(DomainModel\_corresp\_Component)$}

 \hspace*{0.in} \textcolor{labelcolor}{\small{\texttt{grd2}}}: $dom(DomainModel\_corresp\_Component) \neq{} \emptyset{}$
 
  \hspace*{0.in} \textcolor{labelcolor}{\small{\texttt{grd3}}}: $PDM \in{} dom(DomainModel\_corresp\_Component)$
  
   \hspace*{0.in} \textcolor{labelcolor}{\small{\texttt{grd4}}}: $DomainModel\_parent\_DomainModel(DM)=PDM$
   
    \hspace*{0.in} \textcolor{labelcolor}{\small{\texttt{grd5}}}: $Component\_Set \setminus{} Component \neq{}\emptyset{}$
    
     \hspace*{0.in} \textcolor{labelcolor}{\small{\texttt{grd6}}}: $o\_DM \in{} Component\_Set \setminus{} Component $

 \hspace*{-0.05in} \textcolor{keycolor}{\textbf{then}}

 \hspace*{0.in} \textcolor{labelcolor}{\small{\texttt{act1}}}: $Refinement \bcmeq{}  Refinement \bunion{}  \{o\_DM\}$

 \hspace*{0.in} \textcolor{labelcolor}{\small{\texttt{act1}}}: $Component \bcmeq{}  Component \bunion{}  \{o\_DM\}$
 
  \hspace*{0.in} \textcolor{labelcolor}{\small{\texttt{act1}}}: $Refinement\_refines\_Component(o\_DM) \bcmeq{} DomainModel\_corresp\_Component(PDM)$
  
   \hspace*{0.in} \textcolor{labelcolor}{\small{\texttt{act1}}}: $DomainModel\_corresp\_Component(DM)\bcmeq{}o\_DM$

 \hspace*{-0.05in} \textcolor{keycolor}{\textbf{END}}
 
 \noindent \textcolor{keycolor}{\textbf{END}}

\end{scriptsize}

\noindent The  refinement component must be the one refining the component corresponding  to the parent domain model.
Guard \texttt{grd1} is the main guard of the event. It is used to ensure that the event will only handle instances of \textsf{DomainModel} with parent and only instances which have not yet been translated. 
It also guarantee that the event will be enabled until all these instances are translated.
Action \texttt{act3} states that \textit{o\_DM} refines the correspondent of \textit{PDM}. To discharge, for this event, the proof obligation related to  the invariant \texttt{inv0\_6}, it is necessary to guarantee 
that, given  a domain model \textit{m} not translated yet, and its parent \textit{pm} that  has been translated into component \textit{o\_pm}, then \textit{o\_pm} has no refinement yet.  
The  invariant \texttt{inv0\_7} then appears accordingly to encode this constraint.


\subsubsection{Translating a Concept with Parent (line  4 of  table \ref{tableau_recapitulatif_correspondances})}
This rule leads to two events : the first one for when the parent concept corresponds to an abstract set (the parent concept does not have a parent : line 3 of table \ref{tableau_recapitulatif_correspondances})  and the second one for when the parent concept corresponds to a constant (the parent concept has a parent : line 4 of table \ref{tableau_recapitulatif_correspondances}). Below is the specification  of the first event\footnote{Some guards and actions have been removed for the sake of concision}.

\begin{scriptsize}

 
\hspace*{-0.10in}{\footnotesize \textcolor{keycolor}{\textbf{Event}}} concept\_with\_parent\_to\_constant\_1  \textcolor{keycolor}{\small{$\langle$convergent$\rangle$}} $\defi$

 \hspace*{-0.05in} \textcolor{keycolor}{\textbf{any}} CO o\_CO PCO o\_lg o\_PCO

 \hspace*{-0.05in}\textcolor{keycolor}{\textbf{where}}

 \hspace*{0.in} \textcolor{labelcolor}{\small{\texttt{grd1}}}:  $CO \in{} dom(Concept\_parentConcept\_Concept) \setminus{} dom(Concept\_corresp\_Constant)$
 
  \hspace*{0.in} \textcolor{labelcolor}{\small{\texttt{grd2}}}: $PCO \in{} dom(Concept\_corresp\_AbstractSet)$
  
   \hspace*{0.in} \textcolor{labelcolor}{\small{\texttt{grd3}}}: $Concept\_parentConcept\_Concept(CO)=PCO$
   
    \hspace*{0.in} \textcolor{labelcolor}{\small{\texttt{grd4}}}: $Concept\_definedIn\_DomainModel(CO) \in{} dom(DomainModel\_corresp\_Component)$

  \hspace*{0.in} \textcolor{labelcolor}{\small{\texttt{grd5}}}: $o\_CO \in{} Constant\_Set \setminus{} Constant$

  \hspace*{0.in} \textcolor{labelcolor}{\small{\texttt{grd6}}}: $o\_lg \in{} LogicFormula\_Set \setminus{} LogicFormula$
                             
  \hspace*{0.in} \textcolor{labelcolor}{\small{\texttt{grd7}}}: $o\_PCO = Concept\_corresp\_AbstractSet(PCO)$
                      
 \hspace*{-0.05in} \textcolor{keycolor}{\textbf{then}}

 \hspace*{0.in} \textcolor{labelcolor}{\small{\texttt{act1}}}: $Constant \bcmeq{}  Constant \bunion{}  \{o\_CO\}$

 \hspace*{0.in} \textcolor{labelcolor}{\small{\texttt{act2}}}: $Concept\_corresp\_Constant(CO)\bcmeq{}o\_CO$
 
  \hspace*{0.in} \textcolor{labelcolor}{\small{\texttt{act3}}}: $Constant\_definedIn\_Component(o\_CO) \bcmeq{} DomainModel\_corresp\_Component(\\
 \noindent 	\hspace*{0.7in}  
  Concept\_definedIn\_DomainModel(CO))$
  
   \hspace*{0.in} \textcolor{labelcolor}{\small{\texttt{act4}}}: $Property \bcmeq{} Property \bunion{} \{o\_lg\}$
   
   \hspace*{0.in} \textcolor{labelcolor}{\small{\texttt{act5}}}: $LogicFormula \bcmeq{} LogicFormula \bunion{} \{o\_lg\}$
   
     \hspace*{0.in} \textcolor{labelcolor}{\small{\texttt{act6}}}: $LogicFormula\_uses\_Operators(o\_lg) \bcmeq{} \{1\mapsto{}Inclusion\_OP\}$
     
     \hspace*{0.in} \textcolor{labelcolor}{\small{\texttt{act7}}}: $Constant\_isInvolvedIn\_LogicFormulas(o\_CO) \bcmeq{} \{1\mapsto{}o\_lg\}$
     
     \hspace*{0.in} \textcolor{labelcolor}{\small{\texttt{act8}}}: $LogicFormula\_involves\_Sets(o\_lg) \bcmeq{} \{2\mapsto{}o\_PCO\}$   
     
     \hspace*{0.in} \textcolor{labelcolor}{\small{\texttt{act9}}}: $Constant\_typing\_Property(o\_CO) \bcmeq{}  o\_lg$

 \hspace*{-0.05in} \textcolor{keycolor}{\textbf{END}}
 

\end{scriptsize}

The rule asserts that any  concept,  associated with another one, with the \textsf{parentConcept} association, gives rise to a constant, typed as a subset of the \textit{B System} element corresponding to the parent concept. 
We use an instance of \textsf{LogicFormula}, named \textit{o\_lg}, to capture this constraint linking the concept  and its parent correspondents (\textit{o\_CO} and \textit{o\_PCO}). 
Guard \texttt{grd2} constrains the parent correspondent to be an instance of \textsf{AbstractSet}. Guard \texttt{grd4} ensures that the event will not be triggered until the translation of the  domain model containing the definition of the concept.  Action \texttt{act3}  ensures that  \textit{o\_CO} is defined in the component corresponding to the domain model where \textit{CO} is defined.
 Action \texttt{act6}  defines the operator used by \textit{o\_lg}.
Because the parent concept corresponds to an abstract set,  \textit{o\_CO} is the only constant involved in \textit{o\_lg} (\texttt{act7}); \textit{o\_PCO}, the second operand, is a set  (\texttt{act8}). 
   Finally, action \texttt{act9} defines \textit{o\_lg} as the typing predicate of \textit{o\_CO}.

    \textbf{\underline{Example : }}

\begin{scriptsize}
\begin{center}
\begin{longtable}{|p{6.cm}|p{6.cm}|}
\hline
\textbf{SysML/KAOS domain model} &\textbf{B System specification} \\
\hline
\textcolor{darkblue}{ \textbf{\texttt{concept}}}\texttt{ pco}\newline\newline
\noindent \textcolor{darkblue}{ \textbf{\texttt{concept}}} \texttt{co} \textcolor{darkblue} {\textbf{\texttt{parent concept}}} \texttt{pco}
& 
\hspace*{-0.05in} \textcolor{keycolor}{\textbf{SETS}}
\hspace*{0.20in} $pco$\newline
\noindent\textcolor{keycolor}{\textbf{CONSTANTS}}
\hspace*{0.20in}  $co$\newline
\noindent\textcolor{keycolor}{\textbf{PROPERTIES}}
 \hspace*{0.1in} co $\subseteq$ pco 
 \\
 \hline
\end{longtable}
\end{center}
\end{scriptsize}

  The specification of the second event   (when the parent concept corresponds to a constant) is different from the specification of the first one in some points. The three least trivial differences appear at guard \texttt{grd2} and at actions \texttt{act7}  and \texttt{act8}.
Guard \texttt{grd2} constrains the parent correspondent to be an instance of \textsf{Constant}  :  $PCO \in{} dom(Concept\_corresp\_Constant)$.  Thus,  the first and the second operands involved in \textit{o\_lg} are constants :

 \begin{scriptsize}
  \hspace*{-0.25in}  \textcolor{labelcolor}{\small{\texttt{act7}}}: $Constant\_isInvolvedIn\_LogicFormulas\bcmeq{}Constant\_isInvolvedIn\_LogicFormulas\ovl{}\{\\
\hspace*{0.30in} (o\_CO\mapsto{} 
\{1\mapsto{}o\_lg\}),\\
\hspace*{0.30in} o\_PCO\mapsto{}Constant\_isInvolvedIn\_LogicFormulas(o\_PCO)\bunion{}\{2\mapsto{}o\_lg\}\}$

	\noindent	    \textcolor{labelcolor}{\small{\texttt{act8}}}: $LogicFormula\_involves\_Sets(o\_lg) \bcmeq{} \emptyset{}$

 \end{scriptsize}

   This approach to modeling logic formulas allows us to capture all the information conveyed by the predicate which can then be used to make inferences and semantic analysis.
   It is especially useful when we deal
with rules
to propagate changes made to a generated \textit{B System} specification back to the domain model (ie, propagate changes made to the target into the source).

\subsection{Event-B Specification of Back Propagation Rules}
We have modeled back propagation rules  as \textit{Event-B}  \textit{convergent} events;
each  execution of an event  propagates the addition of an element.

\subsection{Addition of a Constant, Subset of the Correspondence of an Instance of \textsf{Concept} (line 5 of  table \ref{tableau_recapitulatif_correspondances})}
This rule leads to two events: the first one is applied for a superset that is an abstract set   and the second one for a superset that is a constant. Below is the specification  of the first event.

\begin{scriptsize}

 
\hspace*{-0.10in}{\footnotesize \textcolor{keycolor}{\textbf{Event}}} constant\_subset\_concept\_1  \textcolor{keycolor}{\small{$\langle$convergent$\rangle$}} $\defi$

 \hspace*{-0.05in} \textcolor{keycolor}{\textbf{any}} CO b\_CO PCO b\_lg b\_PCO

 \hspace*{-0.05in}\textcolor{keycolor}{\textbf{where}}

 
 \textcolor{labelcolor}{\small{\texttt{grd1}}}: $b\_CO \in{} dom(Constant\_typing\_Property) \setminus{} \\ \noindent 	\hspace*{0.7in}ran(Concept\_corresp\_Constant)$
  
   \textcolor{labelcolor}{\small{\texttt{grd2}}}: $b\_lg = Constant\_typing\_Property(b\_CO)$
   
     \textcolor{labelcolor}{\small{\texttt{grd3}}}: $LogicFormula\_uses\_Operators(b\_lg) = \\
 \noindent 	\hspace*{0.7in}\{1\mapsto{}Inclusion\_OP\}$


   \textcolor{labelcolor}{\small{\texttt{grd4}}}: $(2\mapsto{}b\_PCO)\in{}LogicFormula\_involves\_Sets(b\_lg)$
                             
   \textcolor{labelcolor}{\small{\texttt{grd5}}}: $b\_PCO \in{} ran(Concept\_corresp\_AbstractSet)$
  
   \textcolor{labelcolor}{\small{\texttt{grd6}}}: $PCO = Concept\_corresp\_AbstractSet^{-1} (b\_PCO)$
  
   \textcolor{labelcolor}{\small{\texttt{grd7}}}: $CO \in{} Concept\_Set \setminus{} Concept$
  
   \textcolor{labelcolor}{\small{\texttt{grd8}}}: $Constant\_definedIn\_Component(b\_CO) \in{} ran(\\
 \noindent 	\hspace*{0.7in}DomainModel\_corresp\_Component)$
                      
 \hspace*{-0.1in} \textcolor{keycolor}{\textbf{then}}

 \textcolor{labelcolor}{\small{\texttt{act1}}}: $Concept \bcmeq{}  Concept \bunion{}  \{CO\}$

  \textcolor{labelcolor}{\small{\texttt{act2}}}: $Concept\_corresp\_Constant(CO)\bcmeq{}b\_CO$
 
   \textcolor{labelcolor}{\small{\texttt{act3}}}: $Concept\_definedIn\_DomainModel(CO) \bcmeq{} DomainModel\_\-
 \noindent 	\hspace*{0.3in}corresp\_Component^{-1}(Constant\_definedIn\_Component(b\_CO))$
  
    \textcolor{labelcolor}{\small{\texttt{act4}}}: $Concept\_parentConcept\_Concept(CO) \bcmeq{} PCO$
   
    \textcolor{labelcolor}{\small{\texttt{act5}}}: $Concept\_isVariable(CO) \bcmeq{}  FALSE$

 \hspace*{-0.1in} \textcolor{keycolor}{\textbf{END}}
 

\end{scriptsize}

The rule asserts that in order to propagate the addition of a constant, we need to evaluate its typing predicate. When it is typed as a subset of the correspondence of an instance of \textsf{Concept}, then it gives rise to an instance of \textsf{Concept}.
We use an instance of \textsf{LogicFormula}, named \textit{b\_lg}, to
represent the  typing predicate (\textcolor{labelcolor}{\small{\texttt{grd2}}}) of \textit{b\_CO}, defined with \textcolor{labelcolor}{\small{\texttt{grd1}}}. Guards \textcolor{labelcolor}{\small{\texttt{grd3}}} and \textcolor{labelcolor}{\small{\texttt{grd4}}} ensure that \textit{b\_CO} is typed as a subset. 
Guard \textcolor{labelcolor}{\small{\texttt{grd5}}}  ensures that  the superset, \textit{b\_PCO},   is an abstract set corresponding to an instance of \textsf{Concept}.
Guard \textcolor{labelcolor}{\small{\texttt{grd6}}} constrains \textit{PCO} to be the correspondence of  \textit{b\_PCO}.
\textit{CO}, an instance of \textsf{Concept}, is then elicited and \textcolor{labelcolor}{\small{\texttt{act2}}} defines \textit{b\_CO}  as its correspondence. Finally, \textcolor{labelcolor}{\small{\texttt{act4}}} defines \textit{PCO} as its parent concept.
 Guard \textcolor{labelcolor}{\small{\texttt{grd8}}} ensures that the event will 
 be triggered only if the \textit{B System} component, where \textit{b\_CO} is defined, corresponds to an existing domain model.
  Action \textcolor{labelcolor}{\small{\texttt{act3}}}  ensures that  \textit{CO} is defined in that
domain model.

The specification of the second event   (when the superset is a constant) is different from the specification of the first one in five points:

 \begin{scriptsize}

	\noindent	     \textcolor{labelcolor}{\small{\texttt{grd4}}}: $b\_PCO \in{} dom(Constant\_isInvolvedIn\_LogicFormulas)$
	
	\noindent	     \textcolor{labelcolor}{\small{\texttt{grd5}}}: $(2\mapsto{}b\_lg) \in{} Constant\_isInvolvedIn\_LogicFormulas(b\_PCO)$
	
	\noindent	     \textcolor{labelcolor}{\small{\texttt{grd6}}}: $b\_PCO \in{} ran(Concept\_corresp\_Constant)$
	
	\noindent	     \textcolor{labelcolor}{\small{\texttt{grd7}}}: $PCO = Concept\_corresp\_Constant^{-1}(b\_PCO)$

 \end{scriptsize}

Guard  \textcolor{labelcolor}{\small{\texttt{grd4}}} constrains the superset, \textit{b\_PCO},  to be a constant involved in a logic formula. Guard  \textcolor{labelcolor}{\small{\texttt{grd5}}} ensures that \textit{b\_PCO} is involved as the second operand of \textit{b\_lg}. Finally, guards \textcolor{labelcolor}{\small{\texttt{grd6}}} and \textcolor{labelcolor}{\small{\texttt{grd7}}} constrain the domain model element corresponding to \textit{b\_PCO}.

\subsection{Addition of a Variable, Subset of the Correspondence of an Instance of \textsf{Concept} (line 2 of  table \ref{tableau_recapitulatif_correspondances})}\label{addition_variable_formalisation}
Like the previous rule, this rule leads to two events: the first one for when the superset is an abstract set   and the second one for when the superset is a constant. Below is the most relevant part of the specification  of the first event.

\begin{scriptsize}

 
\hspace*{-0.10in}{\footnotesize \textcolor{keycolor}{\textbf{Event}}} variable\_subset\_concept\_1  \textcolor{keycolor}{\small{$\langle$convergent$\rangle$}} $\defi$

 \hspace*{-0.05in} \textcolor{keycolor}{\textbf{any}} x\_CO CO b\_lg b\_CO

 \hspace*{-0.05in}\textcolor{keycolor}{\textbf{where}}

 \textcolor{labelcolor}{\small{\texttt{grd1}}}: $x\_CO \in{} dom(Variable\_typing\_Invariant) \\
 \noindent 	\hspace*{0.7in}\setminus{} ran(Concept\_corresp\_Variable)$
  
   \textcolor{labelcolor}{\small{\texttt{grd2}}}: $b\_lg = Variable\_typing\_Invariant(x\_CO)$
   
     \textcolor{labelcolor}{\small{\texttt{grd3}}}: $LogicFormula\_uses\_Operators(b\_lg) \\
 \noindent 	\hspace*{0.7in}= \{1\mapsto{}Inclusion\_OP\}$

   \textcolor{labelcolor}{\small{\texttt{grd4}}}: $(2\mapsto{}b\_CO)\in{}LogicFormula\_involves\_Sets(b\_lg)$

   \textcolor{labelcolor}{\small{\texttt{grd5}}}: $b\_CO \in{} ran(Concept\_corresp\_AbstractSet)$
                             
   \textcolor{labelcolor}{\small{\texttt{grd6}}}: $CO = Concept\_corresp\_AbstractSet^{-1}(b\_CO)$
   
    \textcolor{labelcolor}{\small{\texttt{grd7}}}: $CO \notin dom(Concept\_corresp\_Variable)$
                      
 \hspace*{-0.1in} \textcolor{keycolor}{\textbf{then}}

 \textcolor{labelcolor}{\small{\texttt{act1}}}: $Concept\_isVariable(CO) \bcmeq{}  TRUE$

  \textcolor{labelcolor}{\small{\texttt{act2}}}: $Concept\_corresp\_Variable(CO) \bcmeq{} x\_CO$

 \hspace*{-0.1in} \textcolor{keycolor}{\textbf{END}}
 

\end{scriptsize}

In order to propagate the addition of a variable, we need to evaluate its typing invariant. When it is typed as a subset of an abstract set,  correspondence of an instance of \textsf{Concept}, then the \textit{isVariable} property of the concept has to be set to \textit{TRUE}.

\subsection{Other Event-B Specifications of Back Propagation Rules}

\paragraph{Addition of a new abstract set}

\MACHINE{event\_b\_specs\_from\_ontologies\_ref\_1}{event\_b\_specs\_from\_ontologies}{EventB\_Metamodel\_Context,Domain\_Metamodel\_Context}{}
\EVT{rule\_101}{false}{ordinary}{}{\\handling the addition of a new abstract set (correspondence to a concept)}{
	\ANY{
		\Param{CO}{true}{}
		\Param{o\_CO}{true}{}
	}
	\GUARDS{true}{
		\Guard{grd0}{false}{$AbstractSet \setminus{} (ran(Concept\_corresp\_AbstractSet) \bunion{} ran(DataSet\_corresp\_Set)) \neq{}\emptyset{}$}{true}{}
		\Guard{grd1}{false}{$o\_CO \in{} AbstractSet \setminus{} (ran(Concept\_corresp\_AbstractSet) \bunion{} ran(DataSet\_corresp\_Set))$}{true}{}
		\Guard{grd2}{false}{$Set\_definedIn\_Component(o\_CO) \in{} ran(DomainModel\_corresp\_Component)$}{true}{}
		\Guard{grd3}{false}{$Concept\_Set \setminus{} Concept \neq{}\emptyset{}$}{true}{}
		\Guard{grd4}{false}{$CO \in{} Concept\_Set \setminus{} Concept$}{true}{}
	}
	\ACTIONS{true}{
		\Action{act1}{$Concept \bcmeq{}  Concept \bunion{}  \{CO\}$}{true}{}
		\Action{act2}{$Concept\_corresp\_AbstractSet(CO)\bcmeq{}o\_CO$}{true}{}
		\Action{act3}{$Concept\_definedIn\_DomainModel(CO) \bcmeq{} DomainModel\_corresp\_Component\converse{}(\\Set\_definedIn\_Component(o\_CO))$}{true}{}
		\Action{act4}{$Concept\_isVariable(CO) \bcmeq{}  FALSE$}{true}{}
	}
}
\EVT{rule\_102}{false}{ordinary}{}{\\handling the addition of a new abstract set (correspondence to a custom data set)}{
	\ANY{
		\Param{DS}{true}{}
		\Param{o\_DS}{true}{}
	}
	\GUARDS{true}{
		\Guard{grd0}{false}{$AbstractSet \setminus{} (ran(Concept\_corresp\_AbstractSet) \bunion{} ran(DataSet\_corresp\_Set)) \neq{}\emptyset{}$}{true}{}
		\Guard{grd1}{false}{$o\_DS \in{} AbstractSet \setminus{} (ran(Concept\_corresp\_AbstractSet) \bunion{} ran(DataSet\_corresp\_Set))$}{true}{}
		\Guard{grd2}{false}{$Set\_definedIn\_Component(o\_DS) \in{} ran(DomainModel\_corresp\_Component)$}{true}{}
		\Guard{grd3}{false}{$DataSet\_Set \setminus{} DataSet \neq{}\emptyset{}$}{true}{}
		\Guard{grd4}{false}{$DS \in{} DataSet\_Set \setminus{} DataSet$}{true}{}
		\Guard{grd5}{false}{$DS \notin{} \{\_NATURAL,\_INTEGER,\_FLOAT,\_BOOL,\_STRING\}$}{true}{}
	}
	\ACTIONS{true}{
		\Action{act1}{$CustomDataSet \bcmeq{}  CustomDataSet \bunion{}  \{DS\}$}{true}{}
		\Action{act2}{$DataSet \bcmeq{}  DataSet \bunion{}  \{DS\}$}{true}{}
		\Action{act3}{$CustomDataSet\_corresp\_AbstractSet(DS)\bcmeq{}o\_DS$}{true}{}
		\Action{act4}{$DataSet\_definedIn\_DomainModel(DS) \bcmeq{} DomainModel\_corresp\_Component\converse{}(\\Set\_definedIn\_Component(o\_DS))$}{true}{}
		\Action{act5}{$DataSet\_corresp\_Set(DS) \bcmeq{}  o\_DS$}{true}{}
	}
}
\END

\paragraph{Addition of an enumerated set}

\MACHINE{event\_b\_specs\_from\_ontologies\_ref\_1}{event\_b\_specs\_from\_ontologies}{EventB\_Metamodel\_Context,Domain\_Metamodel\_Context}{}
\EVT{rule\_103}{false}{ordinary}{}{\\handling  the addition of an enumerated set}{
	\ANY{
		\Param{EDS}{true}{}
		\Param{o\_EDS}{true}{}
		\Param{elements}{true}{}
		\Param{o\_elements}{true}{}
		\Param{mapping\_elements\_o\_elements}{true}{}
	}
	\GUARDS{true}{
		\Guard{grd0}{false}{$EnumeratedSet \setminus{} ran(DataSet\_corresp\_Set) \neq{}\emptyset{}$}{true}{}
		\Guard{grd1}{false}{$o\_EDS \in{} EnumeratedSet \setminus{} ran(DataSet\_corresp\_Set)$}{true}{}
		\Guard{grd2}{false}{$Set\_definedIn\_Component(o\_EDS) \in{} ran(DomainModel\_corresp\_Component)$}{true}{}
		\Guard{grd3}{false}{$DataSet\_Set \setminus{} DataSet \neq{}\emptyset{}$}{true}{}
		\Guard{grd4}{false}{$EDS \in{} DataSet\_Set \setminus{} DataSet$}{true}{}
		\Guard{grd5}{false}{$DataValue\_Set \setminus{} DataValue \neq{}\emptyset{}$}{true}{}
		\Guard{grd6}{false}{$elements \subseteq{}  DataValue\_Set \setminus{} DataValue$}{true}{}
		\Guard{grd7}{false}{$o\_elements = SetItem\_itemOf\_EnumeratedSet\converse{}[\{o\_EDS\}]$}{true}{}
		\Guard{grd8}{false}{$card(o\_elements) = card(elements)$}{true}{}
		\Guard{grd9}{false}{$mapping\_elements\_o\_elements \in{} elements \tbij{} o\_elements$}{true}{}
		\Guard{grd10}{false}{$ran(DataValue\_corresp\_SetItem)\binter{}o\_elements=\emptyset{}$}{true}{}
		\Guard{grd11}{false}{$EDS \notin{} \{\_NATURAL,\_INTEGER,\_FLOAT,\_BOOL,\_STRING\}$}{true}{}
	}
	\ACTIONS{true}{
		\Action{act1}{$EnumeratedDataSet \bcmeq{}  EnumeratedDataSet \bunion{}  \{EDS\}$}{true}{}
		\Action{act2}{$DataSet \bcmeq{}  DataSet \bunion{}  \{EDS\}$}{true}{}
		\Action{act3}{$EnumeratedDataSet\_corresp\_EnumeratedSet(EDS)\bcmeq{}o\_EDS$}{true}{}
		\Action{act4}{$DataSet\_definedIn\_DomainModel(EDS) \bcmeq{} DomainModel\_corresp\_Component\converse{}(\\Set\_definedIn\_Component(o\_EDS))$}{true}{}
		\Action{act5}{$DataValue \bcmeq{} DataValue \bunion{}  elements$}{true}{}
		\Action{act6}{$DataValue\_elements\_EnumeratedDataSet \bcmeq{}  DataValue\_elements\_EnumeratedDataSet \bunion{} \{(xx\mapsto{}yy) |xx\in{}elements \land{}yy=EDS\}$}{true}{}
		\Action{act7}{$DataValue\_corresp\_SetItem \bcmeq{}  DataValue\_corresp\_SetItem \bunion{} mapping\_elements\_o\_elements$}{true}{}
		\Action{act8}{$DataSet\_corresp\_Set \bcmeq{}  DataSet\_corresp\_Set \ovl{} \{EDS\mapsto{}o\_EDS\}$}{true}{}
		\Action{act9}{$DataValue\_valueOf\_DataSet \bcmeq{}  DataValue\_valueOf\_DataSet \bunion{} \{(xx\mapsto{}yy) |xx\in{}elements \land{}yy=EDS\}$}{true}{}
		\Action{act10}{$CustomDataSet \bcmeq{}  CustomDataSet \bunion{}  \{EDS\}$}{true}{}
	}
}
\END

\paragraph{Addition of a set item}

\MACHINE{event\_b\_specs\_from\_ontologies\_ref\_1}{event\_b\_specs\_from\_ontologies}{EventB\_Metamodel\_Context,Domain\_Metamodel\_Context}{}

\EVT{rule\_104}{false}{ordinary}{}{\\handling  the addition of a new element in an existing enumerated set}{
	\ANY{
		\Param{EDS}{true}{}
		\Param{o\_EDS}{true}{}
		\Param{element}{true}{}
		\Param{o\_element}{true}{}
	}
	\GUARDS{true}{
		\Guard{grd0}{false}{$dom(SetItem\_itemOf\_EnumeratedSet) \setminus{} ran(DataValue\_corresp\_SetItem) \neq{}\emptyset{}$}{true}{}
		\Guard{grd1}{false}{$o\_element \in{} dom(SetItem\_itemOf\_EnumeratedSet) \setminus{} ran(DataValue\_corresp\_SetItem)$}{true}{}
		\Guard{grd2}{false}{$o\_EDS = SetItem\_itemOf\_EnumeratedSet(o\_element)$}{true}{}
		\Guard{grd3}{false}{$o\_EDS \in{} ran(EnumeratedDataSet\_corresp\_EnumeratedSet)$}{true}{}
		\Guard{grd4}{false}{$EDS = EnumeratedDataSet\_corresp\_EnumeratedSet\converse{}(o\_EDS)$}{true}{}
		\Guard{grd5}{false}{$DataValue\_Set \setminus{} DataValue \neq{}\emptyset{}$}{true}{}
		\Guard{grd6}{false}{$element \in{} DataValue\_Set \setminus{} DataValue$}{true}{}
	}
	\ACTIONS{true}{
		\Action{act1}{$DataValue \bcmeq{}  DataValue \bunion{}  \{element\}$}{true}{}
		\Action{act2}{$DataValue\_elements\_EnumeratedDataSet(element) \bcmeq{}  EDS$}{true}{}
		\Action{act3}{$DataValue\_corresp\_SetItem(element) \bcmeq{}  o\_element$}{true}{}
		\Action{act4}{$DataValue\_valueOf\_DataSet(element) \bcmeq{}  EDS$}{true}{}
	}
}

\END

 \paragraph{Addition of a constant, sub set of an instance of Concept (full)}

\MACHINE{event\_b\_specs\_from\_ontologies\_ref\_1}{event\_b\_specs\_from\_ontologies}{EventB\_Metamodel\_Context,Domain\_Metamodel\_Context}{}

\EVT{rule\_105\_1}{false}{ordinary}{}{\\handling  the addition of a constant, sub set of an instance of Concept (case where the concept corresponds to an abstract set)}{
	\ANY{
		\Param{CO}{true}{}
		\Param{o\_CO}{true}{}
		\Param{PCO}{true}{}
		\Param{o\_lg}{true}{}
		\Param{o\_PCO}{true}{}
	}
	\GUARDS{true}{
		\Guard{grd0}{false}{$dom(Constant\_typing\_Property) \setminus{} ran(Concept\_corresp\_Constant) \neq{}\emptyset{}$}{true}{}
		\Guard{grd1}{false}{$o\_CO \in{} dom(Constant\_typing\_Property) \setminus{} ran(Concept\_corresp\_Constant)$}{true}{}
		\Guard{grd2}{false}{$o\_lg = Constant\_typing\_Property(o\_CO)$}{true}{}
		\Guard{grd3}{false}{$LogicFormula\_uses\_Operators(o\_lg) = \{1\mapsto{}Inclusion\_OP\}$}{true}{}
		\Guard{grd4}{false}{$LogicFormula\_involves\_Sets(o\_lg) \neq{} \emptyset{}$}{true}{}
		\Guard{grd5}{false}{$(2\mapsto{}o\_PCO)\in{}LogicFormula\_involves\_Sets(o\_lg)$}{true}{}
		\Guard{grd6}{false}{$o\_PCO \in{} ran(Concept\_corresp\_AbstractSet)$}{true}{}
		\Guard{grd7}{false}{$PCO = Concept\_corresp\_AbstractSet\converse{}(o\_PCO)$}{true}{}
		\Guard{grd8}{false}{$Concept\_Set \setminus{} Concept \neq{}\emptyset{}$}{true}{}
		\Guard{grd9}{false}{$CO \in{} Concept\_Set \setminus{} Concept$}{true}{}
		\Guard{grd10}{false}{$Constant\_definedIn\_Component(o\_CO) \in{} ran(DomainModel\_corresp\_Component)$}{true}{}
	}
	\ACTIONS{true}{
		\Action{act1}{$Concept \bcmeq{}  Concept \bunion{}  \{CO\}$}{true}{}
		\Action{act2}{$Concept\_corresp\_Constant(CO)\bcmeq{}o\_CO$}{true}{}
		\Action{act3}{$Concept\_definedIn\_DomainModel(CO) \bcmeq{} DomainModel\_corresp\_Component\converse{}(\\Constant\_definedIn\_Component(o\_CO))$}{true}{}
		\Action{act4}{$Concept\_parentConcept\_Concept(CO) \bcmeq{} PCO$}{true}{}
		\Action{act5}{$Concept\_isVariable(CO) \bcmeq{}  FALSE$}{true}{}
	}
}
\EVT{rule\_105\_2}{false}{ordinary}{}{\\handling  the addition of a constant, sub set of an instance of Concept (case where the concept corresponds to a constant)}{
	\ANY{
		\Param{CO}{true}{}
		\Param{o\_CO}{true}{}
		\Param{PCO}{true}{}
		\Param{o\_lg}{true}{}
		\Param{o\_PCO}{true}{}
	}
	\GUARDS{true}{
		\Guard{grd0}{false}{$dom(Constant\_typing\_Property) \setminus{} ran(Concept\_corresp\_Constant) \neq{}\emptyset{}$}{true}{}
		\Guard{grd1}{false}{$o\_CO \in{} dom(Constant\_typing\_Property) \setminus{} ran(Concept\_corresp\_Constant)$}{true}{}
		\Guard{grd2}{false}{$o\_lg = Constant\_typing\_Property(o\_CO)$}{true}{}
		\Guard{grd3}{false}{$LogicFormula\_uses\_Operators(o\_lg) = \{1\mapsto{}Inclusion\_OP\}$}{true}{}
		\Guard{grd4}{false}{$LogicFormula\_involves\_Sets(o\_lg) = \emptyset{}$}{true}{}
		\Guard{grd5}{false}{$o\_PCO \in{} dom(Constant\_isInvolvedIn\_LogicFormulas)$}{true}{}
		\Guard{grd6}{false}{$(2\mapsto{}o\_lg) \in{} Constant\_isInvolvedIn\_LogicFormulas(o\_PCO)$}{true}{}
		\Guard{grd7}{false}{$o\_PCO \in{} ran(Concept\_corresp\_Constant)$}{true}{}
		\Guard{grd8}{false}{$PCO = Concept\_corresp\_Constant\converse{}(o\_PCO)$}{true}{}
		\Guard{grd9}{false}{$Concept\_Set \setminus{} Concept \neq{}\emptyset{}$}{true}{}
		\Guard{grd10}{false}{$CO \in{} Concept\_Set \setminus{} Concept$}{true}{}
		\Guard{grd11}{false}{$Constant\_definedIn\_Component(o\_CO) \in{} ran(DomainModel\_corresp\_Component)$}{true}{}
	}
	\ACTIONS{true}{
		\Action{act1}{$Concept \bcmeq{}  Concept \bunion{}  \{CO\}$}{true}{}
		\Action{act2}{$Concept\_corresp\_Constant(CO)\bcmeq{}o\_CO$}{true}{}
		\Action{act3}{$Concept\_definedIn\_DomainModel(CO) \bcmeq{} DomainModel\_corresp\_Component\converse{}(\\Constant\_definedIn\_Component(o\_CO))$}{true}{}
		\Action{act4}{$Concept\_parentConcept\_Concept(CO) \bcmeq{} PCO$}{true}{}
		\Action{act5}{$Concept\_isVariable(CO) \bcmeq{}  FALSE$}{true}{}
	}
}

\END

\paragraph{Addition of an individual }

\MACHINE{event\_b\_specs\_from\_ontologies\_ref\_1}{event\_b\_specs\_from\_ontologies}{EventB\_Metamodel\_Context,Domain\_Metamodel\_Context}{}

\EVT{rule\_106\_1}{false}{ordinary}{}{\\handling  the addition of an individual (case where the concept corresponds to an abstract set)}{
	\ANY{
		\Param{ind}{true}{}
		\Param{o\_ind}{true}{}
		\Param{CO}{true}{}
		\Param{o\_lg}{true}{}
		\Param{o\_CO}{true}{}
	}
	\GUARDS{true}{
		\Guard{grd0}{false}{$dom(Constant\_typing\_Property) \setminus{} ran(Individual\_corresp\_Constant) \neq{}\emptyset{}$}{true}{}
		\Guard{grd1}{false}{$o\_ind \in{} dom(Constant\_typing\_Property) \setminus{} ran(Individual\_corresp\_Constant)$}{true}{}
		\Guard{grd2}{false}{$o\_lg = Constant\_typing\_Property(o\_ind)$}{true}{}
		\Guard{grd3}{false}{$LogicFormula\_uses\_Operators(o\_lg) = \{1\mapsto{}Belonging\_OP\}$}{true}{}
		\Guard{grd4}{false}{$LogicFormula\_involves\_Sets(o\_lg) \neq{} \emptyset{}$}{true}{}
		\Guard{grd5}{false}{$(2\mapsto{}o\_CO)\in{}LogicFormula\_involves\_Sets(o\_lg)$}{true}{}
		\Guard{grd6}{false}{$o\_CO \in{} ran(Concept\_corresp\_AbstractSet)$}{true}{}
		\Guard{grd7}{false}{$CO = Concept\_corresp\_AbstractSet\converse{}(o\_CO)$}{true}{}
		\Guard{grd8}{false}{$Individual\_Set \setminus{} Individual \neq{}\emptyset{}$}{true}{}
		\Guard{grd9}{false}{$ind \in{} Individual\_Set \setminus{} Individual$}{true}{}
	}
	\ACTIONS{true}{
		\Action{act1}{$Individual \bcmeq{}  Individual \bunion{}  \{ind\}$}{true}{}
		\Action{act2}{$Individual\_individualOf\_Concept(ind) \bcmeq{}  CO$}{true}{}
		\Action{act3}{$Individual\_corresp\_Constant(ind) \bcmeq{}  o\_ind$}{true}{}
	}
}
\EVT{rule\_106\_2}{false}{ordinary}{}{\\handling  the addition of an individual (case where the concept corresponds to a constant)}{
	\ANY{
		\Param{ind}{true}{}
		\Param{o\_ind}{true}{}
		\Param{CO}{true}{}
		\Param{o\_lg}{true}{}
		\Param{o\_CO}{true}{}
	}
	\GUARDS{true}{
		\Guard{grd0}{false}{$dom(Constant\_typing\_Property) \setminus{} ran(Individual\_corresp\_Constant) \neq{}\emptyset{}$}{true}{}
		\Guard{grd1}{false}{$o\_ind \in{} dom(Constant\_typing\_Property) \setminus{} ran(Individual\_corresp\_Constant)$}{true}{}
		\Guard{grd2}{false}{$o\_lg = Constant\_typing\_Property(o\_ind)$}{true}{}
		\Guard{grd3}{false}{$LogicFormula\_uses\_Operators(o\_lg) = \{1\mapsto{}Belonging\_OP\}$}{true}{}
		\Guard{grd4}{false}{$LogicFormula\_involves\_Sets(o\_lg) = \emptyset{}$}{true}{}
		\Guard{grd5}{false}{$o\_CO\in{}dom(Constant\_isInvolvedIn\_LogicFormulas)$}{true}{}
		\Guard{grd6}{false}{$(2\mapsto{}o\_lg) \in{} Constant\_isInvolvedIn\_LogicFormulas(o\_CO)$}{true}{}
		\Guard{grd7}{false}{$o\_CO \in{} ran(Concept\_corresp\_Constant)$}{true}{}
		\Guard{grd8}{false}{$CO = Concept\_corresp\_Constant\converse{}(o\_CO)$}{true}{}
		\Guard{grd9}{false}{$Individual\_Set \setminus{} Individual \neq{}\emptyset{}$}{true}{}
		\Guard{grd10}{false}{$ind \in{} Individual\_Set \setminus{} Individual$}{true}{}
	}
	\ACTIONS{true}{
		\Action{act1}{$Individual \bcmeq{}  Individual \bunion{}  \{ind\}$}{true}{}
		\Action{act2}{$Individual\_individualOf\_Concept(ind) \bcmeq{}  CO$}{true}{}
		\Action{act3}{$Individual\_corresp\_Constant(ind) \bcmeq{}  o\_ind$}{true}{}
	}
}

\END

\paragraph{Addition of a data value}

\MACHINE{event\_b\_specs\_from\_ontologies\_ref\_1}{event\_b\_specs\_from\_ontologies}{EventB\_Metamodel\_Context,Domain\_Metamodel\_Context}{}

\EVT{rule\_107}{false}{ordinary}{}{\\handling  the addition of a data value}{
	\ANY{
		\Param{dva}{true}{}
		\Param{o\_dva}{true}{}
		\Param{DS}{true}{}
		\Param{o\_lg}{true}{}
		\Param{o\_DS}{true}{}
	}
	\GUARDS{true}{
		\Guard{grd0}{false}{$dom(Constant\_typing\_Property) \setminus{} ran(DataValue\_corresp\_Constant) \neq{}\emptyset{}$}{true}{}
		\Guard{grd1}{false}{$o\_dva \in{} dom(Constant\_typing\_Property) \setminus{} ran(DataValue\_corresp\_Constant)$}{true}{}
		\Guard{grd2}{false}{$o\_lg = Constant\_typing\_Property(o\_dva)$}{true}{}
		\Guard{grd3}{false}{$LogicFormula\_uses\_Operators(o\_lg) = \{1\mapsto{}Belonging\_OP\}$}{true}{}
		\Guard{grd4}{false}{$LogicFormula\_involves\_Sets(o\_lg) \neq{} \emptyset{}$}{true}{}
		\Guard{grd5}{false}{$(2\mapsto{}o\_DS)\in{}LogicFormula\_involves\_Sets(o\_lg)$}{true}{}
		\Guard{grd6}{false}{$o\_DS \in{} ran(DataSet\_corresp\_Set)$}{true}{}
		\Guard{grd7}{false}{$DS = DataSet\_corresp\_Set\converse{}(o\_DS)$}{true}{}
		\Guard{grd8}{false}{$DataValue\_Set \setminus{} DataValue \neq{}\emptyset{}$}{true}{}
		\Guard{grd9}{false}{$dva \in{} DataValue\_Set \setminus{} DataValue$}{true}{}
	}
	\ACTIONS{true}{
		\Action{act1}{$DataValue \bcmeq{}  DataValue \bunion{}  \{dva\}$}{true}{}
		\Action{act2}{$DataValue\_valueOf\_DataSet(dva) \bcmeq{}  DS$}{true}{}
		\Action{act3}{$DataValue\_corresp\_Constant(dva) \bcmeq{}  o\_dva$}{true}{}
	}
}

\END

\paragraph{Addition of a variable, sub set of an instance of Concept (full)}

\MACHINE{event\_b\_specs\_from\_ontologies\_ref\_1}{event\_b\_specs\_from\_ontologies}{EventB\_Metamodel\_Context,Domain\_Metamodel\_Context}{}

\EVT{rule\_108\_1}{false}{ordinary}{}{\\handling  the addition of a variable, sub set of an instance of Concept (case where the concept corresponds to an abstract set)}{
	\ANY{
		\Param{x\_CO}{true}{}
		\Param{CO}{true}{}
		\Param{o\_lg}{true}{}
		\Param{o\_CO}{true}{}
	}
	\GUARDS{true}{
		\Guard{grd0}{false}{$dom(Variable\_typing\_Invariant) \setminus{} ran(Concept\_corresp\_Variable) \neq{}\emptyset{}$}{true}{}
		\Guard{grd1}{false}{$x\_CO \in{} dom(Variable\_typing\_Invariant) \setminus{} ran(Concept\_corresp\_Variable)$}{true}{}
		\Guard{grd2}{false}{$o\_lg = Variable\_typing\_Invariant(x\_CO)$}{true}{}
		\Guard{grd3}{false}{$LogicFormula\_uses\_Operators(o\_lg) = \{1\mapsto{}Inclusion\_OP\}$}{true}{}
		\Guard{grd4}{false}{$LogicFormula\_involves\_Sets(o\_lg) \neq{} \emptyset{}$}{true}{}
		\Guard{grd5}{false}{$(2\mapsto{}o\_CO)\in{}LogicFormula\_involves\_Sets(o\_lg)$}{true}{}
		\Guard{grd6}{false}{$o\_CO \in{} ran(Concept\_corresp\_AbstractSet)$}{true}{}
		\Guard{grd7}{false}{$CO = Concept\_corresp\_AbstractSet\converse{}(o\_CO)$}{true}{}
	}
	\ACTIONS{true}{
		\Action{act1}{$Concept\_isVariable(CO) \bcmeq{}  TRUE$}{true}{}
		\Action{act2}{$Concept\_corresp\_Variable(CO) \bcmeq{} x\_CO$}{true}{}
	}
}
\EVT{rule\_108\_2}{false}{ordinary}{}{\\handling  the addition of a variable, sub set of an instance of Concept (case where the concept corresponds to a constant)}{
	\ANY{
		\Param{x\_CO}{true}{}
		\Param{CO}{true}{}
		\Param{o\_lg}{true}{}
		\Param{o\_CO}{true}{}
	}
	\GUARDS{true}{
		\Guard{grd0}{false}{$dom(Variable\_typing\_Invariant) \setminus{} ran(Concept\_corresp\_Variable) \neq{}\emptyset{}$}{true}{}
		\Guard{grd1}{false}{$x\_CO \in{} dom(Variable\_typing\_Invariant) \setminus{} ran(Concept\_corresp\_Variable)$}{true}{}
		\Guard{grd2}{false}{$o\_lg = Variable\_typing\_Invariant(x\_CO)$}{true}{}
		\Guard{grd3}{false}{$LogicFormula\_uses\_Operators(o\_lg) = \{1\mapsto{}Inclusion\_OP\}$}{true}{}
		\Guard{grd4}{false}{$LogicFormula\_involves\_Sets(o\_lg) = \emptyset{}$}{true}{}
		\Guard{grd5}{false}{$o\_CO\in{}dom(Constant\_isInvolvedIn\_LogicFormulas)$}{true}{}
		\Guard{grd6}{false}{$(2\mapsto{}o\_lg) \in{} Constant\_isInvolvedIn\_LogicFormulas(o\_CO)$}{true}{}
		\Guard{grd7}{false}{$o\_CO \in{} ran(Concept\_corresp\_Constant)$}{true}{}
		\Guard{grd8}{false}{$CO = Concept\_corresp\_Constant\converse{}(o\_CO)$}{true}{}
	}
	\ACTIONS{true}{
		\Action{act1}{$Concept\_isVariable(CO) \bcmeq{}  TRUE$}{true}{}
		\Action{act2}{$Concept\_corresp\_Variable(CO) \bcmeq{} x\_CO$}{true}{}
	}
}

\END

\subsection{Discussion and Experience}
The  rules that we propose allow the automatic translation of  domain properties, modeled as  ontologies, to \textit{B System} specifications, in order to fill the gap between the system textual description and the formal specification.
It is thus possible to benefit from all the advantages of a high-level modeling approach within the framework of the formal  specification of systems : 
decoupling between formal specification handling difficulties and system modeling; better reusability and readability of models; strong traceability between the system structure and stakeholder needs. 
Applying the approach on case studies \cite{sysml_kaos_domain_models_case_studies_link}
allowed us to quickly build the refinement hierarchy of the system and to determine and express the safety invariants, without having to manipulate the formal specifications. Furthermore, it allows us to limit our formal specification to the perimeter defined by the expressed needs.
This step also allowed us to enrich the domain modeling language expressiveness.

Formally defining  the SysML/KAOS domain modeling language, using \textit{Event-B},   allowed us to completely fulfill the criteria for it to be an ontology modeling formalism \cite{DBLPjournals/soco/AmeurBBJS17}.  Furthermore, 
formally defining the rules in \textit{Event-B} and discharging the associated proof obligations  allowed us
 to prove their consistency, to animate them using \textit{ProB}  and to  reveal several constraints (guards and invariants) that were missing   when designing the rules informally or when specifying the metamodels. For instance: (1)  if an instance of \textsf{Concept} \textit{x}, with parent \textit{px} does not have a correspondent yet and if \textit{px} does, then, the correspondent of  \textit{px} should not be refined by any  instance of \textsf{Component} (\texttt{inv0_7} defined in\texttt{Ontologies\_BSys\-tem\_specs\_translation} and  described in Sect. \ref{sect_domain_model_with_parent}); (2) elements of an enumerated data set should have correspondents if and only if the enumerated data set does; (3) if a concept, given as the domain of an attribute (instance of \textsf{Attribute}), is variable, then the attribute must also be variable; the same constraint is needed for the domain and the range of a relation. 
In case of absence of this last constraint, it is possible to reach a state where an attribute maplet (instance of \textsf{AttributeMaplet}) is defined for a non-existing individual (because the individual has been dynamically removed). These  constraints have been integrated in the SysML/KAOS domain modeling language in order to  strengthen its semantics.
 
 There are two essential properties that the specification of the rules must ensure and that we have proved using Rodin. 
The first one is that the rules are isomorphisms and it guarantees that  established links  between elements of the ontologies are preserved between the corresponding elements in the \textit{B System} specification and vice versa. To do this, we have introduced, for each link between elements, an invariant guaranteeing the preservation of the corresponding link between the correspondences and we have discharged the associated proof obligations. This leads to fifty  invariants.  For example, to ensure that
for each domain model \textit{pxx}, parent of \textit{xx}, the correspondent of \textit{xx} refines the correspondent of \textit{pxx} and vice versa, we have defined the following invariants: 

\begin{scriptsize}

  \noindent  \textcolor{labelcolor}{\small{\texttt{inv0\_8}}}: $\forall{}xx,pxx\qdot{}(~ 		(xx\in{}dom(DomainModel\_parent\_DomainModel)~ 		\land{}~pxx=DomainModel\_\-parent\_DomainModel(xx)~ 		\land{} \{xx,pxx\} \subseteq{} dom(DomainModel\_corresp\_\-Component))~ 		\limp{}(
 Domain\-Model\_corresp\_Component(xx) \in dom(Refinement\_refines\_\-Component) \land{} 
    Refinement\_refines\_\-Component(DomainModel\_corresp\_Component(xx))=DomainModel\_corresp\_Component(pxx))~ 		)$
    
    \noindent\textcolor{labelcolor}{\small{\texttt{inv0\_9}}}: $\forall{}o\_xx,o\_pxx\qdot{}(~ 		(o\_xx\in{}dom(Refinement\_refines\_Component)~ 		\land{} ~ o\_pxx=Refinement\_\-refines\_Component(o\_xx)~ 		\land{} \{o\_xx,o\_pxx\} \subseteq{} ran(DomainModel\_corresp\_\-Component))~ 		\limp{} (
    Domain\-Model\_corresp\_Component^{-1}(o\_xx) \in dom(DomainModel\_\-parent\_DomainModel) \land{} 
    DomainModel\_\-parent\_DomainModel(Domain\-Model\_corresp\_\-Component^{-1}(o\_xx))=\\DomainModel\_corresp\_Component^{-1}(o\_pxx))~ 		)$

\end{scriptsize}
%
%
%
%
%

\noindent  The second essential property is to
demonstrate that the system will always reach a state where all translations have been established.   
%
%
%
%
%
%
%
%
%
 To automatically prove it, 
we have introduced, within each machine, a \textit{variant} defined as the  difference between the set of elements to be translated and the set of elements already translated.  
Then, each event representing a translation rule has been marked as \textit{convergent} and we have discharged the proof obligations ensuring that each of them decreases the \textit{variant}.
 For example, in the machine \texttt{Ontologies\_BSystem\_specs\_translation} containing the definition of translation rules from domain models to \textit{B System} components, the variant was defined as $DomainModel \setminus{} dom(DomainModel\_corresp\_Component)$. 
Thus, at the end of system execution, we will have $dom(DomainModel\_corresp\_Component) = DomainModel$, which will reflect the fact that each domain model has been translated into a component.

There is no predefined type for ordered sets in \textit{Event-B}. 
 This problem led us to the definition of composition of functions in order to define relations on ordered sets. Moreover, because of the size of our model (about one hundred invariants and about fifty events for each machine), we noted a rather significant performance reduction of \textit{Rodin} during some operations such as the execution of auto-tactics or  proof replay on undischarged proof obligations that have to be done after each update in order to discharge all previously discharged proofs. 
  Table \ref{tableau_recapitulatif_specification_eventb} summarises the key characteristics of the Rodin project corresponding to the \textit{Event-B} specification of metamodels and rules (translation and back propagation rules). The automatic provers seemed  least comfortable with functions ($\pfun, \pinj, \tfun, \psurj$) and  become almost useless when those  operators are combined in  definitions as for ordered associations ($r \in (A \tfun{} (\natn{}\pfun{} B))$).\\

\begin{table}[htb]
\caption{\label{tableau_recapitulatif_specification_eventb} Key characteristics of the \textit{Event-B} specification of  rules }
\begin{scriptsize}
\begin{tabular}{|p{5.2cm}|p{2.cm}|p{3.5cm}|}
\hline
\textbf{Characteristics} & \textbf{Root level} & \textbf{First refinement level} \\
\hline
\textbf{Events} & \textbf{3} & \textbf{50} \\
\hline
\textbf{Invariants} & \textbf{11} & \textbf{104} \\
\hline
\textbf{Proof Obligations (PO)} & \textbf{37} & \textbf{1123} \\
\hline
\textbf{Automatically Discharged POs} & \textbf{27} & \textbf{257}  \\
\hline
\textbf{Interactively Discharged POs} & \textbf{10} & \textbf{866} \\
\hline

\end{tabular}
\end{scriptsize}
\end{table}

\section{Specification of the Hybrid ERTMS/ETCS Level 3  Standard}
\subsection{Main Characteristics of the Standard}
The Hybrid ERTMS/ETCS level 3 protocol (HEEL3) has been proposed to optimize the use and occupation of railways \cite{hoang_hybrid_2018,ertms_etcs_principles,ertms_l3_game_changer}.
It thus proposes the division of the track into separate entities, each named Trackside Train Detection (TTD).
In addition, each TTD is subdivided into sub-entities called Virtual Sub-Sections  (VSS).
A TTD has two possible states: \textit{free} and \textit{occupied} with a  safety invariant stating that  if a train is located on a TTD, then the state of the TTD must be set to \textit{occupied}.
In addition to these two states, a VSS may have the \textit{unknown} or the  \textit{ambiguous} state.
The \textit{ambiguous} state is used when the information available to the system suggest that two trains are potentially present on the VSS. The \textit{unknown} state is used when the system can guarantee neither the presence nor the absence of a train on the VSS.
For an optimal safety, Movement Authorities (MA) are evaluated and assigned to each connected train.
The MA of a train designates a portion of the track on which it is guaranteed to move safely.
   ERTMS (European Rail Traffic Management System)
   designates a protocol and a set of tools that allow a train to know and report its position.
   Similarly, TIMS (Train Integrity Monitoring System) 
      designates the component that allows a train to know and report its integrity and its size.
HEEL3 
 considers three  train categories : those equipped with ERTMS and TIMS  called \textit{INTEGER}; those that are just equipped with a ERTMS which
allows them to broadcast their position (connected trains);
  and finally, those that are  equipped neither with a ERTMS  nor with a TIMS called unconnected trains.\\

\begin{figure*}[!h]
\begin{center}
\includegraphics[width=0.9\textwidth]{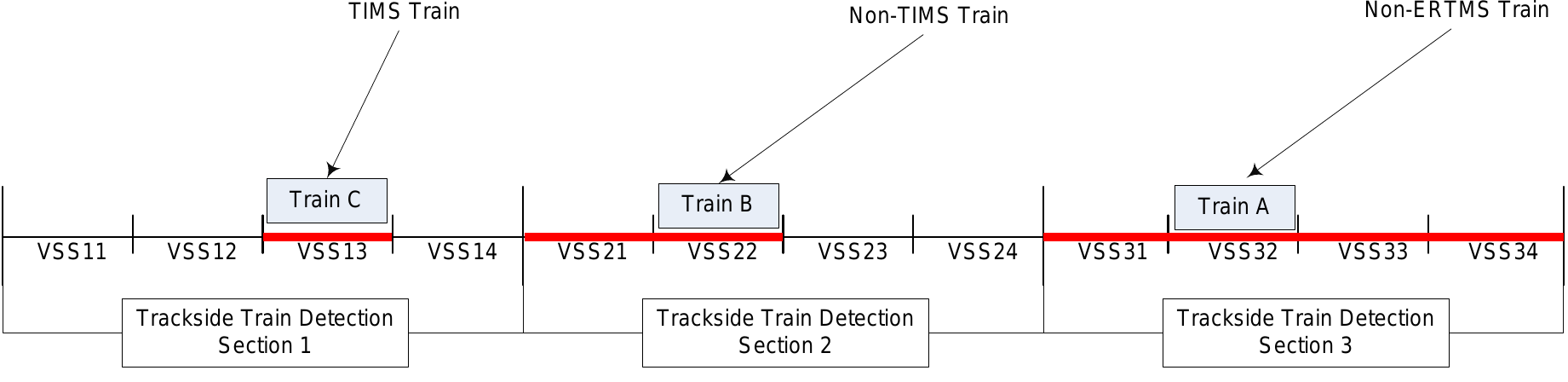}
\end{center}
\caption{\label{ertms_etcs_l3_hybrid_capacity_exploitation}  
Overview of the dependence between the 
capacity exploitation and the
presence of  ERTMS  and TIMS \cite{ertms_l3_game_changer}}
\end{figure*}
\paragraph{}
Figure \ref{ertms_etcs_l3_hybrid_capacity_exploitation} is an overview of   the influence of the presence of ERTMS and TIMS on the track capacity exploitation \cite{ertms_l3_game_changer}. 
A TIMS train (INTEGER) is considered to occupy a whole VSS. A non-TIMS train (connected train) is considered to occupy all the VSSs from its front to the end of the TTD section where it is located. Finally, a non-ERTMS train (unconnected train) is considered to occupy the whole TTD section where the system guess it is.

\begin{figure}[!h]
\begin{center}
\includegraphics[width=0.9\textwidth]{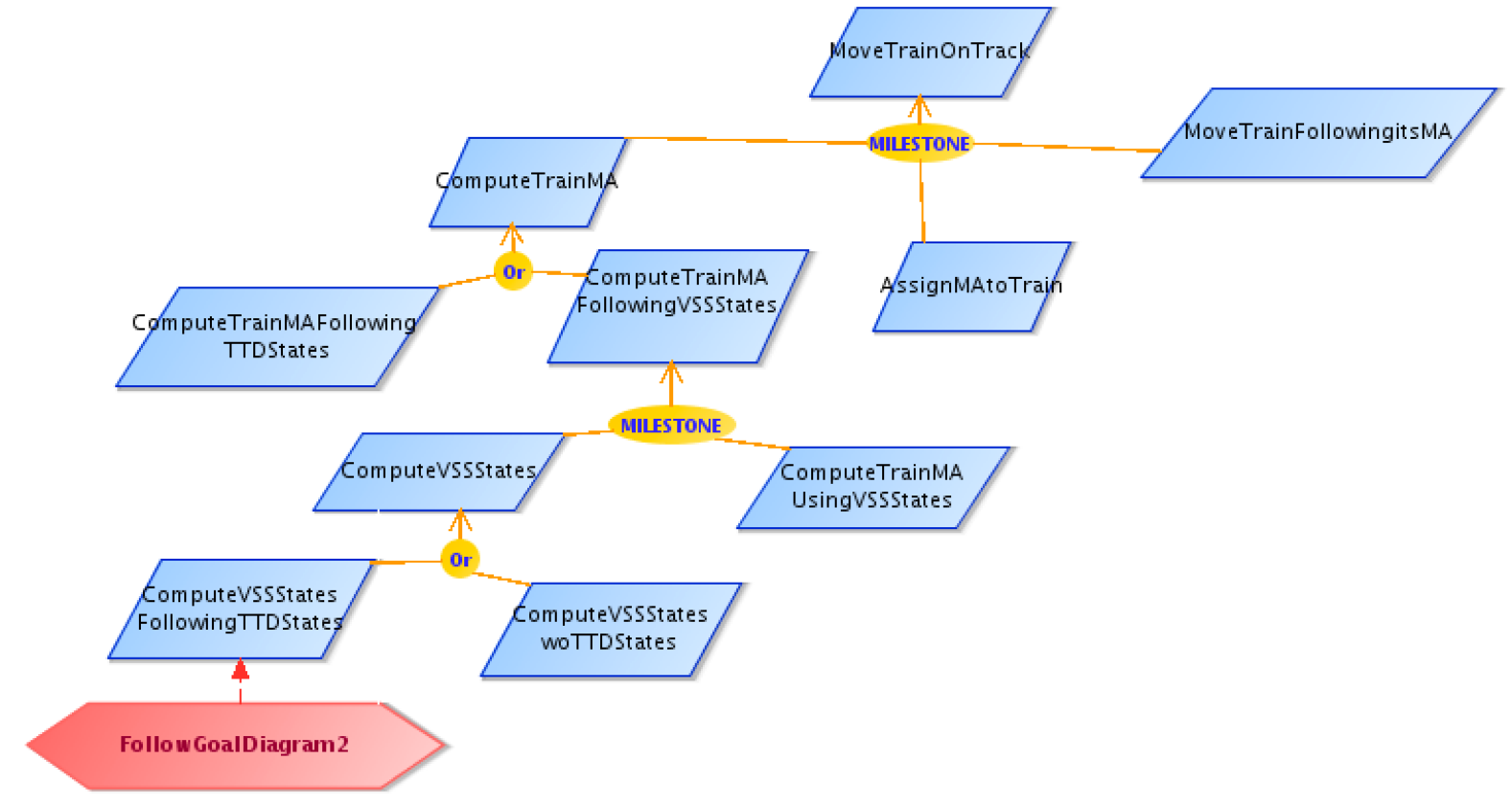}
\end{center}
\caption{\label{ERTMS_L3_Hybrid_ABZ_2018_vue5} The SysML/KAOS goal diagram}
\paragraph{}
\end{figure}

\subsection{The Goal Diagram}
The SysML/KAOS requirements engineering method allows the progressive construction of system requirements from refinements of stakeholder needs. Thus, even if the management of VSSs is the purpose of the case study, we need to put it into perspective with more abstract objectives that will explain what VSSs are useful for.  
Figure \ref{ERTMS_L3_Hybrid_ABZ_2018_vue5} is an excerpt from the SysML/KAOS functional goal diagram focused on the main system purpose :
move trains on the track (\texttt{MoveTrainOnTrack}).
 To achieve it, the system must ensures that the train has a valid MA  (\texttt{ComputeTrainMA}). 
If the MA has been recomputed, then the system must assign the new MA to the train  (\texttt{AssignMAtoTrain}). Finally, the train has to move following its assigned MA (\texttt{MoveTrainFollowingItsMA}).
The second refinement level of the SysML/KAOS goal diagram  focuses on the informations needed to determine the MA of a train : the MA computation  can be  based only on TTD states (\texttt{ComputeTrainMAFollowingTTDStates}) or following VSS states (\texttt{ComputeTrainMAFollowingVSSStates}) \cite{ertms_etcs_principles}. When the computation is only based on TTD states, it corresponds to the \textit{ERTMS/ETCS Level 2} protocol. When VSS states are involved, it corresponds to the \textit{ERTMS/ETCS Level 3} protocol.
The MA computation based on  VSS states  requires the update of the states of  VSSs (\texttt{ComputeVSSStates})  and the computation of the MA (\texttt{ComputeTrainMAUsingVSSStates}). 
Finally, depending on the type of the ERTMS/ETCS level 3 implementation, it is possible to use or not the TTD states when computing the VSS states (table 1 of \cite{ertms_l3_game_changer}). If TTD states are not required (\textit{virtual (without train detection)} level 3 type), it corresponds to   \texttt{ComputeVSSStateswoTTD\-States}, with the disadvantage of only allowing the circulation of trains equipped with TIMS.  If TTD states are used (\textit{hybrid} level 3 type), it corresponds to   \texttt{ComputeVSSStatesFollowingTTDStates}.

\begin{figure}[!h]
\begin{center}
\includegraphics[width=0.9\textwidth]{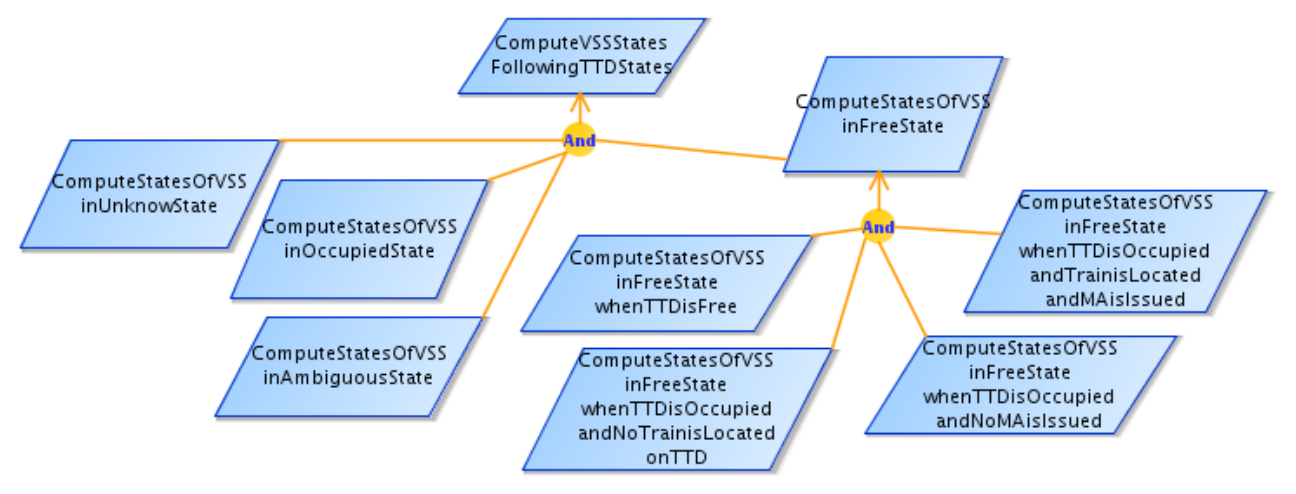}
\end{center}
\caption{\label{ERTMS_L3_Hybrid_ABZ_2018_vue2} SysML/KAOS goal diagram of the VSS state computation purposes}
\paragraph{}
\end{figure}

Figure \ref{ERTMS_L3_Hybrid_ABZ_2018_vue2} is an excerpt from the SysML/KAOS functional goal diagram focused on the  purpose of VSS state computation with the use of TTD states (\texttt{ComputeVSSStatesFollowingTTDStates}).
The computation of the current VSS states  can be splitted into the determination of the current states of  VSSs previously in the unknown state (\texttt{ComputeStatesOfVSSinUnknownState}), in the occupied state (\texttt{ComputeStatesOfVSSinOccupied\-State}), in the ambiguous state (\texttt{ComputeStatesOfVSSinAmbiguousState}) and in the free state (\texttt{ComputeStates\-OfVSSinFreeState}) (Figure 7 of \cite{ertms_etcs_principles}).
The last refinement level is focused on VSSs previously in the free state. Its goals come from the requirements of the transition \textit{ \#1A } of Table 2 of \cite{ertms_etcs_principles}. When the TTD is free, then the VSSs remain free (\texttt{ComputeStatesOfVSSinFreeStateWhenTTDisFree}). When the TTD is occupied and no train is located on it or no MA is issued, then the VSSs move in the unknown state (\texttt{ComputeStatesOfVSSinFreeStateWhenTTDisOccupiedandNoTrain\-isLocatedonTTD}, \texttt{ComputeStatesOfVSS\-inFreeStateWhenTTDisOccupiedandNoMA\-isIssued}). The other transitions are the purpose of \texttt{ComputeStates\-OfVSSinFree\-StateWhenTTDisOccupiedandTrainisLocatedandMAisIssued}.

The rest of this section consists of a presentation of the SysML/KAOS domain models associated with the most relevant  refinement levels   of  the goal diagrams and of a description of the \textit{B System} specifications obtained from goals and ontologies.
From the goal model, we distinguish seven refinement levels which are translated into seven \textit{B System} components.
The formal specification has been verified  using  \textit{Rodin}  \cite{DBLP:conf/rodin/2006}, an  industrial-strength tool supporting  the \textit{Event-B} method \cite{DBLP:books/daglib/0024570}. We have in particular discharged all the proof obligations associated with the safety invariants that we have identified and with the SysML/KAOS refinement operators that appear in the goal diagram.
For the sake of concision, we  will  present here only the first three refinement levels.
The full specification can be found in \cite{sysml_kaos_domain_models_ertms_link}.

\subsection{The Root Level}

\begin{figure}[!h]

\begin{scriptsize}
\lstset{language=MPS}
\begin{lstlisting}
domain model ertms_etcs_case_study {
	concepts:					
		concept TRAIN  is variable: false
	attributes:
		attribute connectedTrain domain: Train  range: BOOL {
			is variable: true		
			is functional: true			
			is total: false
		}
		attribute front domain: dom(connectedTrain)  range: TRACK {
			is variable: true		
			is functional: true			
			is total: true
		}
		attribute rear domain: dom(connectedTrain)  range: TRACK {
			is variable: true		
			is functional: true			
			is total: false
		}
	data sets:		
		custom data set TRACK
	data values:		
		data value a type: NATURAL 		
		data value b type: NATURAL
	predicates:
		p0.1: a<b																										p0.2: TRACK=a..b
		p0.3: !tr. (tr : dom(rear) => rear(tr) < front(tr))
}
\end{lstlisting}
\end{scriptsize}

\caption{\label{domain_model_root_level} SysML/KAOS domain modeling of the goal diagram root level}
\paragraph{}
\end{figure}

Figure \ref{domain_model_root_level} represents the domain model associated with the root level of the SysML/KAOS goal diagram of Figure \ref{ERTMS_L3_Hybrid_ABZ_2018_vue5}. The concept \texttt{TRAIN} models the set of trains. The attribute \texttt{connectedTrain} models the subset of \texttt{TRAIN} that  broadcast their location at least once and for each, the  current connection status. The attribute \texttt{front} models the estimated position of the front of each connected train. For each connected train equipped with a TIMS, the attribute \texttt{rear} models the estimated position of its rear\footnote{ the rear is deduced from the front and length of the train, since a train equipped with a TIMS broadcast its length and its integrity}. Thus, $dom(front)\setminus{}dom(rear)$ represents the set of trains equipped with a ERTMS and not equipped with a TIMS.
Predicates represent constraints on domain model elements. Each predicate is prefixed with an identifier. For example, the predicate \texttt{p0.2} defines  \texttt{TRACK}  as the data range $a..b$.

\begin{figure}[!h]
\begin{scriptsize}
\begin{multicols}{2}

    \textcolor{keycolor}{\textbf{SYSTEM}}
 ertms\_etcs\_case\_study

\textcolor{keycolor}{\textbf{SETS}}  TRAIN

\textcolor{keycolor}{\textbf{CONSTANTS}}
  a b TRACK

\textcolor{keycolor}{\textbf{PROPERTIES}}

 \hspace*{0.1in} \textcolor{labelcolor}{\small{\texttt{axm1}}}:  $a \in \nat{}$
 \hspace*{0.1in} \textcolor{labelcolor}{\small{\texttt{axm2}}}:  $b \in \nat{}$
  \hspace*{0.1in} \textcolor{labelcolor}{\small{\texttt{p0.1}}}: $a < b$

 \hspace*{0.1in} \textcolor{labelcolor}{\small{\texttt{p0.2}}}: $TRACK=a\upto{}b$

\textcolor{keycolor}{\textbf{VARIABLES}}
  connectedTrain front rear

\textcolor{keycolor}{\textbf{INVARIANT}}

 \hspace*{0.1in} \textcolor{labelcolor}{\small{\texttt{inv1}}}: $connectedTrain \in{} TRAIN \pfun{} BOOL$

 \hspace*{0.1in} \textcolor{labelcolor}{\small{\texttt{inv2}}}: $front \in{} \\
 \hspace*{0.3in}dom(connectedTrain) \tfun{} TRACK$
 
 \hspace*{0.1in} \textcolor{labelcolor}{\small{\texttt{inv3}}}: $rear \in{} \\
 \hspace*{0.3in}dom(connectedTrain) \pfun{} TRACK$
  
 \hspace*{0.1in} \textcolor{labelcolor}{\small{\texttt{p0.3}}}: $\forall{}tr\qdot{}(tr\in{}dom(rear)\\
 \hspace*{0.3in}\limp{}rear(tr)<front(tr))$

\hspace*{-0.10in}{ \textcolor{keycolor}{\textbf{Event}}} MoveTrainOnTrack  $\defi$

 \hspace*{0.1in} \textcolor{keycolor}{\textbf{any}} tr len

 \hspace*{0.1in}\textcolor{keycolor}{\textbf{where}}

 \hspace*{0.2in} \textcolor{labelcolor}{\small{\texttt{grd1}}}: $tr \in{} connectedTrain^{-1}[\{TRUE\}]$

 \hspace*{0.2in} \textcolor{labelcolor}{\small{\texttt{grd2}}}: $len \in{} \natn{}$

 \hspace*{0.2in} \textcolor{labelcolor}{\small{\texttt{grd3}}}: $front(tr)+len \in{} TRACK$

 \hspace*{0.1in} \textcolor{keycolor}{\textbf{then}}

 \hspace*{0.2in} \textcolor{labelcolor}{\small{\texttt{act1}}}: $front(tr) \bcmeq{} front(tr)+len$

 \hspace*{0.2in} \textcolor{labelcolor}{\small{\texttt{act2}}}: $rear \bcmeq{}  (\{TRUE\mapsto{}rear\ovl{}\{tr\mapsto{}\\
 \hspace*{0.3in}rear(tr)+len\}, FALSE\mapsto{}rear\\
 \hspace*{0.3in}\})(bool(tr\in{}dom(rear)))$

 \hspace*{0.1in} \textcolor{keycolor}{\textbf{END}}
 
\textcolor{keycolor}{\textbf{END}}

\end{multicols}
\end{scriptsize} 
\caption{\label{BSystem_root_level} \textit{B System} specification of the root level of the  goal diagram of Figure \ref{ERTMS_L3_Hybrid_ABZ_2018_vue5}}
\end{figure}
\paragraph{}
Figure \ref{BSystem_root_level} represents the \textit{B System} model obtained from the translation of the root level of the  goal diagram of Figure \ref{ERTMS_L3_Hybrid_ABZ_2018_vue5} and of the associated domain model of Figure \ref{domain_model_root_level}. The domain model gives rise to sets, constants, properties, variables and invariants of the formal specification. Predicates involving variables give rise to invariants and the others to properties. The \textit{isFunctional} and \textit{isTotal} characteristics of attributes, are used to guess if an attribute should be  translated into a partial or total function. The root goal is translated into an event for which the body has been manually specified: the movement of a connected train (\texttt{grd1}) results in the incrementation of the position of its front (\texttt{act1}) and its rear (\texttt{act2} in the case of an \textit{INTEGER} train) of the value corresponding to the movement. Of course, the movement can only be done if the train stays on the track (\texttt{grd3}).
\subsection{The First Refinement Level}
\begin{figure}[!h]
\lstset{language=MPS}
\begin{scriptsize}
\begin{lstlisting}
domain model ertms_etcs_case_study_ref_1 parent domain model ertms_etcs_case_study {
	attributes:		
		attribute MA domain: dom(connectedTrain)  range: POW(TRACK) {
			is variable: true		
			is functional: true			
			is total: false
		}
	predicates:
		p1.1: !tr. (tr : dom(MA) => #p,q.(p..q<:TRACK & p<=q & MA(tr)=p..q)))
		p1.2: !tr. (tr : dom(MA) => (front(tr) : MA(tr)))
		p1.3: !tr. (tr : dom(rear) & tr : dom(MA) => rear(tr) : MA(tr))
		p1.4: !tr1,tr2. ((tr1  :  dom(MA) &  tr2  :  dom(MA) &  tr1  /= 
			tr2)=>MA(tr1) /\ MA(tr2)={})
}
\end{lstlisting}
\end{scriptsize}
\caption{\label{domain_model_first_refinement_level} SysML/KAOS domain modeling of the goal diagram first refinement level}
\end{figure}
\paragraph{}
Figure \ref{domain_model_first_refinement_level} represents the domain model associated with the first refinement level of the SysML/KAOS goal diagram of Figure \ref{ERTMS_L3_Hybrid_ABZ_2018_vue5}. 
It refines the one associated with the root level and introduces an attribute named \texttt{MA} representing the MA assigned to a connected train.
The MA of a train is modeled as a contiguous part of the track (\texttt{p1.1}), containing the train (\texttt{p1.2} and \texttt{p1.3}). Finally, the predicate \texttt{p1.4} asserts that the MA assigned to two different trains must be disjoint. The predicates \texttt{p1.2} and \texttt{p1.3} are gluing invariants, linking the concrete variable \texttt{MA} with the abstract variables \texttt{front} and \texttt{rear}.

\begin{figure}[!h]
\begin{scriptsize}
\begin{multicols}{2}

    \textcolor{keycolor}{\textbf{REFINEMENT}}
 ertms\_etcs\_case\_study\_ref\_1
 
     \textcolor{keycolor}{\textbf{REFINES}}
  ertms\_etcs\_case\_study

\textcolor{keycolor}{\textbf{VARIABLES}}
  connectedTrain front rear MA MAtemp

\textcolor{keycolor}{\textbf{INVARIANT}}

 \hspace*{0.1in} \textcolor{labelcolor}{\small{\texttt{inv1}}}: $MA \in{} \\
 \hspace*{0.3in}dom(connectedTrain) \pfun{} \pow{}(TRACK)$

 \hspace*{0.1in} \textcolor{labelcolor}{\small{\texttt{p1.1}}}: $\forall{}tr\qdot{}(tr\in{}dom(MA)\limp{}(\exists{}p,q\qdot{}(p\upto{}q\\
 \hspace*{0.3in}\subseteq{}TRACK \land{} p\leq{}q\land{}MA(tr)=p\upto{}q )))$
 
 \hspace*{0.1in} \textcolor{labelcolor}{\small{\texttt{p1.2}}}: $\forall{}tr\qdot{}(tr\in{}dom(MA)\limp{}\\
 \hspace*{0.3in}front(tr)\in{}MA(tr))$
  
 \hspace*{0.1in} \textcolor{labelcolor}{\small{\texttt{p1.3}}}: $\forall{}tr\qdot{}(tr\in{}dom(rear)\binter{}dom(MA)\limp{} \\
 \hspace*{0.3in} rear(tr)\in{}MA(tr))$
 
  \hspace*{0.1in} \textcolor{labelcolor}{\small{\texttt{p1.4}}}: $\forall{}tr1,tr2\qdot{}((\{tr1,tr2\}\subseteq{}dom(MA)\land{}\\
 \hspace*{0.3in}tr1\neq{}tr2)\limp{} MA(tr1)\binter{}MA(tr2)=\emptyset{})$
  
   \hspace*{0.1in} \textcolor{labelcolor}{\small{\texttt{inv6}}}: $MAtemp \in{} \\
 \hspace*{0.3in}dom(connectedTrain) \pfun{}  \pow{}(TRACK)$
   
    \hspace*{0.1in} \textcolor{labelcolor}{\small{\texttt{inv7}}}: $\forall{}tr\qdot{}(tr\in{}dom(MAtemp)\limp{}(\exists{}p,q\qdot{}(\\p\upto{}q\subseteq{}    TRACK \land{} p\leq{}q\land{}MAtemp(tr)=p\upto{}q )))$
    
     \hspace*{0.1in} \textbf{\textcolor{labelcolor}{\small{theorem}}} \textcolor{labelcolor}{\small{\texttt{s1}}}: $ComputeTrainMA\_Guard  \\
     \hspace*{0.3in} \Rightarrow  MoveTrainOnTrack\_Guard$
     
      \hspace*{0.1in} \textbf{\textcolor{labelcolor}{\small{theorem}}} \textcolor{labelcolor}{\small{\texttt{s2}}}: $ComputeTrainMA\_Post  
      \\ \hspace*{0.3in}\Rightarrow AssignMAtoTrain\_Guard$
      
      \hspace*{0.1in} \textbf{\textcolor{labelcolor}{\small{theorem}}} \textcolor{labelcolor}{\small{\texttt{s3}}}: $AssignMAtoTrain\_Post  
      \\ \hspace*{0.3in} \Rightarrow MoveTrainFollowingItsMA\_Guard$
      
      \hspace*{0.1in} \textbf{\textcolor{labelcolor}{\small{theorem}}} \textcolor{labelcolor}{\small{\texttt{s4}}}: $MoveTrainFollowingIts\-
      \hspace*{0.3in} MA\_Post \Rightarrow MoveTrainOnTrack\_Post$

\hspace*{-0.10in}{ \textcolor{keycolor}{\textbf{Event}}}

 ComputeTrainMA  $\defi$

 \hspace*{0.1in} \textcolor{keycolor}{\textbf{any}} tr p q len

 \hspace*{0.1in}\textcolor{keycolor}{\textbf{where}}

 \hspace*{0.2in} \textcolor{labelcolor}{\small{\texttt{grd1}}}: $tr \in{} connectedTrain^{-1}[\{TRUE\}]$

 \hspace*{0.2in} \textcolor{labelcolor}{\small{\texttt{grd2}}}: $p\upto{}q\subseteq{}TRACK \land{} p\leq{}q$

 \hspace*{0.2in} \textcolor{labelcolor}{\small{\texttt{grd3}}}: $front(tr)\in{}p\upto{}q$
 
 \hspace*{0.2in} \textcolor{labelcolor}{\small{\texttt{grd4}}}: $tr\in{}dom(rear)\limp{}rear(tr)\in{}p\upto{}q$
 
 \hspace*{0.2in} \textcolor{labelcolor}{\small{\texttt{grd5}}}: $p\upto{}q \binter{} union(ran(\{tr\}\domsub{}MA))=\emptyset{}$
 
 \hspace*{0.2in} \textcolor{labelcolor}{\small{\texttt{grd6}}}: $len \in{} \natn{}$
 
 \hspace*{0.2in} \textcolor{labelcolor}{\small{\texttt{grd7}}}: $front(tr)+len \in{} TRACK$

 \hspace*{0.1in} \textcolor{keycolor}{\textbf{then}}

 \hspace*{0.2in} \textcolor{labelcolor}{\small{\texttt{act1}}}: $MAtemp(tr) \bcmeq{} p\upto{}q$

 \hspace*{0.1in} \textcolor{keycolor}{\textbf{END}}\\

  AssignMAtoTrain  $\defi$

 \hspace*{0.1in} \textcolor{keycolor}{\textbf{any}} tr  len

 \hspace*{0.1in}\textcolor{keycolor}{\textbf{where}}

 \hspace*{0.2in} \textcolor{labelcolor}{\small{\texttt{grd1}}}: $tr \in{} connectedTrain^{-1}[\{TRUE\}]\\
\hspace*{0.3in}   \binter{}dom(MAtemp)$ 

\hspace*{0.2in} • • •

 \hspace*{0.2in} \textcolor{labelcolor}{\small{\texttt{grd6}}}: $front(tr)+len \in{} MAtemp(tr)$

 \hspace*{0.1in} \textcolor{keycolor}{\textbf{then}}

 \hspace*{0.2in} \textcolor{labelcolor}{\small{\texttt{act1}}}: $MA(tr) \bcmeq{} MAtemp(tr)$

 \hspace*{0.1in} \textcolor{keycolor}{\textbf{END}} \\

  MoveTrainFollowingItsMA  $\defi$

 \hspace*{0.1in} \textcolor{keycolor}{\textbf{any}} tr  len

 \hspace*{0.1in}\textcolor{keycolor}{\textbf{where}}

 \hspace*{0.2in} \textcolor{labelcolor}{\small{\texttt{grd1}}}: $tr \in{} connectedTrain^{-1}[\{TRUE\}]\\
 \hspace*{0.5in}\binter{}dom(MA)$

 \hspace*{0.2in} \textcolor{labelcolor}{\small{\texttt{grd2}}}: $len \in{} \natn{}$

 \hspace*{0.2in} \textcolor{labelcolor}{\small{\texttt{grd3}}}: $front(tr)+len \in{} MA(tr)$

 \hspace*{0.1in} \textcolor{keycolor}{\textbf{then}}

 \hspace*{0.2in} \textcolor{labelcolor}{\small{\texttt{act1}}}: $front(tr) \bcmeq{} front(tr)+len$
 
  \hspace*{0.2in} \textcolor{labelcolor}{\small{\texttt{act2}}}: $rear \bcmeq{}  (\{TRUE\mapsto{}rear\ovl{}\{tr\mapsto{}\\
 \hspace*{0.5in}rear(tr) +len\}, FALSE\mapsto{}rear\\
 \hspace*{0.5in}\})(bool(tr\in{}dom(rear)))$

 \hspace*{0.1in} \textcolor{keycolor}{\textbf{END}}

\textcolor{keycolor}{\textbf{END}}

\end{multicols}
\end{scriptsize}
 
\caption{\label{BSystem_first_refinement_level} \textit{B System} specification of the first refinement level of the   diagram of Figure \ref{ERTMS_L3_Hybrid_ABZ_2018_vue5}}
\end{figure}

Figure \ref{BSystem_first_refinement_level} represents the \textit{B System} model obtained from the translation of the first refinement level of the  goal diagram of Figure \ref{ERTMS_L3_Hybrid_ABZ_2018_vue5} and of the associated domain model of Figure \ref{domain_model_first_refinement_level}. 
Each refinement level goal is translated into an event for which the body has been manually specified : the current MA of the train is computed and stored into a variable named \texttt{MAtemp} (event \texttt{ComputeTrainMA}). Because the computation of the MA is out of  the scope of the case study \cite{hoang_hybrid_2018}, the event simply nondeterministically choose an MA, with respect to the safety invariants. This  MA is then assigned to the train by updating the variable \texttt{MA} (event \texttt{AssignMAtoTrain}) and taken into account for the train displacement (event \texttt{MoveTrainFollowingItsMA}). 
Theorems \texttt{s1}, \texttt{s2}, \texttt{s3} and \texttt{s4} represent the proof obligations related to the usage of the \textit{MILESTONE} operator between the root  and the first refinement levels.
 Since each proof obligation has been modeled as an \textit{Event-B} theorem, it has  been proved based on system properties and invariants.
 To deal with the fact that \textit{Event-B} does not currently support the temporal logic, we have used the proof obligation $G1\_Post \Rightarrow  G2\_Guard$ for the invariants  \texttt{s2} and \texttt{s3}, instead of $\square(G1\_Post \Rightarrow \lozenge G2\_Guard)$ (Sect. \ref{goal_model_formalisation}), since $(G1\_Post \Rightarrow  G2\_Guard) \Rightarrow (\square(G1\_Post \Rightarrow \lozenge G2\_Guard))$. 
 The full specification of \texttt{s1} is given below: \\
 \begin{scriptsize}
 \noindent  \textbf{\textcolor{labelcolor}{\small{theorem}}} \textcolor{labelcolor}{\small{\texttt{s1}}}: $\forall{}tr,p,q,len\qdot{}(( 										(tr \in{} connectedTrain^{-1}[\{TRUE\}])~ 										\land{}(p\upto{}q\subseteq{}TRACK \land{} p\leq{}q)~ 										\land{}(front(tr)\in{}p\upto{}q)~ 										\land{}(tr\in{}dom(rear)\limp{}rear(tr)\in{}p\upto{}q)~ 										\land{}(p\upto{}q \binter{} union(ran(\{tr\}\domsub{}MA))=\emptyset{})~ 										\land{}(len \in{} \natn{})~ 										\land{}(front(tr)+len \in{} TRACK)~ 									) \limp{}~ 							  		(~ 										(tr \in{} connectedTrain^{-1}[\{TRUE\}])~ 										\land{}(len \in{} \natn{})~ 										\land{}(front(tr)+len \in{} TRACK)~ 									))$

\end{scriptsize}
\noindent It expresses the fact that the activation of the guard of \texttt{ComputeTrainMA} for  certain parameters is sufficient for the activation  of the guard of  \texttt{MoveTrainOnTrack} for this same group of parameters. 

\setlength\intextsep{3pt}

\subsection{The Second Refinement Level}

\begin{figure}[!h]

\begin{scriptsize}
\lstset{language=MPS}
\begin{lstlisting}
domain model ertms_etcs_case_study_ref_2 parent domain model ertms_etcs_case_study_ref_1 {
	concepts: 			
		concept TTD  is variable: false 					
		concept VSS is variable: false	
	attributes:
		attribute stateTTD domain: TTD  range: TTD_STATES {
			is variable: true		
			is functional: true			
			is total: true
		}
		attribute stateVSS domain: VSS  range: VSS_STATES {
			is variable: true		
			is functional: true			
			is total: true
		}
	data sets:
		enumerated data set VSS_STATES { elements : 
			data value OCCUPIED   		data value FREE
			data value UNKNOWN 		 data value AMBIGUOUS	
		}
		enumerated data set TTD_STATES { elements :  	
			data value OCCUPIED   			data value FREE 
		}
	predicates:
		p2.1: TTD <: POW1(TRACK)					
		p2.2: union(TTD) = TRACK				
		p2.3: inter(TTD) = {}
		p2.4: !ttd. (ttd : TTD => #p,q.(p..q<:TRACK & p<q & ttd=p..q)))
		p2.5: VSS <: POW1(TRACK)				
		p2.6: union(VSS) = TRACK					
		p2.7: inter(VSS) = {}
		p2.8: !vss. (vss : VSS => #p,q,ttd.(ttd : TTD & p..q<:ttd & p<q & vss=p..q)))
		p2.9:  !ttd,tr. ( tr : dom(front) \ dom(rear) &  ttd  :  TTD &  front(tr) : ttd) 
			 => (( ttd |-> OCCUPIED ) : stateTTD)
		p2.10:  !ttd,tr. (tr  :  dom(rear) &  ttd  :  TTD & (rear(tr)..front(tr))/\ttd /= {})  
			=> (( ttd |-> OCCUPIED ) : stateTTD) 
		p2.11:  !tr1,tr2. (tr1  :  dom(rear) &  tr2  :  dom(rear) &  tr1  /=  tr2)  
			=> ( (rear(tr1)..front(tr1))/\(rear(tr2)..front(tr2))={}) 
		p2.12: !tr1,tr2,ttd.(tr1 : dom(rear) & tr2 : dom(front)\dom(rear) & tr1 /= tr2
				& ttd : TTD & front(tr2) : ttd & rear(tr1)..front(tr1))/\ttd /= {})
				=> ( front(tr2)<rear(tr1)) 
		p2.13:  !tr1,tr2,ttd. ( tr1 : dom(front)\dom(rear) &  tr2 : dom(front)\dom(rear)   
			&tr1  /=  tr2 &  ttd  :  TTD &  front(tr1) : ttd)  => ( front(tr2) /: ttd)
}
\end{lstlisting}
\end{scriptsize}
\caption{\label{domain_model_second_refinement_level} SysML/KAOS domain modeling of the goal diagram second refinement level}
\end{figure}

Figure \ref{domain_model_second_refinement_level} represents the domain model associated with the second refinement level of the  diagram of Figure \ref{ERTMS_L3_Hybrid_ABZ_2018_vue5}. 
It refines the one associated with the first refinement level and introduces 
two concepts named \texttt{TTD} and \texttt{VSS}. The attributes \texttt{stateTTD} and \texttt{stateVSS} represent the states of the corresponding concepts. The predicates \texttt{p2.1}\textrm{..}\texttt{p2.8} define each TTD as a contiguous part of the track and each VSS as a contiguous part of a TTD. The predicates \texttt{p2.9} and \texttt{p2.10} are used to state that if a train is located on a TTD, then its state must be occupied: a train $tr \in TRAIN$ is located on   $ttd \in TTD$ if $front(tr) \in ttd$ (\texttt{p2.9}) or if tr is equipped with a TIMS ($tr \in dom(rear)$) and $(rear(tr)..front(tr)) \cap ttd \neq \emptyset$ (\texttt{p2.10}). Finally, the predicates \texttt{p2.11}\textrm{..}\texttt{p2.13} states that two different trains must be in disjoint parts of the track: for two trains \texttt{tr1} and \texttt{tr2}, if they are equipped with TIMS, then the track portions that  they occupy should just be disjointed (\texttt{p2.11}); if they are on the same TTD and one of them, ( \texttt{tr2}), is not equipped with a TIMS, then, the second, ( \texttt{tr1}),  must be equipped with a TIMS and  \texttt{tr2} must be in the rear of \texttt{tr1} (\texttt{p2.12}); if none of them is an INTEGER train, then they must be in two distincts TTDs (\texttt{p2.13}).
The predicates \texttt{p2.9} and \texttt{p2.10} are gluing invariants, linking the concrete variable \texttt{stateTTD} with the abstract variables \texttt{front} and \texttt{rear}.
The \textit{B System} specification raised from the translation of the second refinement level includes the result of the translation of the domain model of Figure \ref{domain_model_second_refinement_level}, two new events (\texttt{ComputeTrainMAFollowingTTDStates}, \texttt{ComputeTrainMAFol\-lowingVSSStates}), an extension of the event \texttt{MoveTrainFollowingItsMA} taking into account the new safety invariants and the theorems representing the proof obligations related to the usage of the \textit{OR} operator  between the first and second refinement levels. 
The specification below (Figure \ref{BSystem_second_refinement_level}) represents the new definition of \texttt{MoveTrainFollowingItsMA} and the theorems related to the refinement operator. The parameter \texttt{ttds} is introduced to capture the TTD requiring an update of their states because of the train movement (\texttt{grd4}, \texttt{grd5} and \texttt{act3}). Guards \texttt{grd6..grd9} ensure  that the train movement  will not lead to   the violation of the safety invariants \texttt{p2.11..p2.13} : \texttt{grd6} stands for \texttt{p2.11}; \texttt{grd7} and  \texttt{grd8} stand for \texttt{p2.12}; \texttt{grd9} stands for \texttt{p2.13}. 

\begin{figure}[!h]
\begin{scriptsize}

    \textcolor{keycolor}{\textbf{REFINEMENT}}
 ertms\_etcs\_case\_study\_ref\_2
 
     \textcolor{keycolor}{\textbf{REFINES}}
  ertms\_etcs\_case\_study\_ref\_1

\textcolor{keycolor}{\textbf{INVARIANT}}

     \hspace*{0.1in} \textbf{\textcolor{labelcolor}{\small{theorem}}} \textcolor{labelcolor}{\small{\texttt{s1}}}: $ComputeTrainMAFollowingTTDStates\_Guard  
      \Rightarrow  ComputeTrainMA\_Guard$
     
          \hspace*{0.1in} \textbf{\textcolor{labelcolor}{\small{theorem}}} \textcolor{labelcolor}{\small{\texttt{s2}}}: $ComputeTrainMAFollowingVSSStates\_Guard  
      \Rightarrow  ComputeTrainMA\_Guard$
     
     \hspace*{0.1in} \textbf{\textcolor{labelcolor}{\small{theorem}}} \textcolor{labelcolor}{\small{\texttt{s3}}}: $ComputeTrainMAFollowingTTDStates\_Post  
     \Rightarrow  ComputeTrainMA\_Post$
     
          \hspace*{0.1in} \textbf{\textcolor{labelcolor}{\small{theorem}}} \textcolor{labelcolor}{\small{\texttt{s4}}}: $ComputeTrainMAFollowingVSSStates\_Post 
      \Rightarrow  ComputeTrainMA\_Post$
     
       \hspace*{0.1in} \textbf{\textcolor{labelcolor}{\small{theorem}}} \textcolor{labelcolor}{\small{\texttt{s5}}}: $ComputeTrainMAFollowingTTDStates\_Post  
      \Rightarrow  not(ComputeTrainMAFollowingVSSStates\_Guard)$
     
          \hspace*{0.1in} \textbf{\textcolor{labelcolor}{\small{theorem}}} \textcolor{labelcolor}{\small{\texttt{s6}}}: $ComputeTrainMAFollowingVSSStates\_Post 
      \Rightarrow  not(ComputeTrainMAFollowingTTDStates\_Guard)$
      
\hspace*{-0.10in}\textcolor{keycolor}{\textbf{Event}}
MoveTrainFollowingItsMA  $\defi$

 \hspace*{0.1in} \textcolor{keycolor}{\textbf{any}} tr len ttds

 \hspace*{0.1in}\textcolor{keycolor}{\textbf{where}}

 \hspace*{0.2in} \textcolor{labelcolor}{\small{\texttt{grd1}}}: $tr \in{} connectedTrain^{-1}[\{TRUE\}]\binter{}dom(MA)$

 \hspace*{0.2in} \textcolor{labelcolor}{\small{\texttt{grd2}}}: $len \in{} \natn{}$

 \hspace*{0.2in} \textcolor{labelcolor}{\small{\texttt{grd3}}}: $front(tr)+len \in{} MA(tr)$
 
 \hspace*{0.2in} \textcolor{labelcolor}{\small{\texttt{grd4}}}: $ttds \subseteq{} stateTTD^{-1}[\{FREE\}]$
 
 \hspace*{0.2in} \textcolor{labelcolor}{\small{\texttt{grd5}}}: $\forall{}ttd\qdot{}(ttd\in{}stateTTD^{-1}[\{FREE\}] \land{} ((front(tr)+len\in{}ttd) \lor{} (tr\in{}dom(rear)\\ 
 \hspace*{0.6in} \land{}((rear(tr)+len\upto{}front(tr)+len)\binter{}ttd\neq{}\emptyset{})))\limp{}ttd\in{}ttds)$
 
 \hspace*{0.2in} \textcolor{labelcolor}{\small{\texttt{grd6}}}: $tr\in{}dom(rear)\limp{}(\forall{}tr1\qdot{}((tr1\in{}dom(rear)\land{}tr1\neq{}tr)\limp{}(rear(tr1)\upto{}front(tr1))\binter{}(rear(tr)+len\upto{}front(tr)+len)=\emptyset{}))$
 
 \hspace*{0.2in} \textcolor{labelcolor}{\small{\texttt{grd7}}}: $tr\in{}dom(rear)\limp{}(\forall{}tr1,ttd\qdot{}((tr1\in{}dom(front)\setminus{}dom(rear)\land{}tr1\neq{}tr \land{} ttd \in TTD \land{} front(tr1) \in ttd \\ 
 \hspace*{0.6in}  \land{} rear(tr)..front(tr)\cap ttd \neq \emptyset)\limp{}front(tr1) < rear(tr)+len))$
 
  \hspace*{0.2in} \textcolor{labelcolor}{\small{\texttt{grd8}}}: $tr\in{}dom(front)\setminus{}dom(rear)\limp{}(\forall{}tr1,ttd\qdot{}((tr1\in{}dom(rear)\land{}tr1\neq{}tr \land{} ttd \in TTD \land{} front(tr)+len \in ttd \\ 
 \hspace*{0.6in}  \land{} rear(tr1)..front(tr1)\cap ttd \neq \emptyset)\limp{}front(tr)+len < rear(tr1)))$
  
   \hspace*{0.2in} \textcolor{labelcolor}{\small{\texttt{grd9}}}:$tr\in{}dom(front)\setminus{}dom(rear)\limp{}(\forall{}tr1,ttd\qdot{}((tr1\in{}dom(front)\setminus{}dom(rear)\land{}tr1\neq{}tr\land{}ttd\in{}TTD\\
   \hspace*{0.6in} \land{}front(tr1)\in{}ttd)\limp{}front(tr)+len \notin{} ttd))$

 \hspace*{0.1in} \textcolor{keycolor}{\textbf{then}}

 \hspace*{0.2in} \textcolor{labelcolor}{\small{\texttt{act1}}}: $front(tr) \bcmeq{} front(tr)+len$
 
  \hspace*{0.2in} \textcolor{labelcolor}{\small{\texttt{act2}}}: $rear \bcmeq{}  (\{TRUE\mapsto{}rear\ovl{}\{tr\mapsto{}rear(tr)+len\}, FALSE\mapsto{}rear\})(bool(tr\in{}dom(rear)))$
  
   \hspace*{0.2in} \textcolor{labelcolor}{\small{\texttt{act3}}}: $stateTTD \bcmeq{} stateTTD \ovl{} (ttds \cprod{} \{OCCUPIED\})$

 \hspace*{0.1in} \textcolor{keycolor}{\textbf{END}}\\

\textcolor{keycolor}{\textbf{END}}

\end{scriptsize}
 
\caption{\label{BSystem_second_refinement_level} B System specification of the second refinement level of the   diagram of Figure \ref{ERTMS_L3_Hybrid_ABZ_2018_vue5}}
\end{figure}

\subsection{The Fifth Refinement Level}
For the fifth refinement level, corresponding to the first refinement level of the goal diagram of Figure \ref{ERTMS_L3_Hybrid_ABZ_2018_vue2}, the B System specification introduces four events raised from the translation of the goals and five theorems representing the proof obligations related to the usage of the \textit{AND} operator between the fourth and the fifth refinement levels. These theorems are : 

\begin{scriptsize}
\noindent\textbf{\textcolor{labelcolor}{\small{theorem}}} \textcolor{labelcolor}{\small{\texttt{s1}}}: $ComputeStatesOfVSSinUnknownState\_Guard  
      \Rightarrow  ComputeVSSStatesFollowingTTDStates\_Guard$
      
\noindent\textbf{\textcolor{labelcolor}{\small{theorem}}} \textcolor{labelcolor}{\small{\texttt{s2}}}: $ComputeStatesOfVSSinOccupiedState\_Guard  
      \Rightarrow  \\ComputeVSSStatesFollowingTTDStates\_Guard$
      
\noindent\textbf{\textcolor{labelcolor}{\small{theorem}}} \textcolor{labelcolor}{\small{\texttt{s3}}}: $ComputeStatesOfVSSinAmbiguousState\_Guard  
      \Rightarrow  ComputeVSSStatesFollowingTTDStates\_Guard$
      
\noindent\textbf{\textcolor{labelcolor}{\small{theorem}}} \textcolor{labelcolor}{\small{\texttt{s4}}}: $ComputeStatesOfVSSinFreeState\_Guard  
      \Rightarrow  ComputeVSSStatesFollowingTTDStates\_Guard$
      
\noindent\textbf{\textcolor{labelcolor}{\small{theorem}}} \textcolor{labelcolor}{\small{\texttt{s5}}}: $ComputeStatesOfVSSinUnknownState\_Post \wedge ComputeStatesOfVSSinOccupied\-State\_Post \wedge\\ ComputeStatesOfVSSinAmbiguousState\_Post \wedge ComputeStatesOfVSSinFreeState\_Post
      \Rightarrow  ComputeVSSStatesFollowingTTDStates\_Post$                                    

\end{scriptsize}

\subsection{Discussion}
This case study allowed us to benefit from the advantages  of a high-level modeling approach within the framework of the formal  specification of the hybrid ERTMS/ETCS level 3 requirements : 
decoupling between formal specification handling difficulties and system modeling; better reusability and readability of models; strong traceability between the system formal specification  and the goal model, which is an abstraction of the case study description. 
Using the SysML/KAOS method, we have quickly build the refinement hierarchy of the system and we have determined  and formally expressed the safety invariants. The approach bridges the gap between the system textual description and its formal specification. 
Its use has made it possible to better present the specifications, excluding predicates, to stakeholders and to better delineate the system boundaries. 
Using Rodin  \cite{DBLP:conf/rodin/2006}, we have formally verified and validated the safety invariants and the  goal diagram refinement hierarchy. Through proB,  we have animated the formal model. The full specification can be found in \cite{sysml_kaos_domain_models_ertms_link}.
 One conclusion of our work is that the description of the standard, as it exists in the documents \cite{hoang_hybrid_2018,ertms_etcs_principles,ertms_l3_game_changer},    does not guarantee the absence of train collisions. Indeed, since the standard allows the movement of unconnected trains  on the track, 
 nothing is specified to guarantee that an unconnected train will not hit another train (connected or not).
The animation of the specification allows the observation of  these states.
 The only guarantee that the safety invariants expressed in \cite{hoang_hybrid_2018,ertms_etcs_principles,ertms_l3_game_changer}  bring is that a connected train will never hit another train.

We have also specified in a companion paper \cite{amel_marc_classic_approach_case_study_abz18} the case study using plain Event-B, in the traditional style. Two distinct specifiers (first author of \cite{amel_marc_classic_approach_case_study_abz18} and first author of this paper) wrote each specification without interacting with each other during specification construction. Critical reviewing by the team was then conducted after  the specifications were built.
The  specification in \cite{amel_marc_classic_approach_case_study_abz18}  includes four refinement levels. The TTDs and trains are introduced in the root level and the VSSs are introduced in the second refinement level, as refinements of TTDs. The MAs and VSS states are introduced in the third refinement level (M3), for train movement supervision.
A strategy is proposed to prove  the determinism of  the transitions of VSS states.
The state variables of \cite{amel_marc_classic_approach_case_study_abz18} are partitioned into environment variables and controller variables, and similarly for events. Environment events only modify environment   variables.  Controller events read environment variables and update controller variables.
In this paper, we only model controller events; state variables represent the controller view of the environment.
The execution ordering and the refinement strategy are enforced using proof obligations expressed as theorems, whereas in \cite{amel_marc_classic_approach_case_study_abz18}  there is no proof about these aspects.
In \cite{amel_marc_classic_approach_case_study_abz18}, the safety properties are introduced in the last refinement level; here, we introduce them in the first (predicate \texttt{p1.4}) and second (predicates \texttt{p2.9}\textrm{..}\texttt{p2.13}) refinements.
In \cite{amel_marc_classic_approach_case_study_abz18}, all trains equipped with ERTMS are equipped with TIMS, so they broadcast their front and rear; here, we consider ERTMS trains with or without TIMS, so a ERTMS train may or may not broadcast its rear.
The SysML/KAOS method   makes it possible to trace the source and justify the need for each formal component and its contents, in relation with the SysML/KAOS goal and domain models.

The expression of theorems representing proof obligations associated to SysML/KAOS refinement operators was difficult because there is no way in Rodin to designate the guard and the post condition of an event within predicates. Table \ref{tableau_recapitulatif_specification_eventb} summarises the key characteristics related to the formal specification. It seemed  that the provers have a lot of trouble with  data ranges such as $p..q$ and with conditional actions such as $rear \bcmeq{}  (\{TRUE\mapsto{}rear\ovl{}\{tr\mapsto{}rear(tr)+len\}, FALSE\mapsto{}rear\})(bool(tr\in{}dom(rear)))$ defined in the component \texttt{ertms\_etcs\_case\_study} to simulate an \textit{if-then-else} in order to avoid the definition of a second event.

\begin{center}
\begin{longtable}{|p{5cm}|p{0.6cm}|p{0.6cm}|p{0.6cm}|p{0.6cm}|p{0.6cm}|p{0.6cm}|p{0.6cm}|}
\caption{\label{tableau_recapitulatif_specification_eventb} Key characteristics related to  the formal specification}\\
\hline
Refinement level & \textbf{L0} & \textbf{L1} & \textbf{L2} & \textbf{L3} & \textbf{L4} & \textbf{L5}& \textbf{L6}\\
\hline
\textbf{Invariants} & \textbf{4} & \textbf{11}& \textbf{13}& \textbf{4}& \textbf{6}& \textbf{5}& \textbf{9}   \\
\hline
\textbf{Proof Obligations (PO)} & \textbf{20} & \textbf{40}& \textbf{50}& \textbf{13}& \textbf{5}& \textbf{5}& \textbf{14}    \\
\hline
\textbf{Automatically Discharged POs} & \textbf{17} & \textbf{30}& \textbf{30}& \textbf{11}& \textbf{0}& \textbf{0}& \textbf{4}   \\
\hline
\textbf{Interactively Discharged POs} & \textbf{3} & \textbf{5}& \textbf{20}& \textbf{2}& \textbf{5}& \textbf{5} & \textbf{10}   \\
\hline
\end{longtable}
\end{center}

\section*{Acknowledgment}
This work is carried out within the framework of the  \textit{FORMOSE} project \cite{anr_formose_reference_link} funded by the French National Research Agency (ANR).
It  is also partly supported by the Natural Sciences and Engineering Research Council of Canada  (NSERC).



\bibliographystyle{IEEEtran}
\bibliography{references}
%



\end{document}